\documentclass[12pt]{article}

\pdfoutput=1

\usepackage[top=80pt,bottom=85pt,left=85pt,right=85pt]{geometry}
\usepackage[utf8]{inputenc}
\usepackage{amssymb,amsmath,amsthm,mathtools}
\usepackage{bbm}
\usepackage{graphicx}
\usepackage{xcolor}
\usepackage{booktabs}
\usepackage{array}
\usepackage{makecell}
\usepackage{caption}
\usepackage{cite}
\usepackage{tikz}
\usetikzlibrary{arrows.meta}
\usetikzlibrary{positioning}

\usepackage[debug,pageanchor=false]{hyperref}
\definecolor{link}{rgb}{.8,.15,.1}
\hypersetup{colorlinks=true,linkcolor=link,citecolor=link,urlcolor=link,linktocpage}
\usepackage{cleveref}

\allowdisplaybreaks
\renewcommand{\hat}{\widehat}

\theoremstyle{remark}

\tikzset{
  sol/.style     = {circle, draw=black!55, fill=black!12,
                    minimum size=5.5pt, inner sep=0pt},
  Sedge/.style   = {blue!70!black, thick, <->, >={Stealth[length=4pt]},
                    shorten <=2pt, shorten >=2pt},
  Tedge/.style   = {red!75!black,  thick, <->, >={Stealth[length=4pt]},
                    shorten <=2pt, shorten >=2pt},
  Sloop/.style   = {blue!70!black, thick, ->, >={Stealth[length=4pt]},
                    shorten <=1pt, shorten >=1pt},
  Tloop/.style   = {red!75!black,  thick, ->, >={Stealth[length=4pt]},
                    shorten <=1pt, shorten >=1pt},
  Tdir/.style    = {red!75!black,  thick, ->, >={Stealth[length=4pt]},
                    shorten <=2pt, shorten >=2pt},
  sollbl/.style  = {font=\small,      inner sep=2pt},
  edgelbl/.style = {font=\scriptsize, inner sep=1.5pt},
}

\newcommand{\alg}{\mathfrak{g}}
\newcommand{\id}{\mathbbm{1}}
\newcommand{\rr}{\mathbb{R}}
\newcommand{\cc}{\mathbb{C}}
\newcommand{\zz}{\mathbb{Z}}

\DeclareMathOperator{\rk}{rk}

\DeclareMathOperator{\SU}{SU}
\DeclareMathOperator{\SO}{SO}

\DeclareMathOperator{\sn}{sn}
\DeclareMathOperator{\cn}{cn}
\DeclareMathOperator{\dn}{dn}

\makeatletter
\@addtoreset{equation}{section}
\makeatother

\begin{document}
\newtagform{normalsize}[\normalsize]{\normalsize(}{\normalsize)}
\usetagform{normalsize} %

\begin{titlepage}

	\begin{center}

		\vskip .3in \noindent

		{\LARGE \textbf{The $\mathcal{N}=4$ Bethe Ansatz beyond $\mathrm{SU}(N)$}}
		\bigskip

		%{\large \textsc{part i}}

		\bigskip

		Marco Fazzi and Kuba Krawczyk

		%\bigskip

		\bigskip
		{\small

			School of Mathematical and Physical Sciences, University of Sheffield, Sheffield S3 7RH, UK \\
		}

		\smallskip
		{\small \tt \href{mailto:m.fazzi@sheffield.ac.uk}{m.fazzi@sheffield.ac.uk}  \hspace{.5cm} \href{mailto:kkrawczyk1@sheffield.ac.uk}{kkrawczyk1@sheffield.ac.uk}}

		\bigskip
		\bigskip
		{\bf Abstract }
		\vskip .1in
	\end{center}

	\noindent We solve the Bethe Ansatz equations (BAEs) to evaluate the superconformal index
of four-dimensional $\mathcal{N}=4$ super-Yang--Mills with rank-two gauge
algebra, at equal angular momentum fugacities $p=q$. The solutions for $A_2$ are known, and those for $D_2\cong A_1\oplus A_1$ follow
readily from the (equally known) $A_1$ ones. The first genuinely new case is
$B_2\cong C_2$, for which we obtain the complete solution set in closed form;
for the exceptional case $G_2$ our results are numerical, with a single fully
rational solution obtained in closed form. To
the best of our knowledge, this is the first time any solution, analytic or
numerical, has been found in  non-$A$-type gauge algebra (for $\mathcal{N}=4$ or any other $\mathcal{N}=1$ theory). Along the way we
uncover several phenomena absent in type $A$: isolated Weyl-fixed solutions
contributing nontrivially to the index; isolated solutions in which at most half of
the holonomies have rational coefficients (in contrast with the $A$-type fully rational
Hong--Liu family); and solutions whose $|\omega|\to0$
limit (with $p=q\eqqcolon e^{2\pi i\omega}$) evades assumptions standardly made in the
Cardy-like limit literature, landing on saddles that the usual analysis does not
capture, for all types $BC\!D$. We also clarify the Bethe origin of the finite logarithmic correction
to the Cardy expansion: it arises from the orbit size of a
BAE solution under the gauge symmetries of the equations, rather than from the
one-form center symmetry of the theory, to which the index is insensitive.

	\vfill

%	\begin{flushright}
%
%	\end{flushright}
%	\eject

\end{titlepage}

\tableofcontents

%%%%%%%%%%%%%%%%%%%%%%%%
\section{Introduction}
\label{sec:intro}
%%%%%%%%%%%%%%%%%%%%%%%%

Black holes are among the most mysterious objects in the universe. Accounting for their macroscopic entropy
\cite{Bekenstein:1973ur,Hawking:1975vcx} is one of the most profound
outstanding questions in theoretical physics.\footnote{See e.g. \cite{Witten:2024upt} for a recent
introduction to black hole thermodynamics.} For this reason,
any attempt at deriving the Bekenstein--Hawking entropy formula from first
(i.e., microscopic) principles is welcome.

In the case of supersymmetric black holes which can be embedded into string
theory, the situation is much more favorable (starting with the celebrated
asymptotically-flat result of \cite{Strominger:1996sh}). Even more so for
supersymmetric asymptotically anti-de Sitter (AdS) black holes, i.e.\ those
which embed into an AdS vacuum \cite{Cvetic:1999xp}. In cases where the AdS/CFT
correspondence is at play, one can hope to count the quantum microstates of the
black hole making up its entropy by enumerating protected states in the dual
CFT. These protected states can be counted by a supersymmetric partition
function known as the superconformal index.

This approach, pioneered in \cite{Kinney:2005ej}, has seen a great revival in
more recent years, starting with the case of (dyonic) AdS$_4$ black holes
\cite{Benini:2015eyy,Benini:2016rke} and then electrically charged, rotating
AdS$_5$ black holes \cite{Cabo-Bizet:2018ehj,Choi:2018hmj,Benini:2018ywd}, first
constructed in
\cite{Gutowski:2004ez,Gutowski:2004yv,Chong:2005da,Chong:2005hr,Kunduri:2006ek}.
(Other dimensions and horizon topologies have been studied as well; for a review
circa 2020, see \cite{Zaffaroni:2019dhb}.) The Bekenstein--Hawking area term of the black hole entropy can be recovered
from the Legendre transform of the index in at least three complementary ways.
The first is to take a so-called Cardy-like limit of the matrix-model
representation of the index, corresponding to large black hole charges \cite{Choi:2018hmj,Honda:2019cio,ArabiArdehali:2019tdm,Cabo-Bizet:2019osg,Kim:2019yrz}.\footnote{Two
refinements of this route are worth recording. A direct saddle-point analysis of
the matrix integral shows the index to be dominated by complex saddles labeled
by coprime pairs $(m,n)$, corresponding to eigenvalue strings winding $(m,n)$
times around the two cycles of a torus
\cite{Cabo-Bizet:2019eaf,Cabo-Bizet:2020ewf}; the black hole saddle is the
$(1,0)$ one, and the construction reproduces the Cardy-like limit as a special
case (see \cite{BenettiGenolini:2023rkq} for the analogous story in AdS$_4$).
Separately, the $\mathrm{SL}(3,\mathbb{Z})$ modular properties of the elliptic
gamma function \cite{Felder:1999vf} allow the index to be computed at rational
values of the angular chemical potentials, $\tau\to n_1-\tfrac1m$ and
$\sigma\to n_2-\tfrac1m$ (see section~\ref{sub:index} for the definitions), rather than at the origin
\cite{Goldstein:2020yvj,Jejjala:2021hlt}; the ordinary limit is the case
$m=n_1=n_2=1$, and the resulting entropy differs from it by an overall factor
$1/m$, interpreted as a $\mathbb{Z}_m$ quotient of the Hopf fiber of the
$S^3$.} The second is to take a large-$N$ limit of a
finite-sum representation of the index \cite{Benini:2018mlo} over certain
solutions \cite{Benini:2018ywd} to a set of auxiliary transcendental equations,
dubbed Bethe Ansatz equations (BAEs) in \cite{Hosseini:2016cyf,Benini:2018mlo}.
(This approach is rooted in the Bethe/Gauge correspondence
\cite{Nekrasov:2009ui,Nekrasov:2009uh}, which identifies the vacuum equations of
supersymmetric gauge theories with the BAEs of integrable spin chains, and whose
four-dimensional incarnation was developed in \cite{Closset:2017bse}.) The third
is the so-called giant-graviton expansion
\cite{Imamura:2021ytr,Gaiotto:2021xce,Murthy:2022ien,Lee:2022vig,Arai:2019aou,Arai:2019wgv,Arai:2019xmp,Arai:2020uwd}, which reorganizes the
finite-$N$ index as a series of corrections to its large-$N$ limit, and from
which the entropy of large black holes has likewise been extracted
\cite{Beccaria:2023hip}.

In this paper we focus mostly on the Bethe Ansatz approach, returning to the
Cardy-like limit in section~\ref{sec:cardy}. For the prototypical
AdS/CFT pair, AdS$_5\times S^5$ with $N$ units of $F_5$ flux through the $S^5$
dual to four-dimensional (4D) $\mathcal{N}=4$ super-Yang--Mills (SYM) with gauge
algebra $\mathfrak{su}(N)$, the area term is shown to be proportional to the $a$
anomaly of the SYM theory evaluated on a set of matter chemical potentials
(rather than their R-charges), and in turn this $a$ is reproduced at large $N$
by a finite set of isolated solutions to the $\mathcal{N}=4$ BAEs
\cite{Benini:2018ywd}. This resolved a puzzle that had stood for over a decade:
the index had been computed at large $N$ already in \cite{Kinney:2005ej}, where
it was found not to grow rapidly enough to account for the entropy, the required
growth being obscured at real fugacities by boson-fermion cancellations and becoming manifest
only once they are taken complex. The correspondence is moreover finer than a
match of the total entropy:\footnote{The agreement extends to subleading order in the semiclassical
expansion, i.e.\ beyond the area term. \cite{GonzalezLezcano:2020yeb}
compares the outputs of the Cardy-like limit and of the Bethe Ansatz approach
(at $p=q$) beyond leading order at large $N$ for $\mathfrak{su}(N)$ SYM,
uncovering subdominant contributions and a finite $\log N$ correction on which
the two agree; \cite{Amariti:2020jyx} exploits the Cardy perspective to do the
same for $BC\!D$ SYM, and \cite{Amariti:2021ubd} extends the analysis of
subleading and logarithmic corrections to any 4D $\mathcal{N}=1$ theory with
gauge algebra of $ABC\!D$ type, whereas \cite{Cassani:2021fyv,%
ArabiArdehali:2021nsx} focus on $\mathcal{N}=1$ $\mathfrak{su}(N)$ theories at
$p\neq q$. On the gravity side, the corresponding corrections to the
thermodynamics of AdS$_5$ black holes have been obtained from higher-derivative
supergravity in
\cite{Bobev:2022bjm,Cassani:2022lrk,Cassani:2023vsa,Cassani:2024tvk}.} individual BAE solutions map one-to-one onto complex
Euclidean black hole solutions, the matching extending to nonperturbative $1/N$
corrections from wrapped D3-branes \cite{Aharony:2021zkr}. An analogous check has been carried
out for a large class of toric and holographic quiver theories
\cite{Lanir:2019abx,Cabo-Bizet:2020nkr,Benini:2020gjh}.

However, two problems emerge with the Bethe Ansatz formula. First, the
finite-sum formula of \cite{Benini:2018mlo,Benini:2018ywd} explicitly assumes
that only isolated solutions to the BAEs contribute to the index. It has been
shown in subsequent work
\cite{GonzalezLezcano:2021nzk,Benini:2021ano,ArabiArdehali:2019orz,Agarwal:2020zwm}
that, at least in the $\mathfrak{su}(N)$ SYM case, while at large $N$ isolated
solutions indeed suffice to reconstruct the black-hole entropy area term, at
finite $N$ there sometimes appear continuous families of solutions living on a
complex torus, whose contribution to the index is generically nonvanishing and
whose role remains incompletely understood.\footnote{It is not presently known how to compute this contribution in general, though
see \cite{Cabo-Bizet:2024kfe} for a computation in $\mathfrak{u}(3)$ and
$\mathfrak{su}(3)$ SYM. That it is nonvanishing can be established by indirect
means; see e.g.\ \cite[p.~16]{Benini:2021ano}. Beyond SYM, a continuum of
solutions was inferred to exist for 4D $\mathcal{N}=1$ $\mathfrak{usp}(2N)$
SQCD with $N+1$ fundamental chirals, for any $N$, in \cite{Closset:2017bse}. We are not aware of other examples.} A complementary, structural perspective on
such finite-$N$ effects is provided by the notion of fortuity
\cite{Chang:2024zqi}, which classifies BPS states as monotone, forming infinite
sequences in the rank $N$, or fortuitous, existing only within finite windows of
consecutive ranks, the latter being conjectured to account for typical black
hole microstates; this is a classification of the BPS cohomology rather than a
method for evaluating the index, and we are not aware of any established
relation to the continuous families discussed here.

Second, and more fundamentally, essentially all existing work on the BAEs (both
for $\mathcal{N}=4$ SYM and for $\mathcal{N}=1$ quiver theories) has been
restricted to gauge algebras of type $A$, i.e.\ $\mathfrak{su}(N)$. To the best
of our knowledge, no solution (analytic or numerical) to the BAEs has
\emph{ever} been found for any non-$A$-type gauge algebra. (Moreover, we
emphasize that the literature has implicitly focused on the role of the gauge
algebra in the definition of the index, as opposed to the gauge group. We will
make explicit in what follows when and why the distinction becomes important.)

A concrete incentive comes from holography. The duals of the $BC\!D$-type theories
are known \cite{Witten:1998xy,Hanany:2000fq}: 4D $\mathcal{N}=4$
$\mathfrak{so}(N)$ and $\mathfrak{usp}(2N)$ SYM live on D3-branes atop an
orientifold three-plane and are dual to Type IIB string theory on the
$\mathbb{Z}_2$ orientifold $\text{AdS}_5\times\mathbb{RP}^5$ of
$\text{AdS}_5\times S^5$, the four types of O3-plane realizing the $BC\!D$
series.\footnote{The four O3-planes are distinguished by the $\mathbb{Z}_2$
discrete torsions of the NS-NS and R-R fluxes on $\mathbb{RP}^5$:
$\mathrm{O3}^-$ yields $\mathfrak{so}(2N)$, $\widetilde{\mathrm{O3}}{}^-$ yields
$\mathfrak{so}(2N{+}1)$, and $\mathrm{O3}^+$, $\widetilde{\mathrm{O3}}{}^+$ both
yield $\mathfrak{usp}(2N)$ (the last with a shifted $\theta$-angle). The
$\mathfrak{so}(2N{+}1)$ case requires a half-D3-brane stuck on the plane, which
shifts its R-R torsion and D3-brane charge; the $\widetilde{\mathrm{O3}}{}^-$ and
$\mathrm{O3}^+$ planes (hence $\mathfrak{so}(2N{+}1)$ and $\mathfrak{usp}(2N)$)
are exchanged by $\mathrm{SL}(2,\mathbb{Z})$
\cite{Witten:1998xy,Hanany:2000fq,Gaiotto:2008ak}.} While the superconformal
index of these theories has been evaluated in the Cardy-like limit,
first for a certain dominant saddle of the matrix-model representation
\cite{Honda:2019cio,Cabo-Bizet:2019osg}, and subsequently for a much larger set
of saddles, including subleading corrections and a check of S-duality
\cite{Amariti:2020jyx,Amariti:2021ubd,Amariti:2023rci},\footnote{We note that the
saddle structure in the orthosymplectic case differs from the $A$-type one in
ways that matter for which configurations dominate in different regions of the
chemical-potential parameter space; see \cite{Amariti:2020jyx,Amariti:2023rci} and section~\ref{sec:cardy}.}
and via a giant-graviton expansion \cite{Fujiwara:2023bdc}, both the large-$N$
Bethe Ansatz analysis and its gravitational counterpart are still missing: the
supersymmetric black holes embedded in $\text{AdS}_5\times\mathbb{RP}^5$, whose
microstates the index ought to enumerate, have to our knowledge never been
studied in these vacua. Since the orientifold acts only on the internal $S^5$,
the $\text{AdS}_5$ black holes of \cite{Gutowski:2004ez,Gutowski:2004yv} are
expected to persist unchanged;\footnote{The orientifold quotients by the central
$\mathbb{Z}_2\subset \SO(6)_R$ generated by $-\id$ (the antipodal map on $S^5$),
which acts trivially on the metric and on the $\mathbf{15}$ of graviphotons; the
minimal (and STU) gauged supergravity supporting these black holes is therefore
the same truncation as on $S^5$, with only the massive Kaluza--Klein tower and
the flux quantization over $\mathbb{RP}^5$ affected.} what remains open is their
microscopic counting in the $BC\!D$ index (with its orientifold-specific
finite-$N$ data and the
$\mathfrak{so}(2N{+}1)\leftrightarrow\mathfrak{usp}(2N)$ S-duality), for which
solving the BAEs beyond type $A$ is the prerequisite.

We set out here to address a more modest version of this problem. Two caveats
temper the holographic motivation at low rank. First, the duals above are
large-$N$ statements. The first rank at which we encounter a non-$A$-type gauge
algebra is two, which is far from being large (i.e., there is no dual
semiclassical black hole), so the incentive is ``structural'': establishing
whether the BAEs admit solutions at all beyond type $A$, and of what nature,
rather than a direct entropy match. Second, as we will see, the analytic
$B_2\cong C_2$ solutions show \emph{no} obvious route to large $N$, contrary to
naive expectations: unlike for $\mathfrak{su}(N)$, where the Hong--Liu solutions
\cite{Hong:2018viz} (labeled by integers $\{m,n,r\}$ with $mn=N$ and
$r=0,\ldots,n-1$) form a family whose members compete at large $N$, with the
dominant one varying across domains of analyticity separated by Stokes lines
\cite{Benini:2018ywd} (for black holes of large charge it is the ``basic''
solution of \cite{Hosseini:2016cyf}, i.e.\ $\{1,N,0\}$ in the above notation), the rank-two orthosymplectic solutions do not visibly belong to any such
family. Whether the same holds in type $D$ (i.e., $\mathfrak{so}(2N)$ SYM) we do
not know, though there is reason for optimism: $D$ is simply-laced, hence closer
in character to $A$ than to $BC$, and it is the only remaining family for which
the question is well-posed at large $N$, the exceptional algebras having fixed
rank and therefore no semiclassical gravity dual against which to compare.
Settling it requires access to higher rank, which we pursue via a polynomial
method in a companion paper \cite{polypaper}. Here we instead
present the first nontrivial solutions to the $\mathcal{N}=4$ BAEs beyond type
$A$. For $B_2\cong C_2$ we obtain, in closed form, the \emph{complete} solution
set; the remaining rank-two classical cases are already settled, $A_2$ being
known \cite{Benini:2021ano,GonzalezLezcano:2021nzk} and $D_2\cong A_1\oplus A_1$
following rather straightforwardly from $A_1$ \cite{ArabiArdehali:2019orz,Benini:2021ano,GonzalezLezcano:2021nzk}. (Note that, even in type $A$, at
general $N$ the Hong--Liu family is conjectured but not known to exhaust the
isolated solutions \cite{Benini:2021ano}, completeness having been established
only for $N=2,3$, as said.) For the exceptional case $G_2$ our results are
numerical: we find $29$ contributing Weyl orbits of isolated solutions, of which
exactly one is fully rational and Weyl-fixed, and can be obtained in closed
form. We make no claim of completeness there, and in particular cannot exclude
continuous families.

The structure of the paper is as follows. In section~\ref{sec:baes} we revisit
the classic derivation of the BAEs in \cite{Benini:2018mlo}, making explicit the
choices of global form of the gauge group and primitive basis of the gauge algebra that are usually left implicit, and
discussing the symmetries of the equations and the resulting notion of physical
solution. In section~\ref{sec:su2su3} we explain how the ``reduced'' BAEs in
type $A$ fit in this more general formalism, and present the known solutions for
$A_1$ (the only rank-one simple, equivalently semisimple, algebra,
$A_1\cong B_1\cong C_1$) and $A_2$ (including the continuous family) of
\cite{ArabiArdehali:2019orz,Benini:2021ano,GonzalezLezcano:2021nzk}. In section~\ref{sec:D2} we whet
the reader's appetite with the first non-simple case, $D_2\cong A_1\oplus A_1$,
whose solutions can be obtained from those of $A_1$. In section~\ref{sec:B2} we
move to $B_2\cong C_2$ SYM, exhibiting all the new solutions, and in
section~\ref{sec:G2} we treat the remaining rank-two semisimple algebra, the
exceptional $G_2$. Section~\ref{sec:cardy} is devoted to revisiting the Cardy-like limit of the index. We first show that
the BAEs, written as finite-difference equations, reduce to the usual
saddle-point equations as $|\omega|\to0$, so that the Cardy method may be
regarded as an infinitesimal version of the Bethe Ansatz one; the leading terms
of the two expansions agree, and the resulting map from BAE solutions to Cardy
saddles is non-injective, its fibers labeled by the branch integers of the
logarithm. We then track our orthosymplectic solutions numerically along
$|\omega|\to0$. Having access to the full set of solutions, rather than to a
family posited in advance, we find that their images populate both the dominant
and the subleading saddles of the known classification
\cite{Amariti:2020jyx,Amariti:2023rci}, and that some of them land on
configurations outside the reach of the standard Ansatz $a+b\omega$, with $a,b\in\rr$, for which the usual
asymptotic expansion of the elliptic gamma function does not apply. We conclude
with future directions in section~\ref{sec:conc}.

%%%%%%%%%%%%%%%%%%%%%%%%%%%%%%%%%%%%%%%%%%%%%%%%%%%%%%%%%%%%%%%%%%%%%%%%
\section{Revisiting the Bethe Ansatz Equations}
\label{sec:baes}
%%%%%%%%%%%%%%%%%%%%%%%%%%%%%%%%%%%%%%%%%%%%%%%%%%%%%%%%%%%%%%%%%%%%%%%%

\subsection{The superconformal index}
\label{sub:index}

For a 4D $\mathcal{N}=1$ gauge theory, one can define a supersymmetric index 
(or supersymmetric partition function on $S^1\times S^3$) that counts, with 
sign $(-1)^F$, supersymmetric ground states on $S^3$, i.e.\ states in the 
kernel of $\{\mathcal{Q},\mathcal{Q}^\dagger\}$ for a chosen complex 
supercharge $\mathcal{Q}$, without reference to conformal symmetry. Whenever 
the gauge theory flows to a CFT in the infrared (as is the case for 
$\mathcal{N}=4$ SYM), this index coincides with the superconformal 
index of the IR SCFT, which counts operators in short ($\tfrac{1}{4}$-BPS) 
representations of the superconformal algebra annihilated by $\mathcal{Q}$. 
This coincidence follows from the fact that the index receives contributions 
only from $\mathcal{Q}$-closed states and is therefore invariant under renormalization group flow.

Henceforth we specialize to $\mathcal{N}=4$ SYM, which is a CFT for any gauge
algebra $\alg$. In $\mathcal{N}=1$ language, $\mathcal{N}=4$ SYM with
gauge algebra $\alg$ consists of a vector multiplet and three chiral
multiplets $X_1,X_2,X_3$, all in the adjoint representation of $\alg$,
with superpotential $W=\mathrm{Tr}\,X_1[X_2,X_3]$. As we will see momentarily, the matter content and $W$
depend only on $\alg$; the global form of the gauge group $G$, with
$\operatorname{Lie}G=\alg$, enters only when we specify the holonomies
$u$ summed over in the index, and it is there that the algebra v. group distinction becomes
operative.

The R-symmetry is $\mathrm{SU}(4)_R$; we select Cartan generators
$R_1,R_2,R_3$ such that $R_a$ assigns charge $2$ to $X_a$ and charge $0$ to the
remaining two chiral multiplets, with R-charges satisfying $r_1+r_2+r_3=2$ (as
required by marginality of $W$). States are further labeled by angular momenta
$J_1,J_2$ generating rotations in two orthogonal planes of $\mathbb{R}^4$.

The superconformal index of $\mathcal{N}=4$ SYM is defined as \cite{Kinney:2005ej,Romelsberger:2005eg,Romelsberger:2007ec}
\begin{equation}
\mathcal{I}(p,q,y_1,y_2) = \mathrm{Tr}_{S^3}\,(-1)^F\, 
e^{-\beta\{\mathcal{Q},\mathcal{Q}^\dagger\}}\, 
p^{J_1+\frac{1}{2}R_3}\, q^{J_2+\frac{1}{2}R_3}\, 
y_1^{\frac{1}{2}(R_1-R_3)}\, y_2^{\frac{1}{2}(R_2-R_3)}\ ,
\label{eq:SCI_def}
\end{equation}
where $p,q$ are fugacities for the two angular momenta (dressed by $R_3$), 
$y_1,y_2$ are flavor fugacities for two Cartan generators of the 
$\mathrm{SU}(3)\subset\mathrm{SU}(4)_R$ that commutes with $\mathcal{Q}$, and 
the trace is over the Hilbert space on $S^3$, with $|p|,|q|<1$ required for 
convergence. By standard arguments \cite{Witten:1981nf,Witten:1982df}, only states 
annihilated by both $\mathcal{Q}$ and $\mathcal{Q}^\dagger$ contribute, so 
the index is independent of the $S^1$ circumference $\beta$. It is convenient to introduce chemical 
potentials $\sigma,\tau,\Delta_1,\Delta_2$ via $p=:e^{2\pi i\sigma}$, 
$q=:e^{2\pi i\tau}$, $y_a=:e^{2\pi i\Delta_a}$, together with an auxiliary 
fugacity $y_3=:e^{2\pi i\Delta_3}$ defined by the constraint
\begin{equation}
\Delta_1+\Delta_2+\Delta_3 - \sigma - \tau \;\in\; \mathbb{Z}\ ,
\label{eq:Delta_constraint}
\end{equation}
which restores the manifest permutation symmetry among $a=1,2,3$ (the Weyl 
group of the $\mathrm{SU}(3)\subset\mathrm{SU}(4)_R$ that commutes with 
$\mathcal{Q}$) \cite{Benini:2021ano}.

For later convenience we record the relation to the original parameterization 
of \cite{Kinney:2005ej}, in terms of fugacities $(t,y,v_1,v_2)$:
\begin{equation}
p =: t^3 y\ ,\quad q =: t^3/y\ ,\quad 
y_1 =: t^2 v_1\ ,\quad y_2 =: t^2 v_2\ ,\quad y_3 =: t^2/(v_1 v_2)\ ,
\label{eq:fugacity_dictionary}
\end{equation}
so that the auxiliary fugacity is fixed by $y_1 y_2 y_3 = pq$ --- or 
equivalently, at the level of chemical potentials, by the constraint 
\eqref{eq:Delta_constraint}.

The index is also independent of the exactly marginal gauge coupling
$\tau_\text{YM}=\tfrac{\theta}{2\pi}+\tfrac{4\pi i}{g^2}$, since the latter is a $\mathcal{Q}$-exact
(D-term) deformation. It can be computed exactly via supersymmetric localization
\cite{Romelsberger:2005eg,Dolan:2008qi} as a contour integral over the maximal
torus $\mathbb{T}^{\mathrm{rk}(\alg)}$ of the gauge group $G$ (with
$\operatorname{Lie}G=\alg$). Following \cite{Benini:2018ywd}, we further specialize to $p=q$ (i.e. 
$\sigma=\tau$, or equivalently $J_1=J_2$), which is the natural specialization for the black hole entropy problem of \cite{Gutowski:2004ez,Gutowski:2004yv}; see \cite{Benini:2020gjh,Aharony:2024ntg} for the Bethe Ansatz at unequal angular momenta.
The integral formula for \eqref{eq:SCI_def} then reads
\begin{equation}
\mathcal{I}(q;v) = \frac{(q;q)^{2\,\mathrm{rk}(\alg)}_\infty}
{|\mathcal{W}_{\alg}|}
\oint_{\mathbb{T}^{\mathrm{rk}(\alg)}} 
\frac{\displaystyle\prod_{a=1}^3\prod_{\rho\in\mathcal{R}}
\Gamma_\text{el}\!\left(q^{r_a/2}z^\rho v^{\nu_a};q,q\right)}
{\displaystyle\prod_{\alpha\in\Phi} \Gamma_\text{el} \!\left(z^\alpha;q,q\right)}
\prod_{i=1}^{\mathrm{rk}(\alg)}\frac{dz^i}{2\pi i z^i}\ ,
\label{eq:index_integral}
\end{equation}
where the elliptic gamma function $\Gamma_\text{el}(z;p,q)$ is defined by \cite{Ruijsenaars:1997,Felder:1999vf}
\begin{equation}
\Gamma_\text{el}(z;p,q) \coloneqq \prod_{j,k=0}^\infty 
\frac{1-z^{-1}p^{j+1}q^{k+1}}{1-z\,p^j q^k}\ ,
\quad |p|,|q|<1\ ,
\label{eq:elliptic_gamma}
\end{equation}
the $q$-Pochhammer symbol is $(q;q)_\infty\coloneqq\prod_{k=0}^\infty(1-q^{k+1})$, 
$\Phi$ is the set of nonzero roots of $\alg$, $\mathcal{R}$ is the 
set of weights of the adjoint representation (i.e.,\ roots together with 
$\mathrm{rk}(\alg)$ zero weights), $\mathcal{W}_{\alg}$ is the 
Weyl group, the holonomies $z^i$ take values in 
$\mathbb{T}^{\mathrm{rk}(\alg)}$, and $v^{\nu_a}$ are flavor fugacities associated to the 
flavor weights $\nu_a$. The denominator encodes the vector multiplet, while 
the numerator encodes the three adjoint chiral multiplets of $\mathcal{N}=4$ 
SYM.

It is convenient to work additively along the $S^1$. Writing $z^i=:e^{2\pi i u^i}$ for the 
holonomies (now viewed as additive variables in the Cartan subalgebra 
$\mathfrak{h}$), $p=q=:e^{2\pi i\omega}$ for the common modular parameter (i.e., $\sigma=\tau =:\omega$), and 
absorbing the R-charge and flavor data into shifted arguments 
$q^{r_a/2}z^\rho v^{\nu_a}=e^{2\pi i(\langle\rho,u\rangle+\Delta_a)}$, we 
recast \eqref{eq:index_integral} in a purely Lie-algebraic form, as an 
integral over a fundamental cell of the cocharacter lattice in $\mathfrak{h}$ 
induced by the Haar measure. Concretely:
\begin{equation}\label{eq:int_formula}
    \mathcal{I}(\omega;\Delta_a) = \kappa_{\alg}
    \int_{[0,1]^{\mathrm{rk}(\alg)}} \mathcal{Z}(u;\Delta_a,\omega) 
    \prod_{i=1}^{\mathrm{rk}(\alg)}d u^i\ ,
\end{equation}
with
\begin{equation}\label{eq:ZZ}
    \mathcal{Z}(u;\Delta_a,\omega)\coloneqq \frac{\prod_{a=1}^3 \prod_{\rho\in \mathcal{R}} 
    \tilde{\Gamma}\!\left( \langle \rho,u \rangle+\Delta_a;\omega,\omega 
    \right)}{\prod_{\alpha\in \Phi} \tilde{\Gamma}\!\left(\langle \alpha,u 
    \rangle;\omega,\omega \right)}\ .
\end{equation}
Here $\langle\cdot,\cdot\rangle:\mathfrak{h}^*\times\mathfrak{h}\to\mathbb{C}$
is the canonical pairing of roots/weights with elements of the Cartan;
$\Delta_a$ ($a=1,2,3$) are the chiral-multiplet chemical potentials, subject
to $\Delta_1+\Delta_2+\Delta_3-2\omega\in\mathbb{Z}$ (the $p=q$ form of
\eqref{eq:Delta_constraint}); the prefactor is
$\kappa_{\alg}\coloneqq(q;q)_\infty^{2\rk\alg}/|\mathcal{W}_{\alg}|$; and
$\tilde{\Gamma}(u;\omega,\omega)\coloneqq\Gamma_\text{el}\!\left(e^{2\pi iu};
e^{2\pi i\omega},e^{2\pi i\omega}\right)$ is the elliptic gamma function in
additive variables. The integration domain $[0,1]^{\rk\alg}$ is a fundamental
cell of the cocharacter lattice $L=\bigoplus_i\mathbb{Z}v_i$, with $\{v_i\}$ a
primitive basis of $L$;\footnote{A primitive basis of a 
lattice consists of elements whose parallelogram is exactly one fundamental 
cell; it thereby provides a unit parameterization of the cell for integration.} it parameterizes the
compact maximal torus $\mathbb{T}^{\rk\alg}=\mathfrak{h}_{\mathbb{R}}/L$. The complexified holonomies appearing below live
instead on the complex torus $\mathfrak{h}/(L+\omega L)$, on which the elliptic
functions are (quasi-)periodic; we distinguish the two throughout.

A remark on notation. Before a primitive basis is fixed, $u\in\mathfrak{h}$ has
contravariant components $u^i$, while roots $\alpha\in\mathfrak{h}^*$ have
covariant components $\alpha_i=\langle\alpha,v_i\rangle$, so that the pairing
$\alpha\cdot u=\sum_i\alpha_iu^i$ requires no metric. In the explicit rank-two
computations of sections~\ref{sec:su2su3}--\ref{sec:cardy} we work in a fixed basis (which we always specify), where the distinction is immaterial, and sometimes write $u_i$ for
$u^i$ by a standard abuse of notation.

\subsubsection{Choice of basis and global structure}
\label{subsub:index-choice}
Before proceeding, it is worth pausing on a simple but crucial observation: 
the index of a general 4D $\mathcal{N}=1$ gauge theory (or CFT) is sensitive 
only to the gauge algebra and matter species in representations of said 
algebra. This follows from two facts. First, by definition the index traces 
over gauge-invariant states, so the center of the gauge group acts trivially 
and is invisible to the index --- independently of the gauge group's global 
structure. (For instance, in type $A$ the simply connected form 
$\mathrm{SU}(N)$ has center $\mathbb{Z}_N$, and the above reduces to the 
well-known statement that $N$-ality is not detected by gauge-invariant 
operators; the adjoint form $\mathrm{SU}(N)/\mathbb{Z}_N$, which has trivial 
center, gives the same index.) Second, on $S^1\times S^3$ there are no spatial 
cycles to support Wilson or 't~Hooft lines, so the index cannot distinguish 
gauge groups sharing the same algebra but differing in global structure --- 
unlike on lens spaces $S^1\times L(k,1)$ (i.e.\ $S^1\times S^3/\mathbb{Z}_k$; 
see, e.g., \cite{Benini:2011nc,Razamat:2013opa,Amariti:2022cxh}) or on other 
backgrounds $S^1\times M_3$ with nontrivial spatial cycles, i.e. with
$H_1(M_3,\mathbb{Z})\neq \emptyset$.

To make this precise: given a gauge algebra $\alg$, global forms of 
the gauge group are in bijection with subgroups $\Gamma\subseteq Z(\widetilde 
G)$ of the center of the simply connected cover $\widetilde G$, via 
$G=\widetilde G/\Gamma$. A further, strictly finer, refinement 
\cite{Aharony:2013hda} is that of global variant of SYM: for fixed $G$ one 
may additionally choose a maximal isotropic sublattice of the charge lattice 
$\Gamma\times\Gamma^\vee$ (with respect to the Dirac pairing), encoding the 
allowed spectrum of line operators together with a discrete theta angle. The 
set of global variants is acted on by the S-duality group $\mathrm{SL}(2,\mathbb{Z})$ (or a 
subgroup thereof for non-simply-laced $\alg$), and two variants with 
the same $G$ can be physically inequivalent. Nevertheless, all global variants 
share the same index on $S^1\times S^3$: local operators on $S^3$ carry no 
topological charge, and line operators require a spatial cycle to contribute 
\cite{Aharony:2013hda}. We therefore do not distinguish between global forms 
or variants henceforth, speaking simply of ``$\mathcal{N}=4$ SYM with gauge 
algebra $\alg$'' (or $\alg$ SYM for short).

Despite this independence of the final answer, writing \eqref{eq:int_formula} 
still requires \emph{choosing} a global structure (i.e., a group). The reason is that the 
formula integrates over (a maximal torus of) a group: to project the integrand 
onto gauge singlets we integrate over $G$ with the Haar measure, and the Weyl 
integration formula then reduces this to an integral over the maximal torus 
$\mathbb{T}^{\rk(\alg)}$, with the root-product in the denominator of $\mathcal{Z}$ supplying the 
vector-multiplet measure (the elliptic analogue of the Weyl--Vandermonde 
factor). By standard Lie theory, choosing a global structure is equivalent to 
choosing a lattice in $\mathfrak{h}$ between the coroot and coweight lattices 
(the cocharacter lattice $L$); to write the measure explicitly we must further 
fix a primitive basis $\{v_i\}$ of $L$.  
The final result is, of course, independent of these choices.

We record in table \ref{tab:bases} a few standard choices of global structure and associated primitive basis.
\begin{table}[ht]
\centering
\renewcommand{\arraystretch}{1.3}
\resizebox{\textwidth}{!}{%
\begin{tabular}{llll}
\toprule
\textbf{Name of basis} & \textbf{Root basis} & \textbf{Cartan basis} & \textbf{Global structure} \\
\midrule
\textbf{Orthogonal}   & Orthonormal         & Dual              & Simply connected ($A$, $C$), \\
                      & of Euclidean space  & of Euclidean space                      & $\SO(n)$ ($B$, $D$) \\
\addlinespace
$\boldsymbol{\omega}$ & Fundamental weights & Simple coroots        & Simply connected \\
$\boldsymbol{\alpha}$ & Simple roots        & Fundamental coweights & Adjoint \\
\bottomrule
\end{tabular}}
\caption{Standard choices of primitive basis for $\alg$, and the global structure $G$ (with $\mathrm{Lie}(G)=\alg$) each
induces. The orthogonal basis is the orthonormal basis of the Euclidean space
in which the roots are realized, the Cartan basis being its dual; it is the one
induced by the defining representation. In type $A$,
\cite{Benini:2018ywd} works with $\mathrm{SU}(N)$ holonomies in this basis,
subject to $\sum_{i=1}^Nu^i=0$.}
\label{tab:bases}
\end{table}

\subsection{The equations for general gauge algebra}
\label{sub:choices}

In \cite{Benini:2018mlo} it was proven that the index can be rewritten as a finite sum over solutions $\hat{u}$ to a set of equations (the BAEs), assuming only isolated solutions exist (i.e., isolated points, rather than curves, on the torus $\cc/(\zz+\omega\zz)$):
\begin{equation}\label{eq:index-finite}
     \mathcal{I}(\omega;\Delta_a) = \kappa_{\alg} \sum_{\hat{u}\, \in\, \text{BAEs}} \mathcal{Z}(\hat{u};\Delta_a,\omega) H^{-1}(\hat{u};\Delta_a,\omega)\ ,\quad H\coloneqq\det_{ij} \left[\frac{1}{2\pi i }\frac{\partial Q_i(\hat{u};\Delta_a,\omega)}{\partial u^j} \right]
\end{equation}
where $H$ is the Jacobian of the Bethe Ansatz operators $Q_i$ defined below in \eqref{eq:BAops}. To write down the BAEs for $\alg$ SYM at $p=q$ we follow closely \cite{Benini:2018mlo,Benini:2018ywd}, but make explicit the 
choices (such as those in table~\ref{tab:bases}) that are typically 
left implicit; this will prove essential for non-$A$-type algebras. 

Start by fixing a primitive basis $\{v_i\}_{i=1}^{\mathrm{rk}(\alg)}$ of the 
cocharacter lattice $L$, in which the holonomy is expanded as $u=\sum_i u^i v_i$, 
and write $\alpha_i\coloneqq\langle\alpha,v_i\rangle$ for the $i$-th component of a 
root $\alpha$ in the dual basis. (Unless stated otherwise, all components of 
roots and holonomies are understood in these dual bases.) We define the Bethe Ansatz operators
\begin{equation}\label{eq:BAops}
    Q_i(u;\Delta_a,\omega) \coloneqq (-1)^{\sum_{\alpha>0}\langle\alpha,v_i\rangle}
    \prod_{a=1}^{3}\prod_{\rho\,\in\,\mathcal{R}} 
    P(\langle\rho,u\rangle+\Delta_a;\omega)^{\langle\rho,v_i\rangle}\ , 
    \quad i=1,\ldots,\mathrm{rk}(\alg)\ ,
\end{equation}
where
\begin{equation}
    P(u;\omega)\coloneqq\frac{e^{-\pi i\,u^2/\omega+\pi i u}}{\theta_0(u;\omega)}\ ,
\end{equation}
and $\theta_0(u;\omega)$ is the standard theta function defined in \eqref{eq:theta-def}.
The BAEs are then
\begin{equation}
    Q_i(u;\Delta_a,\omega)=1\ ,\quad i=1,\ldots,\mathrm{rk}(\alg)\ .
\end{equation}
The explicit form of the equations \emph{depends} on the choice of primitive basis 
(and hence on the global structure) through the $Q_i$,\footnote{Below we suppress the arguments $(\Delta_a,\omega)$ when no confusion arises, writing simply $Q_i$.} a point which has not been stressed before in the literature. As a rather concrete illustration of this point, note that the prefactor $(-1)^{\sum_{\alpha>0}\langle\alpha,v_i\rangle}$ is absent in 
\cite[Eq.~(3.4)]{Benini:2018mlo}, but is necessary in general. Indeed, the 
exponent is $\sum_{\alpha>0}\langle\alpha,v_i\rangle=\langle 2\rho_\mathcal{W},v_i\rangle$, 
where $\rho_\mathcal{W}\coloneqq\tfrac12\sum_{\alpha>0}\alpha$ is the Weyl vector. In the 
simple-coroot basis $\boldsymbol{\omega}$ (i.e.,\ $v_i=\alpha_i^\vee$) one has 
$\langle\rho_\mathcal{W},\alpha_i^\vee\rangle=1$, hence 
$\langle 2\rho_\mathcal{W},\alpha_i^\vee\rangle=2$, always even, so the sign is $+1$ and 
the prefactor may be dropped (as in \cite{Benini:2018mlo}); for any other basis 
(e.g.\ the orthogonal bases typically used in the physics literature) the exponent 
can be odd, and the sign \emph{must} be retained.

In the case of $\mathcal{N}=4$ SYM it is convenient to fix a set of simple 
roots, dividing the roots into positive and negative; as in the above formulae, we write 
$\{\alpha>0\}$ for the set of positive roots. This brings the Bethe operators $Q_i$ to the 
simpler, familiar form
\begin{equation}\label{eq:BAEs_SYM}
    Q_i=\prod_{\alpha>0}\left(-\prod_{a=1}^3
    \frac{P(\alpha\cdot u+\Delta_a)}{P(-\alpha\cdot u+\Delta_a)}\right)^{\alpha_i}\ .
\end{equation}
Rewriting $P$ in terms of $\theta_1$ via \eqref{eq:theta-def}, using its modular 
properties, imposing the constraint on the matter chemical potentials in the form 
$\Delta_3=2\omega-\Delta_1-\Delta_2$, and simplifying, we obtain the following BAEs:\footnote{This is the form typically appearing in the $\mathfrak{su}(N)$ literature, e.g.\
\cite{Benini:2018ywd}, which predates the ``reduced BAEs'' terminology of
\cite{Benini:2021ano}.}
\begin{equation}\label{eq:theta1_BAE}
    Q_i=\prod_{\alpha>0}\left(-\frac{\theta_1(-\alpha\cdot u+\Delta_1)\,
    \theta_1(-\alpha\cdot u+\Delta_2)\,\theta_1(-\alpha\cdot u-\Delta_1-\Delta_2)}
    {\theta_1(\alpha\cdot u+\Delta_1)\,\theta_1(\alpha\cdot u+\Delta_2)\,
    \theta_1(\alpha\cdot u-\Delta_1-\Delta_2)}\right)^{\alpha_i}=1\ .
\end{equation}
To offload the notation, we sometimes abbreviate the 
per-root factor of \eqref{eq:theta1_BAE} as
\begin{equation}
\label{eq:Fdef}
    \mathcal{F}(x) \coloneqq -\,\frac{\theta_1(-x+\Delta_1)\,\theta_1(-x+\Delta_2)\,
    \theta_1(-x-\Delta_1-\Delta_2)}
    {\theta_1(x+\Delta_1)\,\theta_1(x+\Delta_2)\,\theta_1(x-\Delta_1-\Delta_2)}\ ,
\end{equation}
so that the BAEs become $Q_i=\prod_{\alpha>0}\mathcal{F}(\alpha\cdot u)^{\alpha_i}=1$.

Finally, as a heuristic for selecting the ``best'' global structure and basis,
we introduce the quantity
\begin{equation}\label{eq:min_mod}
    \mathrm{TotalExp}_{\alg}(G,\{v_i\})\coloneqq
    \max_{1\leq i\leq\rk\alg}\ \sum_{\alpha>0}|\alpha_i|\ ,
\end{equation}
which estimates the complexity of the equations as a function of these choices:
the exponents $\alpha_i$ in \eqref{eq:theta1_BAE} vary with the basis, and
dictate how many copies of each root factor enter the equations. We call the
choice minimizing \eqref{eq:min_mod} the \emph{minimal model} of the BAEs.

\subsubsection{Symmetries}
\label{subsub:symm}

Understanding the symmetries of the equations greatly simplifies solving them. Each symmetry can be made to act on the holonomy vector 
$u=(u^1,\ldots,u^{\mathrm{rk}(\alg)})$, and the solutions organize into 
orbits of the total symmetry group. There are two ``levels'' of symmetry at play. At the level of a single Bethe operator:
\begin{enumerate}
    \item[\emph{i)}] $Q_i(u+\lambda_1+\omega\lambda_2)=Q_i(u)$ for any 
    $\lambda_1,\lambda_2\in L$ \cite[Eq. (3.16)]{Benini:2018mlo}: not a symmetry proper, but the statement 
    that the holonomies live on the complex torus $\mathfrak{h}/(L+\omega L)$;
    
  \item[\emph{ii)}] $Z(G)_1\times Z(G)_\omega$: the center of the gauge 
    group, where the subscripts $1$ and $\omega$ label the two cycles of the torus 
    $\mathbb{C}/(\mathbb{Z}+\omega\mathbb{Z})$. It is realized as shifts of $u$ 
    by coweights along each cycle (roots evaluate to integers on coweights, so the Bethe operator
    $P$ is invariant); shifts by coroots are the trivial torus identifications 
    of \emph{i)}, so the genuine action is the quotient 
    (coweight lattice)/(coroot lattice)$\,\cong Z(G)$ for simply connected $G$, and $Z(\widetilde{G})=\pi_1(G)$ for the adjoint form;

\item[\emph{iii)}] $\mathrm{SL}(2,\zz)$,\footnote{This $\mathrm{SL}(2,\zz)$ should not be confused with the S-duality
group of the gauge theory. Ours acts on the modular parameter $\omega$ of the
$S^1\times S^3$ background, whereas S-duality acts on the holomorphic gauge
coupling $\tau_{\mathrm{YM}}$, a parameter on which the index does not depend at
all. The two also differ in structure: for non-simply-laced gauge algebras
S-duality exchanges $\alg$ with its Langlands dual $\alg^\vee$
\cite{Goddard:1976qe} and is realized not by $\mathrm{SL}(2,\zz)$ but by a Hecke
group \cite{Argyres:2006qr}, whereas the action considered here is
$\mathrm{SL}(2,\zz)$ for every $\alg$.} generated by
$T:(\omega,\Delta_a,u)\mapsto(\omega+1,\Delta_a,u)$ and
$S:(\omega,\Delta_a,u)\mapsto(-\tfrac1\omega,\tfrac{\Delta_a}{\omega},\tfrac{u}{\omega})$.
Note that $S^2=:C$ acts as charge conjugation,
$C:(\omega,\Delta_a,u)\mapsto(\omega,-\Delta_a,-u)$, the center of
$\mathrm{SL}(2,\zz)$, under which the Bethe operators are invariant,
$Q_i(-u,-\Delta_a,\omega)=Q_i(u,\Delta_a,\omega)$ \cite[Eq.~(3.7)]{Benini:2018mlo}.
On the other hand, a single sign flip separately inverts them,
$Q_i(-u,\Delta_a,\omega)=Q_i(u,-\Delta_a,\omega)=Q_i(u,\Delta_a,\omega)^{-1}$ (where the last identity is a consequence of $\theta_1$ being odd), so that in particular $u\mapsto-u$ maps solutions to solutions at fixed
$\Delta_a$. Whenever $-\id\in\mathcal{W}_{\alg}$ (as for
$A_1$, $B_N$, $C_N$, $D_{2N}$, and every exceptional algebra except $E_6$, and
in particular for all the rank-two algebras considered here) charge
conjugation is already included in the Weyl quotient, so only
$\mathrm{PSL}(2,\zz)$ acts effectively\footnote{For
$A_N$ with $N\geq2$, $D_{2N+1}$, and $E_6$, instead, $-\id$ equals the longest
Weyl element $w_0$ composed with the nontrivial diagram automorphism, hence
lies in $\mathcal{W}_{\alg}\rtimes\mathrm{Out}(\alg)$ but not in
$\mathcal{W}_{\alg}$; there the full $\mathrm{SL}(2,\zz)$ can act
nontrivially. In type $A$ it nonetheless acts trivially on the Hong--Liu
solutions, whose orbits are accordingly classified by $\mathrm{PSL}(2,\zz)$ in
\cite[Sec. 3.1]{Benini:2018ywd}.} on \emph{physical solutions} (the terminology will be introduced momentarily).

This action on the parameters induces one on the solutions. A solution is a
function $\hat u(\omega,\Delta_a)$ assigning a holonomy to each choice of
parameters; its image is obtained by evaluating $\hat u$ at the transformed
parameters and rescaling, so that
\begin{equation}\label{eq:PSLonsol-bis}
    S:\ \hat u(\omega,\Delta_a)\mapsto
    \omega\,\hat u\!\left(-\tfrac1\omega,-\tfrac{\Delta_a}{\omega}\right),
    \qquad
    T:\ \hat u(\omega,\Delta_a)\mapsto\hat u(\omega+1,\Delta_a)\ ,
\end{equation}
each understood modulo the lattice $\zz+\omega\zz$ of the original torus, i.e.\
reduced to its fundamental cell.\footnote{For $\alg$ of type $A$, and for algebras built from it such as
$D_2\cong A_1\oplus A_1$, the resulting orbit structure of solutions coincides with that of
the $\mathrm{SL}(2,\zz)$ action on the global variants of the theory
\cite{Aharony:2013hda}. This is not the case for non-simply-laced algebras, as
figure~\ref{fig:b2_saddles} shows; we return to this in \cite{polypaper}.}

\end{enumerate}
At the level of the full set of BAEs:
\begin{enumerate}
    \item[\emph{i)}] $\mathcal{W}_{\alg}$: the Weyl group of 
    $\alg$;
    \item[\emph{ii)}] $\mathrm{Out}(\alg)$: the outer automorphisms of $\alg$.
\end{enumerate}
The total symmetry group of the BAEs is therefore:
\begin{equation}
    \mathcal{}\mathcal{S}_\alg := (Z(G)_1\times Z(G)_\omega)\rtimes
    (\mathcal{W}_{\alg}\rtimes\mathrm{Out}(\alg))
    \times\mathrm{PSL}(2,\mathbb{Z})\ .
\end{equation}

\subsection{Solutions of the Bethe Ansatz Equations}
\label{sub:solutions}

\subsubsection{Physical solutions}
\label{subsub:physsol}

We have seen that the choice of global structure and basis affects the form of
the BAEs, and hence their solution set. Any physical quantity (such as the index) must be
independent of these choices. Since each choice is tied to a
symmetry of the equations, the choice-independent data are precisely the
\emph{orbits} of solutions under those symmetries. We call such an orbit a
\emph{physical solution}: it is what remains after quotienting by every symmetry
that merely reflects a choice, and is by construction blind to that choice.

Concretely, this is dictated by the symmetries common to the BAEs and to the
index contribution $\mathcal{Z}/H$: we identify solutions related by the
action of
\begin{equation}\label{eq:Sgauge}
    \mathcal{S}_{\text{gauge}}\coloneqq
    (Z(G)_1\times Z(G)_\omega)\rtimes\mathcal{W}_{\alg}\ ,
\end{equation}
and study the resulting orbit space, on which $\mathrm{PSL}(2,\mathbb{Z})$ (and
potentially $\mathrm{Out}(\alg)$) then acts. The physical rationale for each
quotient is the following:
\begin{itemize}
    \item We quotient by the center $Z(G)_1\times Z(G)_\omega$. This symmetry is
    an artifact of the chosen global structure, which, as emphasized above, is
    auxiliary, needed only to pull back the Haar measure. Therefore it carries
    no physical meaning.
    \item We quotient by the Weyl group $\mathcal{W}_{\alg}$, since Weyl-related
    holonomies are gauge-equivalent points on the torus.
\end{itemize}
The invariance of $\mathcal{Z}$ under $\mathcal{S}_{\text{gauge}}$ is clear from
the product over all roots in \eqref{eq:ZZ}, together with the fact that
$\alpha\cdot u=0$ for any $u$ in the center. To see the invariance of $H$ we can
use the formula \eqref{eq:Hchange} below, specialized to the transformation matrices
of $\mathcal{S}_{\text{gauge}}$, whose determinant is always $\pm1$.

Note that we do \emph{not}, however, quotient by the outer automorphisms
$\mathrm{Out}(\alg)$ (if nontrivial): these are genuine global symmetries of the
theory, and identifying solutions related by them would discard physical
information. The simplest instance appears already for $D_2\cong A_1\oplus A_1$,
whose outer automorphism $\mathbb{Z}_2$ exchanges the two $A_1$ summands: a
solution $(u_1,u_2)$ and its image $(u_2,u_1)$ are physically distinct (see
section~\ref{sec:D2}). By contrast, the $\mathbb{Z}_2$ of $A_{N-1}$ ($N\geq3$),
the diagram automorphism reversing the order of the $\mathfrak{u}(N)$
holonomies, acts trivially on the solutions of section~\ref{sec:su2su3}: there
its effect is undone by a Weyl reflection, so that no new physical solutions
arise. Richer examples, such as the $S_3$ triality of $D_4$ and the
$\mathbb{Z}_2$ of $E_6$, first act nontrivially at higher rank and are discussed
in the companion paper \cite{polypaper}.

Having identified the physical solutions, the BAE evaluation formula
\eqref{eq:index-finite} becomes
\begin{equation}\label{eq:index-finite-new}
    \mathcal{I}(\omega;\Delta_a)=\kappa_{\alg}
    \sum_{\substack{\hat u\,\in\, \text{physical}\\ \text{solutions}}}
    \big|\mathcal{S}_{\text{gauge}}(\hat u)\big|\;
    \mathcal{Z}(\hat u;\Delta_a,\omega)\,H^{-1}(\hat u;\Delta_a,\omega)\ ,
\end{equation}
where $|\mathcal{S}_{\text{gauge}}(\hat u)|$ is the size of the orbit of
$\hat u$ under $\mathcal{S}_{\text{gauge}}$. This choice-dependent prefactor is
what is sometimes called in the literature the ``multiplicity'' of a BAE
solution. We stress that, while it is choice-dependent, it is merely an artifact
of the unphysical presentation choices one has to make, and the whole per-summand
contribution
\begin{equation}\label{eq:full-contribution}
    \big|\mathcal{S}_{\text{gauge}}(\hat u)\big|\;
    \mathcal{Z}(\hat u;\Delta_a,\omega)\,H^{-1}(\hat u;\Delta_a,\omega)
\end{equation}
is, as expected, choice-\emph{independent} (see the proof of this statement in section~\ref{subsub:basis-indep}).

The orbit-size prefactor gives a precise origin for the multiplicity, which does
not require invoking the one-form center symmetry; the index counts local
operators, and so cannot be sensitive to extended-operator notions. It has of
course long been known that the Cardy expansion carries a finite $\log|Z(G)|$
correction, for $\mathcal{N}=4$ SYM with any $ABC\!D$ gauge algebra and for
$\mathcal{N}=1$ SCFTs built on them
\cite{GonzalezLezcano:2020yeb,Amariti:2020jyx,Amariti:2021ubd,Cassani:2021fyv,ArabiArdehali:2021sfg}.
We identify this correction as the image, under the Cardy map of
section~\ref{sec:cardy}, of the prefactor in \eqref{eq:index-finite-new}: both
sides evaluate the same index, and the Bethe Ansatz side makes the group-theory
origin of the number manifest.

\subsubsection{Independence of the basis}
\label{subsub:basis-indep}

To complete our choices-conscious treatment of the BAEs, we prove that the set
of physical solutions for any $\alg$ is independent of those choices, and so are their contributions to the index.

Let us consider a primitive basis $\{v_i\}$ of the coweight lattice of $\alg$
and a primitive basis $\{w_i\}$ of some other (or the same) lattice of $\alg$.
The basis-change matrix $M$ from the $w$-basis to the $v$-basis,
\begin{equation}\label{eq:basischange}
    u^v_i=\sum_jM_{ji}\,u^w_j\ ,
\end{equation}
is guaranteed to have integer coefficients, because any such lattice is a
sublattice of the coweight one; for the same lattice it is even an element of
$\mathrm{GL}(\rk\alg,\mathbb{Z})$. The same $M$ gives
\begin{equation}\label{eq:rootchange}
    \alpha^w_i=\sum_j\alpha^v_j\,M_{ji}\ .
\end{equation}
Then, there is a relation between the Bethe Ansatz operators expressed in the
two bases, which we can write compactly using the shorthand $\mathcal{F}$ introduced in
\eqref{eq:Fdef}:
\begin{equation}\label{eq:Qwv}
    Q^w_i=\prod_{\alpha>0}\mathcal{F}(\alpha\cdot u)^{\alpha^w_i}
     =\prod_{\alpha>0}\mathcal{F}(\alpha\cdot u)^{\sum_jM_{ji}\alpha^v_j}
     =\prod_j\big(Q_j^v\big)^{M_{ji}}\ .
\end{equation}
We use \eqref{eq:Qwv} to define a manifestly injective map from the
$v$-solutions to the $w$-solutions. Let $u^v$ be a $v$-basis solution, i.e.\
$Q_i^v(u^v)=1$. Choosing a concrete presentation of $M^{-1}$ (there can be
several, owing to the periodicity of the torus), the configuration
$u^w\coloneqq M^{-1}u^v$ is a $w$-solution:
\begin{equation}\label{eq:Qvtow}
    Q^w_i(u^w)=\prod_j\big(Q_j^v(u^w)\big)^{M_{ji}}
              =\prod_j\big(Q_j^v(u^v)\big)^{M_{ji}}=1\ ,
\end{equation}
where we used that $\mathcal{F}(\alpha\cdot u)$ involves only basis-independent
quantities.

We now prove surjectivity of the map, up to the center action, by comparing
index contributions. Since the index is itself basis-independent, any two choices
of basis must lead to the same value; we therefore rewrite the contribution of
$u^v$ in terms of $u^w$ and compare. The integrand $\mathcal{Z}(u)$ is
expressed in terms of basis-independent quantities, so
$\mathcal{Z}(u^w)=\mathcal{Z}(u^v)$. For the Jacobian, inserting
\eqref{eq:Qvtow} gives
\begin{equation}\label{eq:Jacchange}
\begin{split}
    J^w_{ab}\coloneqq \frac{\partial\log Q^w_a}{\partial u^w_b}
    &=\sum_kM_{bk}\,\frac{\partial}{\partial u^v_k}
      \Big(\sum_jM_{aj}\log Q^v_j\Big)\\
    &=\sum_{j,k}M_{aj}\,
      \frac{\partial\log Q^v_j}{\partial u^v_k}\,M_{bk}
     =\sum_{j,k}M_{aj}\,J^v_{jk}\,M_{bk}\ ,
\end{split}
\end{equation}
and therefore
\begin{equation}\label{eq:Hchange}
    H^w(u^w)=(\det M)^2\,H^v(u^v)\ .
\end{equation}
It is known that $\det M$ equals $|Z(\{w_i\})|/|Z(\{v_j\})|$, the ratio of the
orders of the centers; in particular, for the $v$-basis corresponding to the
coweight lattice, $\det M=|Z(\{w_i\})|$. Combining the two results,
\begin{equation}\label{eq:ZHchange}
    \left(\frac{\mathcal{Z}}{H}\right)(u^v)
    =|Z(\{w_i\})|^2\left(\frac{\mathcal{Z}}{H}\right)(u^w)\ ,
\end{equation}
that is, changing bases automatically accounts for the sum over all
center-equivalent BAE solutions. In particular, there can be no solution for the
$w$-choices that is not center-equivalent to some $M^{-1}u^v$, which completes
the proof. We note, finally, that the above allows one to go back explicitly:
for any $w$-solution $u^w$, the configuration $u^v\coloneqq Mu^w$ is a
$v$-solution, and this map is surjective.

\subsubsection{Nonzero contribution of Weyl-fixed solutions}
\label{subsub:weyl-contrib}

It is widely believed that, in the Bethe Ansatz approach, Weyl-fixed solutions do not
contribute to the index \cite{Benini:2018mlo}. While this holds in type $A$, we
find for other types Weyl-fixed solutions that do contribute nontrivially, $\mathcal{Z}\neq 0$ (with finite Jacobian $H$).

We call a solution ${u}$ \emph{Weyl-fixed} if it is left invariant on the torus by
some $w\in\mathcal{W}_{\alg}$, i.e.\ if
\begin{equation}
    w({u})-{u} = 0 \mod \mathbb{Z}+\omega\mathbb{Z}\ .
\end{equation}
We first recall the standard argument.\footnote{The reasoning of
\cite[App.~C]{Benini:2018mlo} collapses to the one presented here under the
equal-fugacity condition $\sigma=\tau\eqqcolon\omega$ assumed throughout this
paper.} It proceeds not from Weyl-fixedness directly, but from the existence of
a root annihilating the solution: suppose some $\alpha\in\Delta$ satisfies
$\langle\alpha,u\rangle\in\mathbb{Z}\oplus\omega\mathbb{Z}$. Then the
vector-multiplet contribution
$\tilde\Gamma(\langle\alpha,{u}\rangle;\omega,\omega)$ to the integrand
develops a pole, so that $\mathcal{Z}({u})=0$, the Jacobian $H({u})$ being
finite because ${u}$ is isolated.

The annihilating root is then usually tied to Weyl-fixedness by the standard
Lie-theoretic fact that the Weyl group acts simply transitively on the chambers,
so that a fixed point in the Cartan algebra must lie on a wall, where some root vanishes. That fact,
however, holds for $u$ in the real Cartan; our solutions live on the complex
torus $\mathbb{C}/(\mathbb{Z}\oplus\omega\mathbb{Z})$, and can evade it. 

What is always is true is the inverse implication: a solution annihilated by a root $\alpha$ is guaranteed to be Weyl-fixed by the reflection about $\alpha$. Let us also point out that being Weyl-fixed is a basis-dependent statement, as the torus identification changes with the cocharacter lattice. In particular, if a Weyl-fixed solution translated to a coroot lattice basis is not Weyl-fixed, by the implication, it cannot contribute trivially.

A simple example of ``truly'' Weyl-fixed nontrivial solution, which we develop below and which is in fact a BAE solution, is
$(u_1,u_2)=(\tfrac{\omega}{2},\tfrac12)$ for $G_2$ in the orthogonal basis (which for $G_2$ corresponds to the coroot lattice). It is fixed on
the torus by $-\id\in\mathcal{W}_{G_2}$, yet every root evaluates on it to a
half-period rather than to a lattice point, so no root annihilates it and
$\mathcal{Z}$ does not vanish.

\subsection{Solutions of special form}
\label{sub:special}

For later convenience, we introduce nomenclature for certain special classes of 
solutions. Each class simplifies the BAEs in some way, making the 
corresponding solutions easier to find or to characterize analytically. 
Throughout, $r\coloneqq\mathrm{rk}(\alg)$ denotes the number of holonomy 
components, and we work in the chosen primitive basis $\{v_i\}$.
\begin{itemize}
    \item \emph{Fully rational} (or \emph{Hong--Liu-like}) solutions: those that are 
    independent of the chemical potentials $\Delta_a$ and whose holonomy 
    components are all rational. The ``basic'' and Hong--Liu solutions of type $A$ \cite{Hosseini:2016cyf,Hong:2018viz} are the     prototype.
    \item \emph{$k/r$-partially rational} solutions: those for which only $k$ 
    of the $r$ holonomy components are rational, the remaining ones being 
    generic (in particular, $\Delta_a$-dependent). Fully rational solutions 
    are the special case $k=r$.
    \item \emph{$m$-partially pinned} solutions: those solutions $u$ for which there 
    exist $m$ positive roots $\alpha$ with 
    $\langle\alpha,u\rangle=\tfrac{a+b\omega}{2}$ for some $a,b\in\mathbb{Z}$, 
    i.e.\ lying at a half-period of the torus. Such points are significant 
    because $\theta_1$ is odd and quasi-periodic: at a half-period, the ratio
    \begin{equation*}
        \prod_a\frac{P(\langle\alpha,u\rangle+\Delta_a)}
        {P(-\langle\alpha,u\rangle+\Delta_a)}
    \end{equation*}
    appearing in the Bethe operator \eqref{eq:theta1_BAE} collapses to a sign, so the corresponding factor trivializes. This effectively 
    ``pins'' the equations along those $m$ root directions, reducing the number 
    of genuinely transcendental constraints.
    \item \emph{$0$-partially pinned} (or \emph{generic}) solutions: the 
    $m=0$ case, i.e.\ solutions at which no positive root sits at a 
    half-period, so that none of the $\theta_1$-factors trivializes and every 
    Bethe equation remains fully transcendental. Such solutions first arise, 
    among the rank-two algebras, in the exceptional case $G_2$ 
    (see section~\ref{sec:G2}), where the isolated solutions are generically 
    unconstrained.
\end{itemize}
Note that the indices $k$ and $m$ count different things: $k$ (in 
``$k/r$-partially rational'') counts \emph{holonomy components} that are 
rational, whereas $m$ (in ``$m$-partially pinned'') counts \emph{positive 
roots} at which the holonomy sits at a half-period. A solution may well be 
both partially rational and partially pinned, with $k\neq m$.

%%%%%%%%%%%%%%%%%%%%%%%%%%%%%%%%%%%%%%%%%%%%%%%%%%%%%%%%%%%%%%%%%%%%%%%%%%%%%%%%%%%%%%%%%%%%%%%%%%%%%%%%%%%%%%%%%%%%
\section{\texorpdfstring{The well-known case of ${A_{N-1}=\mathfrak{su}(N)}$ SYM}{The well-known case of A(N-1)=su(N) SYM}}
\label{sec:su2su3}
%%%%%%%%%%%%%%%%%%%%%%%%%%%%%%%%%%%%%%%%%%%%%%%%%%%%%%%%%%%%%%%%%%%%%%%%%%%%%%%%%%%%%%%%%%%%%%%%%%%%%%%%%%%%%%%%%%%%

Up to now, the only semisimple gauge algebras studied in the literature via the Bethe Ansatz have been 
of type $A$. The cases $A_1$ and $A_2$ were solved explicitly in 
\cite{ArabiArdehali:2019orz,Benini:2021ano,GonzalezLezcano:2021nzk}, in \cite{Benini:2021ano} by focusing on a set of ``reduced BAEs'', while a large class 
of solutions valid at any rank --- the Hong--Liu (HL) solutions --- was 
constructed in \cite{Hong:2018viz}; these are conjectured (but not known) to exhaust the 
isolated solutions in type $A$. Below we show that both 
the reduced equations and the ``center-of-mass'' freedom of a HL solution are 
particular manifestations of the general structures introduced in section~\ref{sec:baes}: 
respectively, a choice of global structure and primitive basis, and the action 
of the center.

\subsection{\texorpdfstring{The reduced equations in type ${A}$ as a special choice}{The reduced equations in type A as a special choice}}
\label{sub:reduced}

To derive the reduced BAEs of \cite{Benini:2021ano}, one starts from the Bethe operators of $\mathfrak{u}(N)$ rather than $\mathfrak{su}(N)$,
\begin{equation}
    \tilde{Q}_j(u;\Delta,\omega) = e^{-2\pi i \sum_{k=1}^N u^{jk}} 
    \prod_{a=1}^3 \prod_{k=1}^N 
    \frac{\theta_0(u^{kj} + \Delta_a;\omega)}{\theta_0(u^{jk} + \Delta_a;\omega)}\ ,
    \quad j=1,\ldots,N\ ,
\end{equation}
where $u^{jk}\coloneqq u^j-u^k$ and the $u^i$ are the components of $u$ in the 
orthogonal basis of the Cartan. Since these operators satisfy 
$\prod_{i=1}^N\tilde{Q}_i=1$, the $\mathfrak{su}(N)$ BAEs can be written as
\begin{equation*}
    1 = \frac{\tilde{Q}_i}{\tilde{Q}_N}\ , \qquad i=1,\ldots,N-1\ .
\end{equation*}
Each solution $u$ of this system also satisfies $\tilde{Q}_j(u)=e^{2\pi i\lambda/N}$ 
for some $\lambda\in\{0,\ldots,N-1\}$. It is argued in \cite{Benini:2021ano} 
that only the solutions with $\lambda=\tfrac{N(N-1)}{2}\bmod N$ contribute 
nontrivially to the index, so the equations effectively reduce to
\begin{equation}\label{eq:red_BAE}
    \tilde{Q}_i = (-1)^{N-1}\ , \qquad i=1,\ldots,N-1\ .
\end{equation}
Presented this way, the reduced BAEs look like a computational trick: one 
restricts to a single $\lambda$-sector of solutions, and trades the equations 
$\tilde{Q}_i/\tilde{Q}_N=1$ for the simpler $\tilde{Q}_i=(-1)^{N-1}$.

In the language of the previous section, however, this reduction is nothing more 
than a particular (implicit) choice of global structure for $\mathfrak{su}(N)$ together 
with a primitive basis:
\begin{itemize}
    \item the global form is the adjoint $\mathrm{PSU}(N)$, i.e.\ the 
    cocharacter lattice is the coweight lattice;
    \item the primitive basis $\{v_i\}$ is built from the weights of the 
    defining representation: in terms of the fundamental coweights 
    $\{\omega_i\}$, $v_i=\omega_i-\omega_{i-1}$ (with $\omega_0\coloneqq0$).
\end{itemize}
We call this choice the \emph{reduced basis}. Let us explicitly verify that it reproduces \eqref{eq:red_BAE}. Expressing 
the roots and basis in the standard $N$-dimensional orthogonal basis $\{e_i\}$ of the 
Cartan (i.e., $\alpha^{ij}\coloneqq e^i-e^j$ and $v_i\coloneqq \tfrac{1}{N}(N e_i-\sum_{j=1}^N e_j)$), we can see that they satisfy
\begin{equation}\label{eq:fund_basis_A}
    \langle\alpha^{ij},v_k\rangle=\delta^i_k-\delta^j_k\ .
\end{equation}
The $i$-th Bethe equation \eqref{eq:BAEs_SYM}, with positive roots 
$\{\alpha^{st}:1\leq t<s\leq N\}$, then becomes
\begin{equation}
    1 = \prod_{1\leq t<s\leq N}\left(-\prod_{a=1}^3
    \frac{P(\alpha^{st} \cdot u +\Delta_a)}{P(-\alpha^{st} \cdot u+\Delta_a)}\right)^{\delta^s_i-\delta^t_i}
    = (-1)^{N-1}\prod_{t=1}^{N}\prod_{a=1}^3
    \frac{P(\alpha^{it} \cdot u+\Delta_a)}{P(-\alpha^{it} \cdot u+\Delta_a)}\ .
\end{equation}
Furthermore, we can decompose the holonomy vector $u=\sum_{k=1}^{N-1}u^ke_k$ in the
orthogonal basis, so that
\begin{equation}
    \langle\alpha^{ij},u\rangle=u^i-u^j\ .
\end{equation}
This amounts to rewriting the holonomies, though \emph{not} the equations themselves,
i.e.\ not the exponents $\alpha_i$ of \eqref{eq:BAEs_SYM}, in the simply
connected $\SU(N)$ global form, and thereby reintroduces the center action by
hand. The BAEs then turn into
\begin{equation}
    1 = (-1)^{N-1}\prod_{t=1}^{N}\prod_{a=1}^3
    \frac{P(u^i-u^t+\Delta_a)}{P(-u^i+u^t+\Delta_a)} = e^{-2\pi i \sum_{k=1}^N u^{jk}} 
    \prod_{a=1}^3 \prod_{k=1}^N 
    \frac{\theta_0(u^{kj} + \Delta_a;\omega)}{\theta_0(u^{jk} + \Delta_a;\omega)}\ ,
\end{equation}
which is exactly the reduced form \eqref{eq:red_BAE}. The prefactor $(-1)^{N-1}$ (which was introduced by hand in the original derivation) here arises 
automatically from the sign prefactor in the Bethe operators, evaluated in 
this non-simple-coroot basis (see the discussion below \eqref{eq:theta1_BAE}).

As a second illustration of the principle of section~\ref{subsub:physsol}, 
consider the known isolated solutions in type $A$, i.e. the Hong--Liu solutions \cite{Hong:2018viz}. They are labeled by triples
\begin{equation}
    \{m,n,r\} \quad \text{such that} \qquad N = m n \,, \quad r=0,\ldots,n-1 \ ,
\end{equation}
and are explicitly given by
\begin{equation}\label{eq:HL}
    u_j \eqqcolon  u_{\hat{\jmath}\hat{k}} = \bar{u} + \frac{\hat{\jmath}}{m} + \frac{\hat{k}}{n} \left(\tau + \frac{r}{m}\right)
\end{equation}
with the index $j = 0,\dots,N - 1$ decomposed into the indices $\hat{\jmath} = 0,\dots,m - 1$
and $\hat{k} = 0,\dots,n - 1$. $\bar{u}$ is a constant imposing the $\mathfrak{su}(N)$ constraint $\sum_{j=0}^{N-1}u_j=0$.

We reinterpret their ``center-of-mass'' freedom, that is, the choice of overall constant component 
of the holonomy, precisely as the action of the center of the simply 
connected form $\SU(N)$, i.e.\ the symmetry \emph{ii)} of 
section~\ref{subsub:physsol}. Quotienting by it is exactly the statement that this 
freedom is unphysical. 

We note, finally, the remarkable fact that the Hong--Liu solutions are independent of 
the chemical potentials $\Delta_a$. As we will see in section~\ref{sec:D2}, this 
is \emph{not} a generic feature of isolated BAE solutions --- it is special to 
type $A$.

\subsection{Revisiting the low-rank cases}

For low-rank cases of $A_{N-1}$ gauge algebra, \cite{ArabiArdehali:2019orz,Benini:2021ano,Lezcano:2021qbj} made an important discovery: namely, the Bethe Ansatz formula \eqref{eq:index-finite} is incomplete written as is, as it does not account for the contribution of \emph{continuous} families of solutions to the index. For $N=2$ there are no continuous families, but starting at $N=3$ one can in general expect their presence, and \cite[Sec. 4.3.2]{ArabiArdehali:2019orz} conjectures that there exist in fact $\ell_\cc$-dimensional families, with $(\ell+1)(\ell+2)\leq 2N$ (so, e.g., $N=2$ can only have isolated solutions, i.e. $\ell=0$; $N=3,4,5$ have $\ell=0,1$; $N=6,\ldots,9$ have $\ell=0,1,2$; and $N=10$ is the first case to also allow for $\ell=3$). Note that already for $N=3$ the one-dimensional family (i.e., a codimension-two subspace of the solution set of three equations) will not appear as a common factor of the three BAEs.

A few interesting facts on these continuous families are:
\begin{itemize}
    \item Accounting for their presence is necessary to reconstruct the full index; in other words, evaluating the Bethe Ansatz formula just on isolated Hong--Liu solutions (or, more generally, solutions with rational coefficients only) does not give back the numerical or small R-symmetry fugacity $t$ evaluations of the index one can compute via localization.
    \item It is not known how to evaluate their contributions in general (though see \cite{Cabo-Bizet:2024kfe} for an attempt for $N=3$). A possible approach is based on the continuous version of Grothendieck residue theorem; see our brief discussion in section~\ref{sec:conc} on this point.
    \item It is not known for which $N$ continuous families actually appear in type $A$ (beyond the two explored cases of $N=2,3$). Nor is it known whether they appear for other 4D $\mathcal{N}=1$ theories or $\mathcal{N}=4$ SYM with different gauge algebras.\footnote{A related observation appears in \cite{Amariti:2025vjd}, which
evaluates the Bethe Ansatz formula for some toric 4D $\mathcal{N}=1$ quivers with
$\mathfrak{su}(2)$ gauge algebras. For the conifold, whose effective rank is
$2=1+1$, the isolated solutions alone (which are combinations of Hong--Liu solutions for the two $A_1$ nodes) reproduce the direct evaluation of the
index at generic flavor fugacities, the divergent contributions of individual
solutions canceling in the sum as for $A_1$ SYM \cite{Benini:2021ano,Lezcano:2021qbj}; for the suspended pinch point, of effective
rank $3=1+1+1$, the cancellation fails, which the authors read as a signal that
further solutions, isolated or continuous, are missing.} This has prompted us to investigate SYM with $\mathfrak{su}(4)$ as well as other gauge algebras with more powerful methods based on a polynomial analysis, which will be introduced elsewhere \cite{polypaper}.
    \item As we will see in the next two sections, SYM with any other classical algebra of rank two lacks continuous families, just like $\mathfrak{su}(2)$ does.    
\end{itemize}

\subsubsection{$A_1=\mathfrak{su}(2)$: only isolated solutions}
\label{sub:A1}

Solving the $\mathfrak{su}(2)\cong \mathfrak{usp}(2)\cong \mathfrak{so}(3)$ SYM BAEs is a simple task. All isolated solutions are of Hong--Liu type, labeled by three integers $\{m,n,r\}$, and moreover there are no continuous families of solutions. Let us briefly summarize the derivation of \cite{Benini:2021ano} to introduce some notation that is also relevant for further examples. 

To simplify the formulae, we first solve the constraint \eqref{eq:Delta_constraint}, $\sum_{a=1}^3 \Delta_a = 2\omega \mod \zz+\omega\zz$, by setting $\Delta_3 = -\Delta_1-\Delta_2$ on the torus. We then call $\Delta$ the set $\{\Delta_1,\Delta_2,-\Delta_1-\Delta_2\}$ by slight abuse of notation. Then, the reduced $A_1$ BAE in terms of $u:=u_1-u_2$ (which actually is the coordinate in the reduced basis vector $v_1$, see \eqref{eq:fund_basis_A}) and \eqref{eq:Fdef} , reads
\begin{align}\label{eq:A1eqn}
    Q&=\mathcal{F}(u)= -\displaystyle\prod_{\Delta}
    \frac{\theta_1(-u+\Delta)}{\theta_1(u+\Delta)}=1 \ .
\end{align}
To separate the dependence of the elliptic functions on the holonomy $u$ 
from that on the chemical potentials $\Delta_a$, we
introduce the identity \cite[Eq. (3.4)]{Benini:2021ano}
\begin{equation}\label{eq:fdef}
    f(u)\equiv f(u;\Delta) \coloneqq \theta_1(u)\,\theta_1(u+\Delta_1)\,
    \theta_1(u+\Delta_2)\,\theta_1(u-\Delta_1-\Delta_2)=\sum_j c_j\,\theta_j(2u)\ ,
\end{equation}
where
\begin{equation}\label{eq:cdef}
    c_j\coloneqq\begin{cases}
        -\tfrac12\,\theta_j(\Delta_1)\,\theta_j(\Delta_2)\,\theta_j(\Delta_1+\Delta_2) 
        & j\in\{1,2,4\}\ ,\\[1ex]
        +\tfrac12\,\theta_j(\Delta_1)\,\theta_j(\Delta_2)\,\theta_j(\Delta_1+\Delta_2) 
        & j=3\ .
    \end{cases}
\end{equation}
Inserting the identity $1=-\theta_1(-x)/\theta_1(x)$ lets us replace every 
$\theta_1$ in \eqref{eq:A1eqn} by $f$, so that all dependence on $\Delta_a$ is 
absorbed into the coefficients $c_j$, which we henceforth treat as numerical 
constants. Using this observation, we see that
\begin{equation}\label{eq:Finf}
    \mathcal{F}(x)=\frac{f(-x)}{f(x)}\  ,
\end{equation}
and the $A_1$ BAE takes the simple form
\begin{equation}
    f(u)=f(-u)\ .
\end{equation}
Let us stress here that inserting $1=-\theta_1(-x)/\theta_1(x)$ may introduce spurious solutions satisfying $\theta_1(x)=0$, which one should discard at the end if they do not solve the original BAEs.

Taking advantage of the parity properties of $\theta_r$ functions (see appendix~\ref{app:ellitpic}), our BAE further simplifies to
\begin{equation*}
    c_1\theta_1(2u)=0\ ,
\end{equation*}
whose solutions on the torus are
\begin{equation}
    u \in \left\{\frac{1}{2}, \ \frac{\omega}{2}, \ \frac{1+\omega}{2}\right\}\ .
\end{equation}
In the orthogonal basis, these map exactly to the Hong--Liu solutions \eqref{eq:HL} labeled by $\{2,1,0\}$, $\{1,2,0\}$ (the ``basic'' solution of \cite{Hosseini:2016cyf}), and $\{1,2,1\}$.

Before moving to $A_2$, let us comment on  $A_1$ in the $\boldsymbol{\omega}$-basis as a further illustration of the proven basis-independence of physical solutions. The root in that basis has component $2$ (instead of $1$, as in the orthogonal basis). Therefore, the BAE is
\begin{equation*}
    \mathcal{F}(u)^2=1\ .
\end{equation*}
It would seem like there could be more solutions now, coming from the branch $\mathcal{F}(u)=-1$. Contrary to that, the proof of section \ref{subsub:basis-indep} guarantees that this branch cannot lead to new, physical contributing solutions.

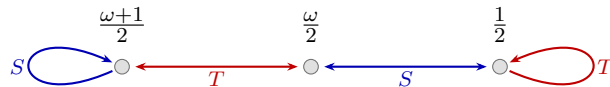
\begin{figure}[htb!]
\centering
\begin{tikzpicture}
    \node[sol, label={[sollbl, label distance=4pt]90:$\tfrac{\omega+1}{2}$}] (m1) at (-2.5,0) {};
    \node[sol, label={[sollbl, label distance=4pt]90:$\tfrac{\omega}{2}$}]     (m2) at (   0,0) {};
    \node[sol, label={[sollbl, label distance=4pt]90:$\tfrac{1}{2}$}]                       (m3) at ( 2.5,0) {};

\begin{scope}[every node/.style={}]
    \draw[Tedge] (m1) -- node[edgelbl, below]{$T$} (m2);
    \draw[Sedge] (m2) -- node[edgelbl, below]{$S$} (m3);
    \draw[Sloop] (m1) to[out=180+24,in=180-24,looseness=50]
                 node[edgelbl, left] {$S$} (m1);
    \draw[Tloop] (m3) to[out=-24,in=+24,looseness=50]
                 node[edgelbl, right]{$T$} (m3);
\end{scope}
\end{tikzpicture}
\caption{The three-element orbit of the $\mathrm{PSL(2, \mathbb{Z})}$ action on $A_1$ solutions.}
\label{fig:A_1}
\end{figure}

The $\mathrm{PSL}(2,\zz)$ orbit of the $A_1$ solutions is presented in figure \ref{fig:A_1}.

\subsubsection{$A_2=\mathfrak{su}(3)$: the continuous family}
\label{sub:A2}

The case of $\mathfrak{su}(3)$ is the first in which the BAEs admit an
$\ell_\cc$-dimensional continuous family of solutions on the torus, depending on
the $\Delta_a$; here $\ell_\cc=1$. Using again the function $f$ of
\eqref{eq:fdef} and the identity it satisfies, the reduced BAEs read, in the
reduced coordinates of section~\ref{sub:reduced},
\begin{subequations}\label{eq:A2}
\begin{align}
    f(u_1)f(u_2)-f(-u_1)f(-u_2)&=0\ ,\\
    f(-u_1)f(u_2-u_1)-f(u_1)f(u_1-u_2)&=0\ ,
\end{align}
\end{subequations}
where, as before, the spurious solutions introduced by passing to $f$ must be
removed. This system reduces \cite{Benini:2021ano} to
\begin{equation}\label{eq:A2_sols_or}
    \theta_1(u_1+u_2)=\theta_1(2u_1-u_2)=0
    \quad\cup\quad h(u_1,u_2,\Delta_a)=0\ .
\end{equation}
The first branch is solved by the isolated Hong--Liu solutions, which in our
primitive basis read
\begin{equation}\label{eq:A2HLsols}
    (u_1,u_2)\in\left\{\Big(\tfrac13,\tfrac23\Big),\
    \Big(\tfrac{\omega}{3},\tfrac{2\omega}{3}\Big),\
    \Big(\tfrac{1+\omega}{3},\tfrac{2+2\omega}{3}\Big),\
    \Big(\tfrac{2+\omega}{3},\tfrac{1+2\omega}{3}\Big)\right\}\ .
\end{equation}
There are also Weyl-fixed solutions on this branch, but since we are in type $A$
these do not contribute; see the discussion in section~\ref{subsub:weyl-contrib}. The second
branch gives the continuous family, defined implicitly by
\begin{equation}\label{eq:A2cont}
    \frac{\theta_4(u_1)\,\theta_4(u_2)\,\theta_4(u_1-u_2)}
         {\theta_2(u_1)\,\theta_2(u_2)\,\theta_2(u_1-u_2)}=C\ ,
\end{equation}
with the $\Delta_a$- and $\omega$-dependent constant
\begin{equation}\label{eq:C}
    C\coloneqq\frac{\theta_4(\Delta_1)\,\theta_4(\Delta_2)\,
      \theta_4(\Delta_1+\Delta_2)}
      {\theta_2(\Delta_1)\,\theta_2(\Delta_2)\,\theta_2(\Delta_1+\Delta_2)}\ .
\end{equation}
The mechanism behind this family is worth isolating. At rank two, a
one-dimensional family is a codimension-one locus, and the two Bethe operators
can vanish simultaneously along such a locus only if they share a common factor:
\eqref{eq:A2_sols_or} exhibits precisely this, the factor $h$ being common to
$Q_1$ and $Q_2$. This is a stringent condition, and $A_2$ turns out to be the
only rank-two algebra whose BAEs meet it, which is why no continuous families
appear in the remaining cases of sections~\ref{sec:D2}--\ref{sec:B2}.

%%%%%%%%%%%%%%%%%%%%%%%%%%%%%%%%%%%%%%%%%%%%%%%%%%%%%%%%%%%%%%%
\section{The simplest rank-two case: $D_2=\mathfrak{so}(4)$ SYM}
\label{sec:D2}
%%%%%%%%%%%%%%%%%%%%%%%%%%%%%%%%%%%%%%%%%%%%%%%%%%%%%%%%%%%%%%%

We begin our search for BAE solutions beyond type $A$ with the simplest 
possible case, $D_2=\mathfrak{so}(4)$. Its simplicity stems from $D_2$ not 
being simple: since $D_2=A_1\oplus A_1$ (the familiar 
$\mathfrak{so}(4)\cong\mathfrak{su}(2)\oplus\mathfrak{su}(2)$), the solutions of 
the $D_2$ SYM BAEs must be pairs of solutions of the $A_1$ BAEs, which, as we have seen in section~\ref{sub:A1}, are 
known. Although we could not find this observation stated explicitly in the 
literature, it is surely known to the experts.\footnote{\label{foot:D2}It was certainly known 
to A. Amariti, C. Mascherpa, P. Glorioso, and A. Zanetti (with whom MF had worked out figure~\ref{fig:D2_orbits_ortho} in November 2024); see the thesis 
\cite{Mascherpa:2022thesis}, where a partial list of $D_2$ solutions (found 
using this very observation) appears, namely 
$(u_1,u_2)\in \big\{ \big(\tfrac12,\tfrac{\omega}{2}\big),\, 
\big(\tfrac12,\tfrac{\omega+1}{2}\big),\, 
\big(\tfrac{\omega}{2},\tfrac{\omega+1}{2}\big)\}$. Here $u_1,u_2$ are the 
holonomies of the two $A_1$ summands. Since we do not quotient by 
the outer automorphism $\mathrm{Out}(D_2)=\mathbb{Z}_2$ exchanging them (see 
section~\ref{subsub:physsol}), the reflected pairs $(u_2,u_1)$ are distinct 
solutions, not identified with $(u_1,u_2)$, though the three lie in a  
$\mathrm{PSL}(2,\zz)$ sextuplet (see figure~\ref{fig:D2_orbits}).}
The simply connected cover of $G$ whose $\mathrm{Lie}(G)=D_2$ is $\widetilde{G}=\mathrm{Spin}(4)$; its center 
$Z(\widetilde{G})=\zz_2\times\zz_2$ has four subgroups (the identity 
$\{e\}$, two copies of $\zz_2$, and the full $\zz_2\times\zz_2$), giving the 
five global forms of table~\ref{tab:D2globalforms}.
\begin{table}[htb!]
\centering\scriptsize
\setlength{\tabcolsep}{3.5pt}
\renewcommand{\arraystretch}{1.5}
\begin{tabular}{cccccc}
\toprule
$\Gamma$ & Global form & Basis of $L$ & $Q_1=1$ & $Q_2=1$ & $\mathrm{TotExp}$ (\ref{eq:min_mod}) \\
\midrule
$\{e\}$ & $\mathrm{Spin}(4)\cong\SU(2)^2$
    & \makecell{$\boldsymbol{\omega}$ (coroots)\\[-3pt]
      $\scriptstyle\{\alpha_1^\vee,\,\alpha_2^\vee\}$}
    & $\mathcal{F}(2u_1)^{2}$ & $\mathcal{F}(2u_2)^{2}$ & $4$ \\
$\zz_2\times\{e\}$ & $\mathrm{PSU}(2)\times\SU(2)$
    & \makecell{mixed\\[-3pt]
      $\scriptstyle\{\tfrac12\alpha_1^\vee,\,\alpha_2^\vee\}$}
    & $-\mathcal{F}(u_1)$ & $\mathcal{F}(2u_2)^{2}$ & $4$ \\
$\{e\}\times\zz_2$ & $\SU(2)\times\mathrm{PSU}(2)$
    & \makecell{mixed\\[-3pt]
      $\scriptstyle\{\alpha_1^\vee,\,\tfrac12\alpha_2^\vee\}$}
    & $\mathcal{F}(2u_1)^{2}$ & $-\mathcal{F}(u_2)$ & $4$ \\
$\zz_2^{\text{diag}}$ & $\SO(4)$
    & \makecell{orthogonal\\[-3pt]
      $\scriptstyle\{\tfrac12(\alpha_1^\vee{+}\alpha_2^\vee),\,
      \tfrac12(\alpha_2^\vee{-}\alpha_1^\vee)\}$}
    & $\mathcal{F}(u_1{+}u_2)\mathcal{F}(u_1{-}u_2)$
    & $\mathcal{F}(u_1{+}u_2)/\mathcal{F}(u_1{-}u_2)$ & $4$ \\
$\zz_2\times\zz_2$ & $\mathrm{PSO}(4)\cong\mathrm{PSU}(2)^2$
    & \makecell{$\boldsymbol{\alpha}$ (coweights)\\[-3pt]
      $\scriptstyle\{\tfrac12\alpha_1^\vee,\,\tfrac12\alpha_2^\vee\}$}
    & $-\mathcal{F}(u_1)$ & $-\mathcal{F}(u_2)$ & $2$ \\
\bottomrule
\end{tabular}
\caption{The five global forms of $D_2\cong A_1\oplus A_1$ as quotients 
$\mathrm{Spin}(4)/\Gamma$ by $\Gamma\subseteq Z(\mathrm{Spin}(4))=\zz_2\times\zz_2$, 
with a primitive basis of the corresponding cocharacter lattice $L$, the resulting 
BAEs in terms of (\ref{eq:Fdef}), and the complexity measure 
(\ref{eq:min_mod}). Here $\alpha_{1,2}^\vee$ are the coroots and 
$\omega_{1,2}^\vee=\tfrac12\alpha_{1,2}^\vee$ the fundamental coweights. An 
explicit overall $-$ sign indicates that the prefactor 
$(-1)^{\sum_{\alpha>0}\langle\alpha,v_i\rangle}$ of (\ref{eq:BAops}) is 
nontrivial (odd exponent) for that equation; where absent, the exponent is even 
and the prefactor is $+1$. Only the adjoint form $\mathrm{PSO}(4)$ in the 
$\boldsymbol{\alpha}$ basis attains $\mathrm{TotalExp}=2$: it is the minimal 
model, and its equations are two decoupled copies of the single-holonomy $A_1$ 
equation. Note that $\mathrm{PSU}(2)\times\SU(2)$ and $\SU(2)\times\mathrm{PSU}(2)$ 
(with $\mathrm{PSU}(2) \cong \SO(3)$) are isomorphic as groups but correspond to distinct subgroups $\Gamma$, 
exchanged by $\mathrm{Out}(D_2)=\zz_2$.}
\label{tab:D2globalforms}
\end{table}

\subsection{The minimal model}
\label{sub:D2basis}

The most natural choice for the $D_N$ family, and the one usually made in 
physics, is the orthogonal basis, corresponding to the $\SO(4)$ global form at $N=2$. 
Its positive roots are $\{(1,1),(1,-1)\}$, so \eqref{eq:theta1_BAE} becomes
\begin{subequations}\label{eq:BAE_D2}
\begin{align}
        Q_1&=\displaystyle\prod_{\Delta}
        \frac{\theta_1(-u_1-u_2+\Delta)\,\theta_1(-u_1+u_2+\Delta)}
        {\theta_1(u_1+u_2+\Delta)\,\theta_1(u_1-u_2+\Delta)}=1\ ,\\[2ex]
        Q_2&=\displaystyle\prod_{\Delta}
        \frac{\theta_1(-u_1-u_2+\Delta)\,\theta_1(u_1-u_2+\Delta)}
        {\theta_1(u_1+u_2+\Delta)\,\theta_1(-u_1+u_2+\Delta)}=1\ .
\end{align}
\end{subequations}
One could solve this by the same manipulations used for $A_2$, or by forming 
the ratio $Q_1/Q_2$.

It is far more efficient, however, to work in the $\boldsymbol{\alpha}$ 
(fundamental-coweight) basis, which for $D_2$ is the minimal model of BAEs and 
corresponds to the adjoint global form $\mathrm{PSO}(4)\cong \mathrm{PSU}(2)\times \mathrm{PSU}(2)\cong \SO(3)\times\SO(3)$. 
Its positive roots are $\{(0,1),(1,0)\}$, and the equations collapse to two 
decoupled copies of the single-variable $A_1$ equation:
\begin{equation}\label{eq:BAE_A1A1}
        Q_1=\displaystyle-\prod_{\Delta}
        \frac{\theta_1(-u_2+\Delta)}{\theta_1(u_2+\Delta)}=1\ ,\quad
        Q_2=\displaystyle-\prod_{\Delta}
        \frac{\theta_1(-u_1+\Delta)}{\theta_1(u_1+\Delta)}=1\ ,
\end{equation}
i.e.\ $Q_1=\mathcal{F}(u_2)=1$ and $Q_2=\mathcal{F}(u_1)=1$ in the notation of 
\eqref{eq:Fdef} --- the concrete realization of the statement that the $D_2$ 
solutions are pairs of $A_1$ solutions. In this basis the Weyl group is 
generated by the two independent sign flips $u_i\mapsto-u_i$,
\begin{equation}
    \mathcal{W}_{D_2} = \left\{
    \begin{psmallmatrix} 1 & 0 \\ 0 & 1 \end{psmallmatrix},\,
    \begin{psmallmatrix} -1 & 0 \\ 0 & 1 \end{psmallmatrix},\,
    \begin{psmallmatrix} 1 & 0 \\ 0 & -1 \end{psmallmatrix},\,
    \begin{psmallmatrix} -1 & 0 \\ 0 & -1 \end{psmallmatrix}
    \right\}\cong\mathbb{Z}_2\times\mathbb{Z}_2\ .
\end{equation}
Borrowing the $A_1$ result in each factor, the solutions are the fully rational
\begin{equation}
\label{eq:D2_sols}
    (u_1,u_2)\ ,\qquad u_1,u_2\in\left\{\tfrac12,\ \tfrac{\omega}{2},\ 
    \tfrac{1+\omega}{2}\right\}\ .
\end{equation}
These sit at the half-periods of the torus and are therefore 
Weyl-fixed in the orthogonal basis: each $u_i$ obeys $-u_i = u_i$ modulo the lattice, so the 
sign-flip generators act trivially. However, these solutions once translated to the coroot basis are no longer Weyl-fixed. Therefore, as discussed in section \ref{subsub:weyl-contrib}, they are guaranteed to be actually contributing and not yet proper examples of Weyl-fixed and contributing solutions (which we will first encounter for $G_2$).

\subsection{\texorpdfstring{$\mathrm{PSL}(2,\mathbb{Z})$}{PSL(2,Z)} orbits}
\label{sub:D2orbits}

It remains to organize the solutions \eqref{eq:D2_sols} into
$\mathrm{PSL}(2,\mathbb{Z})$ orbits. Although the equations decouple over the
two $A_1$ summands, the solutions share the single modular parameter $\omega$,
so it is the \emph{diagonal} $\mathrm{PSL}(2,\mathbb{Z})$, acting simultaneously
on $u_1$ and $u_2$, that acts on them. Applying \eqref{eq:PSLonsol-bis}, the nine
solutions fall into \emph{two} orbits, shown in figure~\ref{fig:D2_orbits}: a
sextuplet of pairs with distinct components, among them the three of
footnote~\ref{foot:D2}, and a triplet of the diagonal pairs $(x,x)$.
Figure~\ref{fig:D2_orbits_ortho} displays the same two orbits in the orthogonal
($\SO(4)$) basis, providing the dictionary between the two coordinate choices.
That the two figures are the same graph with different node labels is precisely
the basis-independence of the physical (orbit) data: although the equations
decouple in the $\SO(3)\times\SO(3)$ form, the diagonal
$\mathrm{PSL}(2,\mathbb{Z})$ binds the two summands' solutions into common
orbits.
\begin{figure}[htb!]
\centering
\begin{tikzpicture}
  \def\R{2.3}
  \node[sol, label={[sollbl]120:$\left(\tfrac12,\ \tfrac{\omega+1}{2}\right)$}] (n1) at (120:\R) {};
  \node[sol, label={[sollbl] 60:$\left(\tfrac12,\ \tfrac{\omega}{2}\right)$}]   (n2) at ( 60:\R) {};
  \node[sol, label={[sollbl]  0:$\left(\tfrac{\omega}{2},\ \tfrac12\right)$}]   (n3) at (  0:\R) {};
  \node[sol, label={[sollbl]-60:$\left(\tfrac{\omega+1}{2},\ \tfrac12\right)$}] (n4) at (-60:\R) {};
  \node[sol, label={[sollbl]240:$\left(\tfrac{\omega+1}{2},\ \tfrac{\omega}{2}\right)$}] (n5) at (240:\R) {};
  \node[sol, label={[sollbl]180:$\left(\tfrac{\omega}{2},\ \tfrac{\omega+1}{2}\right)$}] (n6) at (180:\R) {};

  \draw[Tedge] (n1) -- node[edgelbl, above]      {$T$} (n2);
  \draw[Sedge] (n2) -- node[edgelbl, above right]{$S$} (n3);
  \draw[Tedge] (n3) -- node[edgelbl, below right]{$T$} (n4);
  \draw[Sedge] (n4) -- node[edgelbl, below]      {$S$} (n5);
  \draw[Tedge] (n5) -- node[edgelbl, below left] {$T$} (n6);
  \draw[Sedge] (n6) -- node[edgelbl, above left] {$S$} (n1);

  \begin{scope}[yshift=-4.6cm]
    \node[sol, label={[sollbl, label distance=4pt]90:$\left(\tfrac{\omega+1}{2},\ \tfrac{\omega+1}{2}\right)$}] (m1) at (-2.5,0) {};
    \node[sol, label={[sollbl, label distance=4pt]90:$\left(\tfrac{\omega}{2},\ \tfrac{\omega}{2}\right)$}]     (m2) at (   0,0) {};
    \node[sol, label={[sollbl, label distance=4pt]90:$\left(\tfrac12,\ \tfrac12\right)$}]                       (m3) at ( 2.5,0) {};

    \draw[Tedge] (m1) -- node[edgelbl, below]{$T$} (m2);
    \draw[Sedge] (m2) -- node[edgelbl, below]{$S$} (m3);
    \draw[Sloop] (m1) to[out=180+24,in=180-24,looseness=50]
                 node[edgelbl, left] {$S$} (m1);
    \draw[Tloop] (m3) to[out=-24,in=+24,looseness=50]
                 node[edgelbl, right]{$T$} (m3);
  \end{scope}
\end{tikzpicture}
\caption{The two $\mathrm{PSL}(2,\mathbb{Z})$ orbits of the $D_2$ solutions
(\ref{eq:D2_sols}), in the $\boldsymbol{\alpha}$ (coweight) basis: a
six-element orbit of the pairs with distinct components (top) and a
three-element orbit of the diagonal pairs (bottom). Blue and red edges denote
the generators $S$ and $T$ of (\ref{eq:PSLonsol-bis}).}
\label{fig:D2_orbits}
\end{figure}
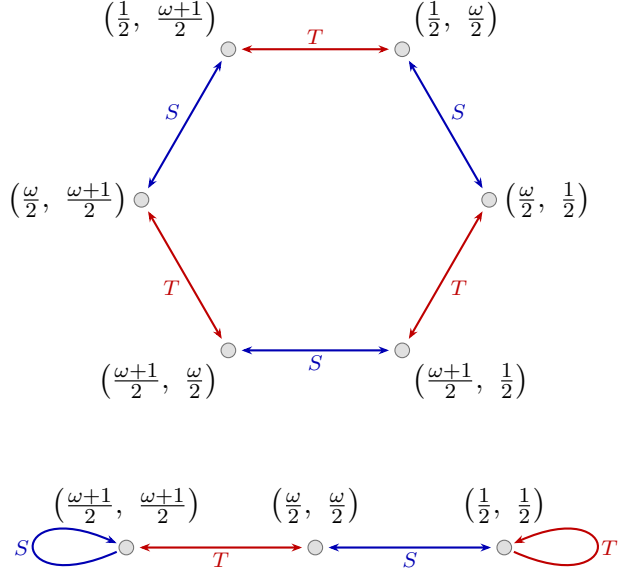
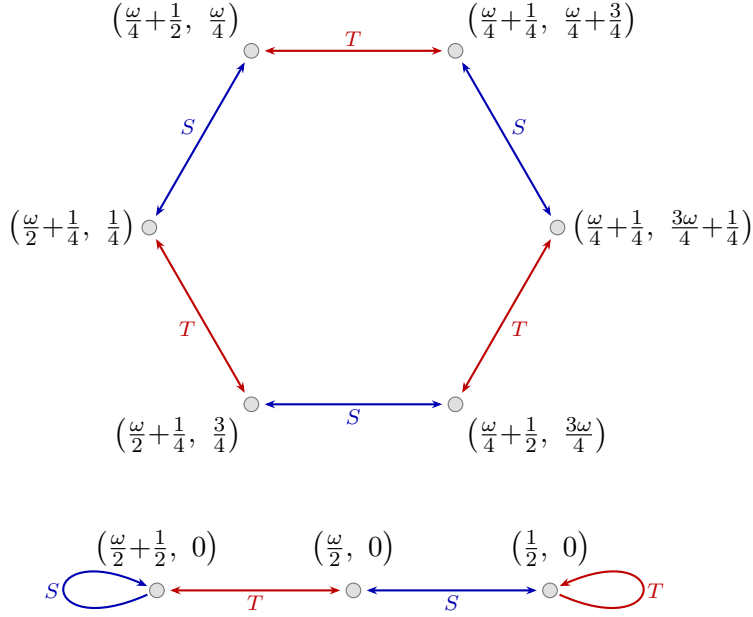
\begin{figure}[htb!]
\centering
\begin{tikzpicture}
  \def\R{2.7}
  \node[sol, label={[sollbl]120:$\left(\tfrac{\omega}{4}{+}\tfrac12,\ \tfrac{\omega}{4}\right)$}] (n1) at (120:\R) {};
  \node[sol, label={[sollbl] 60:$\left(\tfrac{\omega}{4}{+}\tfrac14,\ \tfrac{\omega}{4}{+}\tfrac34\right)$}] (n2) at (60:\R) {};
  \node[sol, label={[sollbl]  0:$\left(\tfrac{\omega}{4}{+}\tfrac14,\ \tfrac{3\omega}{4}{+}\tfrac14\right)$}] (n3) at (0:\R) {};
  \node[sol, label={[sollbl]-60:$\left(\tfrac{\omega}{4}{+}\tfrac12,\ \tfrac{3\omega}{4}\right)$}] (n4) at (-60:\R) {};
  \node[sol, label={[sollbl]240:$\left(\tfrac{\omega}{2}{+}\tfrac14,\ \tfrac34\right)$}] (n5) at (240:\R) {};
  \node[sol, label={[sollbl]180:$\left(\tfrac{\omega}{2}{+}\tfrac14,\ \tfrac14\right)$}] (n6) at (180:\R) {};

  \draw[Tedge] (n1) -- node[edgelbl, above]      {$T$} (n2);
  \draw[Sedge] (n2) -- node[edgelbl, above right]{$S$} (n3);
  \draw[Tedge] (n3) -- node[edgelbl, below right]{$T$} (n4);
  \draw[Sedge] (n4) -- node[edgelbl, below]      {$S$} (n5);
  \draw[Tedge] (n5) -- node[edgelbl, below left] {$T$} (n6);
  \draw[Sedge] (n6) -- node[edgelbl, above left] {$S$} (n1);

  \begin{scope}[yshift=-4.8cm]
    \node[sol, label={[sollbl, label distance=4pt]90:$\left(\tfrac{\omega}{2}{+}\tfrac12,\ 0\right)$}] (m1) at (-2.6,0) {};
    \node[sol, label={[sollbl, label distance=4pt]90:$\left(\tfrac{\omega}{2},\ 0\right)$}]            (m2) at (   0,0) {};
    \node[sol, label={[sollbl, label distance=4pt]90:$\left(\tfrac12,\ 0\right)$}]                     (m3) at ( 2.6,0) {};

    \draw[Tedge] (m1) -- node[edgelbl, below]{$T$} (m2);
    \draw[Sedge] (m2) -- node[edgelbl, below]{$S$} (m3);
    \draw[Sloop] (m1) to[out=180+24,in=180-24,looseness=50] node[edgelbl, left] {$S$} (m1);
    \draw[Tloop] (m3) to[out=-24,in= 24,looseness=50] node[edgelbl, right]{$T$} (m3);
  \end{scope}
\end{tikzpicture}
\caption{The same two orbits in the orthogonal ($\SO(4)$) basis, obtained from
figure~\ref{fig:D2_orbits} by
$(u_1,u_2)\mapsto\big(\tfrac{u_1+u_2}{2},\tfrac{u_2-u_1}{2}\big)$; each node
occupies the same position as its counterpart there. That the two figures are
the same graph with different node labels is precisely the basis-independence
of the physical (orbit) data.}
\label{fig:D2_orbits_ortho}
\end{figure}

%%%%%%%%%%%%%%%%%%%%%%%%%%%%%%%%%%%%%%%%%%%%%%%%%%%%%%%%%%%%%%%%%%%%%%%%%%%%%%%%%%%%%%%%%%%%%%%%%%%%
\section{Analytic results: $B_2 = \mathfrak{so}(5) \cong \mathfrak{usp}(4) =C_2$ SYM}
\label{sec:B2}
%%%%%%%%%%%%%%%%%%%%%%%%%%%%%%%%%%%%%%%%%%%%%%%%%%%%%%%%%%%%%%%%%%%%%%%%%%%%%%%%%%%%%%%%%%%%%%%%%%%%

We now come to the first genuinely new case, and the main result of this paper:
$\mathcal{N}=4$ SYM with gauge algebra $B_2\cong C_2$. Despite being, like
$A_2$, of rank two, it exhibits a range of phenomena absent in type $A$, and
the analysis below shows that guessing a uniform all-rank ansatz, as one can in
type $A$, is hopeless outside it.

Our route to the solutions is indirect. We first search numerically at fixed
fugacities, obtaining a list of solutions and an indication of how they are
distributed (section~\ref{sub:num}); we then use that list as an Ansatz to
solve the equations analytically, recovering every solution in closed form
(section~\ref{sub:sol}). The solutions so obtained fall into three families:
fully rational, half-pinned, and half-rational.

A word on completeness. Every solution we find is isolated: the numerical
searches produced no continuous families, in contrast with $A_2$, cf. \eqref{eq:A2cont}. We have
closed-form expressions for all of them, but no analytic proof that the list is
exhaustive. Strong evidence of this comes from the numerical evaluation of
the index in section~\ref{sub:num-eval}, where the Bethe Ansatz sum over our
solutions is compared with the direct expansion of the localisation integral and
the two agree order by order. A proof will be given in the companion paper
\cite{polypaper}, where the BAEs are recast as polynomial equations and the
solution set is determined by methods of computational commutative algebra. In
that sense, the completeness statement made here is established analytically,
though not in this paper.

Finally, in section~\ref{sub:C2} we discuss the $C_2$ presentation, related to
the above by the change of primitive basis implementing Langlands duality.

\subsection{The equations}

The minimal model \eqref{eq:min_mod} for $B_2$ is realized in the orthonormal 
basis $\{e_1,e_2\}$ of the root lattice, for which the cocharacter lattice is 
$\mathbb{Z}^2$ and the corresponding global form is the adjoint group $\mathrm{SO}(5)$. 
This is the optimal choice both for the complexity of the equations and 
because the center redundancy has already been quotiented out (the orthonormal 
basis spans the full coweight lattice, i.e.\ the adjoint form).

The Weyl group is the set
\begin{equation}
    \mathcal{W}_{B_2}=\left\{
    \begin{psmallmatrix} 1 & 0 \\ 0 & 1 \end{psmallmatrix},\,
    \begin{psmallmatrix} 0 & 1 \\ 1 & 0 \end{psmallmatrix},\,
    \begin{psmallmatrix} 1 & 0 \\ 0 & -1 \end{psmallmatrix},\,
    \begin{psmallmatrix} 0 & -1 \\ 1 & 0 \end{psmallmatrix},\,
    \begin{psmallmatrix} 0 & 1 \\ -1 & 0 \end{psmallmatrix},\,
    \begin{psmallmatrix} -1 & 0 \\ 0 & 1 \end{psmallmatrix},\,
    \begin{psmallmatrix} 0 & -1 \\ -1 & 0 \end{psmallmatrix},\,
    \begin{psmallmatrix} -1 & 0 \\ 0 & -1 \end{psmallmatrix}
    \right\} \cong D_4\,
\end{equation}
i.e. the dihedral group of order $8$. The positive 
roots, in components, are
\begin{equation*}
    \{\alpha>0\}_{B_2}=\{(1,1),\,(1,0),\,(0,1),\,(1,-1)\}\ ,
\end{equation*}
and the BAEs \eqref{eq:theta1_BAE} take the form:\footnote{The $B_2$ solutions were previously sought in \cite{Mascherpa:2022thesis},
where the equations were written in the orthogonal basis without the sign
prefactor $(-1)^{\langle2\rho_{\mathcal{W}},e_i\rangle}$ of (\ref{eq:BAops}). The equations solved there therefore differ from
(\ref{eq:BAE_B2}) by an overall sign. The fully rational sector of that system
was determined completely, but does not coincide with ours. See
section~\ref{sub:choices} for the general form of the prefactor.}
\begin{subequations}\label{eq:BAE_B2}
\begin{align}
        Q_1 &=\displaystyle\prod_{\Delta}
        \frac{\theta_1(-u_1-u_2+\Delta)\,\theta_1(-u_1+\Delta)\,\theta_1(-u_1+u_2+\Delta)}
        {\theta_1(u_1+u_2+\Delta)\,\theta_1(u_1+\Delta)\,\theta_1(u_1-u_2+\Delta)}=1\ ,\\
        Q_2 &=\displaystyle\prod_{\Delta}
        \frac{\theta_1(-u_1-u_2+\Delta)\,\theta_1(-u_2+\Delta)\,\theta_1(u_1-u_2+\Delta)}
        {\theta_1(u_1+u_2+\Delta)\,\theta_1(u_2+\Delta)\,\theta_1(-u_1+u_2+\Delta)}=1\ .
\end{align}
\end{subequations}
Making use of the $f$ defined in \eqref{eq:fdef} and the identify it satisfies, \eqref{eq:BAE_B2}  become
\begin{subequations}\label{eq:BAE_B2_2}
\begin{align}
        f(-u_1-u_2)\,f(-u_1)\,f(-u_1+u_2) &=f(u_1+u_2)\,f(u_1)\,f(u_1-u_2)\ ,\\
        f(-u_1-u_2)\,f(-u_2)\,f(u_1-u_2) &=f(u_1+u_2)\,f(u_2)\,f(-u_1+u_2)\ .
\end{align}
\end{subequations}

\subsection{Numerical analysis}
\label{sub:num}

As a first, exploratory approach, we search for solutions numerically at fixed 
values of the (generically complex) chemical potentials $\{\omega,\Delta_1,\Delta_2\}$, using a 
multidimensional Newton--Raphson method with repeated random initial points. 
For each solution we check numerically that it is nontrivial (i.e., that its evaluation 
on the integrand $\mathcal{Z}(u;\Delta_a,\omega)$ is nonzero) and isolated (the Jacobian 
of the BAE system $H$ is non-degenerate there). Solutions obtained from different 
random seeds are identified as belonging to the same Weyl orbit by comparing 
their images under $\mathcal{W}_{B_2}$ (modulo the torus lattice) to within a 
fixed numerical tolerance, and we retain a solution as new only if it lies 
outside all previously found orbits.

After several thousand runs we found fifteen distinct nontrivial Weyl orbits, all 
of them isolated, leading us to conjecture that $B_2$ has no continuous 
families of solutions. Since we use the numerical solutions as an Ansatz for 
the subsequent analytic computation, we select within each orbit the simplest 
representative, namely the one with the largest number of rational components 
(of the form $a+b\,\omega$ with $a,b\in\mathbb{Q}$). The chosen representatives are collected in table~\ref{tab:solutions}, at generic chemical potentials.
\begin{table}[h!]
\centering
\renewcommand{\arraystretch}{1.0}
    \begin{tabular}{c l l}
    \hline
    No. & \textbf{$u_1$} & \textbf{$u_2$} \\
    \hline
    1 & $\frac{1}{2}\omega$ & $\frac{1}{2}\omega + \frac{1}{2}$ \\
    2 & $\frac{1}{2}$ & $\frac{1}{2}\omega + \frac{1}{2}$ \\
    3 & $\frac{1}{2}$ & $\frac{1}{2}\omega$ \\
    4 & $\frac{1}{2}$ & $0.63787707\omega + 0.74436341$ \\
    5 & $\frac{1}{2}\omega$ & $0.79454554\omega + 0.52769491$ \\
    6 & $\frac{1}{2}\omega + \frac{1}{2}$ & $0.85370706\omega + 0.10939850$ \\
    7 & $\frac{1}{2}$ & $0.00837757\omega + 0.15480862$ \\
    8 & $\frac{1}{2}\omega$ & $0.73725746\omega + 0.61042566$ \\
    9 & $\frac{1}{2}\omega + \frac{1}{2}$ & $0.72029041\omega + 0.50730831$ \\
    10 & $\frac{1}{2}$ & $0.56877170\omega + 0.24991670$ \\
    11 & $\frac{1}{2}\omega$ & $0.85069257\omega + 0.00307284$ \\
    12 & $\frac{1}{2}\omega + \frac{1}{2}$ & $0.79359409\omega + 0.70878178$ \\
    13 & $0.80168878\omega + 0.38388557$ & $0.80168878\omega + 0.88388557$ \\
    14 & $0.60851224\omega + 0.22275114$ & $0.10851224\omega + 0.22275114$ \\
    15 & $0.88218483\omega + 0.19257884$ & $0.38218483\omega + 0.69257884$ \\
    \hline
    \end{tabular}
\caption{Representatives of $B_2$ solutions in ortogonal basis, one per Weyl orbit, at generic chemical potentials 
$\omega=1.82i,\ \Delta_1=0.421+0.411i,\ \Delta_2=0.34+0.33i$.}
\label{tab:solutions}
\end{table}

\subsection{All analytic solutions }
\label{sub:sol}

   We see that all but three solutions are partially rational: $u_1$ takes one of 
the three values $\{\tfrac12,\ \tfrac{\omega}{2},\ \tfrac{1+\omega}{2}\}$. 
Remarkably, the three remaining solutions all satisfy $u_2=u_1+x$ with $x$ 
again one of these three half-periods. In every case, then, recovering the 
numerical solutions analytically reduces to solving a single equation in the 
one variable $u_1$. We treat the cases in turn.

\paragraph{Fully rational solutions (rows 1--3 in table \ref{tab:solutions}).}
Three of the contributing classes are fully rational:
\begin{equation}
    (u_1,u_2)\in \left\{\left(\tfrac{\omega}{2},\tfrac{1+\omega}{2}\right),\
    \left(\tfrac{1}{2},\tfrac{1+\omega}{2}\right),\
    \left(\tfrac{1}{2},\tfrac{\omega}{2}\right)\right\}.
\end{equation}
That these solve \eqref{eq:BAE_B2_2} is immediate from the identity 
$f\!\left(\tfrac{a+b\omega}{2}\right)/f\!\left(-\tfrac{a+b\omega}{2}\right)=1$, 
valid for any $a,b\in\mathbb{Z}$ (since $f$ is even up to a quasi-period that 
cancels in the ratio at half-periods).

\paragraph{The 1-partially pinned solutions (rows 13--15 in table \ref{tab:solutions}) with $u_2=u_1+x$ 
(Weyl-equivalent to $u_1+u_2=x$).}
Using the shift identities of $\theta_1$, one sees from \eqref{eq:BAE_B2} that 
whenever $u_2=u_1+x$ a given $u_1$ solves $Q_1$ if and only if it solves $Q_2$; 
we may therefore restrict to $Q_1$. Moreover, since $f(x)/f(-x)=1$, the 
equation collapses to
\begin{equation}
    f(-2u_1+x)\,f(-u_1)=f(2u_1-x)\,f(u_1)\ .
\end{equation}
This is precisely the $\mathfrak{su}(3)$ equation of \cite{Benini:2021ano}: the 
$1$-partially pinned sector of $B_2$ is governed by $A_2$ dynamics. Borrowing 
the $A_2$ solution \eqref{eq:A2_sols_or} gives
\begin{equation*}
    \theta_1(3u_1-x)=0 \quad \text{or} \quad 
    \frac{\theta_4(2u_1-x)\,\theta_4(u_1)\,\theta_4(u_1-x)}
         {\theta_2(2u_1-x)\,\theta_2(u_1)\,\theta_2(u_1-x)} 
    = C\ .
\end{equation*}
The first branch gives $u_1=x$ (hence $u_2=0$), which contributes trivially to 
the index. The second branch is solved by reducing it to a polynomial:
\begin{enumerate}
    \item rewrite each ratio $\theta_4/\theta_2$ as $\mathrm{cn}$ (see \eqref{eq:theta-ratio-nc}) and apply the 
    shift formulae of appendix~\ref{app:ellitpic}, producing a product of Jacobi functions on the left-hand 
    side;
    \item expand the Jacobi functions of argument $2u_1$ via the 
    double-argument formulae;
    \item use the elliptic identities to convert every $\mathrm{cn}^2$ and 
    $\mathrm{dn}^2$ into $\mathrm{sn}^2$, obtaining a quadric in 
    $\mathrm{sn}^2(u_1)$.
\end{enumerate}
Defining the $\Delta$- and $\omega$-dependent constant
\begin{align}\label{eq:D}
    D&\coloneqq\Big(\frac{1-m}{m}\Big)^{3/4}\frac{1}{C}\ ,
\end{align}
in terms of $C$ from \eqref{eq:C} and with $m$ the elliptic parameter and $K\equiv K(m)$ the quarter-period of
appendix~\ref{app:theta-to-jacobi}, the solutions are
\begin{subequations}\label{eq:B2halfpinnedBAE}
\begin{align}
  u_1&=\tfrac{1}{4K}\sn^{-1}\sqrt{\tfrac{D\big(mD+2(1-m)\big)}{(mD+1-m)^2}}\ , \quad
     u_2=u_1+\tfrac12\ ,\\
  u_1&=\tfrac{1}{4K}\sn^{-1}\sqrt{\tfrac{2mD-m+1}{m\big(1-m(1-D)^2\big)}}\ ,\quad
     u_2=u_1+\tfrac{\omega}{2}\ ,\\
  u_1&=\tfrac{1}{4K}\sn^{-1}\sqrt{1-\Big(\tfrac{m-1}{mD}\Big)^{2}}\ ,\quad
     u_2=u_1+\tfrac{1+\omega}{2}\ .
\end{align}
\end{subequations}
To the best of our knowledge, these are the first closed-form isolated solutions
to the BAEs that are not fully rational.

\paragraph{The 1/2-partially rational solutions (rows 4--12 in table \ref{tab:solutions}) $u_1=x$ 
(Weyl-equivalent to $u_2=x$).}
Here, utilising the usual $f(x)/f(-x)=1$ as well as \eqref{eq:A2_sols_or}, we see that $Q_1$ is automatically satisfied for 
$u_1\in\{\tfrac12,\ \tfrac{\omega}{2},\ \tfrac{1+\omega}{2}\}$, so we 
restrict to $Q_2$, which becomes
\begin{equation*}
   f(-x-u_2)\,f(-u_2)\,f(x-u_2) =f(x+u_2)\,f(u_2)\,f(-x+u_2)\ .
\end{equation*}
The strategy is as before: reduce to a polynomial in $\mathrm{sn}^2(2u_2)$. We 
shift all $\theta_i$ to common argument $2u_1$, divide by a suitable power of 
$\theta_4(2u_2)$ to express the ratios $\theta_i/\theta_4$ as Jacobi functions 
(see \eqref{eq:theta-def}), apply the elliptic Pythagorean identities, and 
square. Carrying this out with computer algebra (\textsc{Mathematica}), and 
eliminating $c_4$ via the Jacobi relation 
$\theta_2(0)\,c_2+\theta_3(0)\,c_3+\theta_4(0)\,c_4=0$, the result is a cubic in 
$S\coloneqq\mathrm{sn}^2(4Ku_2)$ (times an overall factor of $S$, whose root 
$S=0$ recovers the fully rational solutions of rows 1--3 in table \ref{tab:solutions}). Solving the cubic by the standard 
formula yields closed forms for $S$, each producing several candidate solutions $\hat{u}_2$ upon taking inverses:
\begin{equation}
    \pm \hat{u}_2+\sigma_1\,\tfrac12+\sigma_2\,\tfrac{\omega}{2}\ ,
    \qquad \sigma_1,\sigma_2\in\{-1,0,+1\}\ ,
\end{equation}
most of which are Weyl-equivalent. $\hat{u}_2\coloneqq\tfrac{1}{4K}\,
\mathrm{sn}^{-1}(\sqrt{S})$ is the principal-branch value.

Each value of $x$ thus yields three $1/2$-partially rational solutions and one
fully rational solution. The closed forms are lengthy and not especially
illuminating, but we collect the cubics in appendix~\ref{sec:closed-form}. In each case, the cubic reads
\begin{equation}\label{eq:B2cubic}
    A_3\,S^3+A_2\,S^2+A_1\,S+A_0=0\ ,
\end{equation}
with coefficients $A_k\equiv A_k(m,c_1,c_2,c_3)$ given in section~\ref{sub:x12} for $x=1/2$, in section~\ref{sub:xomega2} for $x=\omega/2$, and in section~\ref{sub:x1plusomega2} for $x=(1+\omega)/2$. (Here $m\coloneqq k^2$ and the $c_i$ are as in \eqref{eq:cdef}.)

\begin{figure}[hbt!]
\centering
\begin{tikzpicture}
  \foreach \y [count=\row] in {0,-2.6,-5.2,-7.8,-10.4} {
    \begin{scope}[yshift=\y cm]
      \pgfmathtruncatemacro{\nOne}{(\row-1)*3+1}
      \pgfmathtruncatemacro{\nTwo}{(\row-1)*3+2}
      \pgfmathtruncatemacro{\nThree}{(\row-1)*3+3}

      \node[sol, label={[sollbl]90:$\nOne$}]   (N1) at (0,0) {};
      \node[sol, label={[sollbl]90:$\nTwo$}]   (N2) at (2.8,0) {};
      \node[sol, label={[sollbl]90:$\nThree$}] (N3) at (5.6,0) {};

      \draw[Sloop] (N1) to[out=180+24,in=180-24,looseness=50]
                   node[edgelbl, left] {$S$} (N1);
      \draw[Tedge] (N1) -- node[edgelbl, below]{$T$} (N2);
      \draw[Sedge] (N2) -- node[edgelbl, below]{$S$} (N3);
      \draw[Tloop] (N3) to[out=-24,in=24,looseness=50]
                   node[edgelbl, right]{$T$} (N3);
    \end{scope}
  }
\end{tikzpicture}
\caption{The $\mathrm{PSL}(2,\mathbb{Z})$ orbits of the $B_2$ BAE solutions,
with vertices labeled as in table~\ref{tab:solutions}. All five are copies of
the $A_1$-type orbit: the first consists of the fully rational solutions, the
fifth of the half-pinned ones, and the remaining three of the half-rational
ones. Blue edges denote $S$, red edges $T$.}
\label{fig:b2_saddles}
\end{figure}
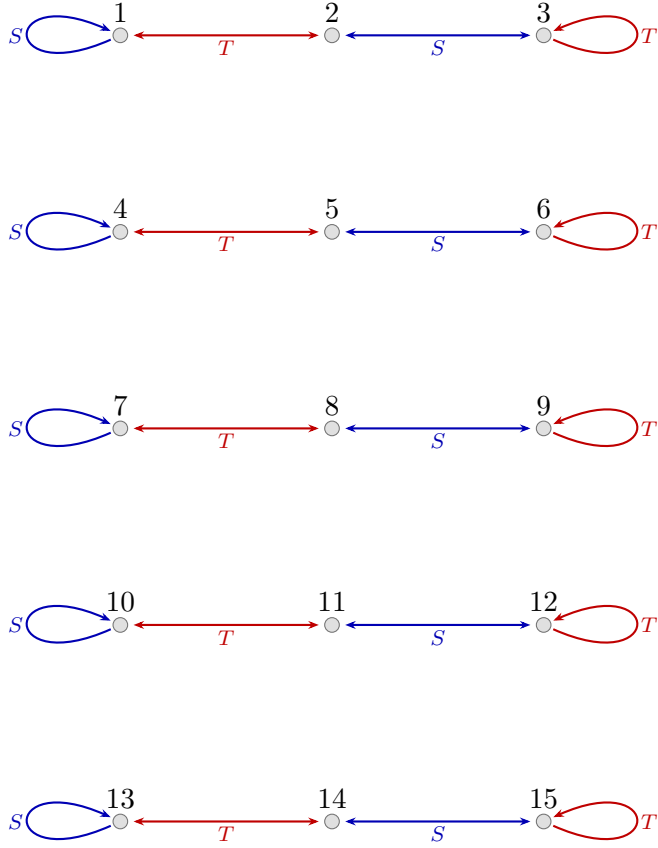

A natural question, and a further check on the completeness of our list, is to
determine the orbits under the $\mathrm{PSL}(2,\mathbb{Z})$ action
\eqref{eq:PSLonsol-bis} of section~\ref{subsub:symm}. The result is shown in
figure~\ref{fig:b2_saddles}: the fifteen solutions fall into five orbits, each a
copy of the $A_1$-type orbit of figure \ref{fig:A_1}.

\subsection{Numerical evaluation of the index}
\label{sub:num-eval}

\begin{figure}[ht!]
    \centering
    \includegraphics[width=0.8\linewidth]{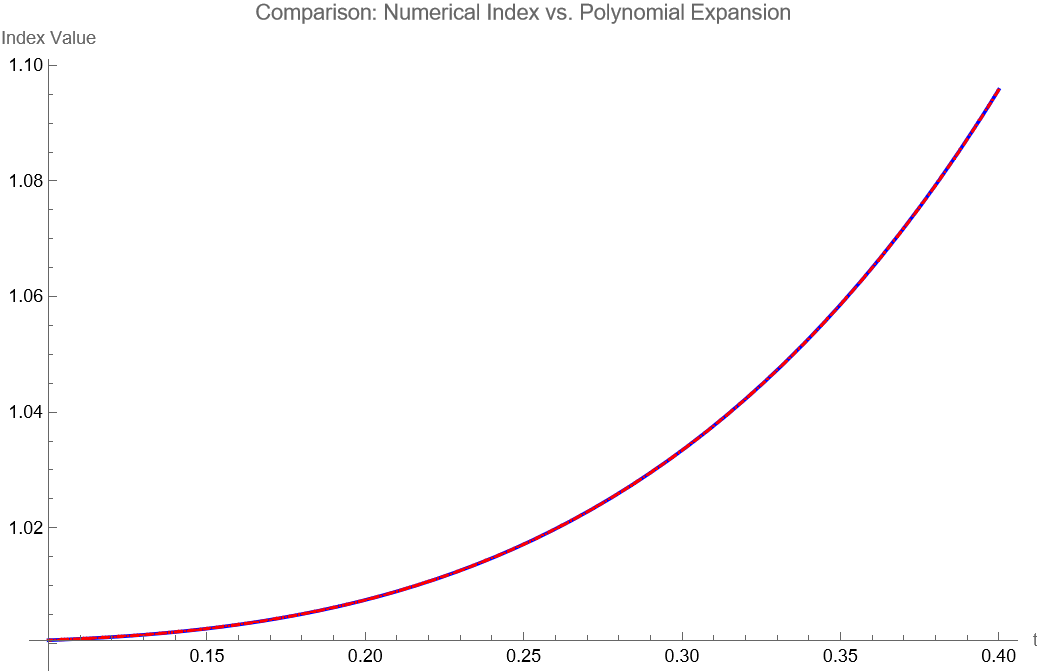}
    \caption{The $t$-expansion of the superconformal index of $B_2$ SYM,
    computed via the Bethe Ansatz sum over the solutions of
    section~\ref{sub:sol} versus the direct localization expansion of
    (\ref{eq:index_integral}). The two agree to the order computed, and converge to 1 as $t\to0$ (on the horizontal axis).}
    \label{fig:index}
\end{figure}

We close with an important check on the solutions found above. Because most of
them are pinned or only partially rational, rather than fully rational as in
type $A$, evaluating the finite-sum formula \eqref{eq:index-finite} in closed
form is cumbersome, so we proceed numerically. We compute the expansion of the
superconformal index in the small fugacity $t$ of \eqref{eq:fugacity_dictionary}
at $p=q$ in two independent ways and compare them: the finite Bethe Ansatz sum
over the solutions of section~\ref{sub:sol}, and the direct expansion of the
localization integral \eqref{eq:index_integral}. The latter gives
{\small
\begin{equation}\label{eq:B2texp}
    \mathcal{I}=1+6t^4-6t^5-7t^6+18t^7+27t^8-72t^9-15t^{10}+198t^{11}-86t^{12}-426t^{13}+441t^{14}+844t^{15}+\mathcal{O}(t^{16})\ ,
\end{equation}}%
and, as shown in figure~\ref{fig:index}, the Bethe Ansatz sum reproduces it (through $\mathcal{O}(t^{16})$). This provides strong
evidence that we have found all solutions contributing to the $B_2\cong C_2$
BAEs.

We do not, however, prove completeness here. A near-complete proof comes from a
polynomial method, in which the BAEs are recast as polynomial equations,
developed in the companion paper \cite{polypaper}. That method applies in
principle to \emph{any} 4D $\mathcal{N}=1$ gauge theory, not only
$\mathcal{N}=4$; we showcase there its application to high-rank SYM, a
long-standing open problem (see, e.g., \cite{Benini:2021ano}).

\paragraph{A curious observation.}
One further feature emerged from the numerical evaluation. Each solution
contributes with a multiplicity: that of its Weyl orbit (the \emph{Weyl
multiplicity}), or that of its orbit under Weyl and center together (the
\emph{full multiplicity}). Accounting for these, we find three pairs of
solutions contributing \emph{equally} to the index, namely solutions Nos. 1 and 13, 2 and 14, 3 and 15 in the numbering of table~\ref{tab:solutions}. In each
pair the two members have full multiplicities in ratio 4:1 and bare
contributions in the inverse ratio. We have no explanation for this degeneracy; it
may signal a hidden symmetry of the $B_2$ BAEs, which we leave to future
investigation.\footnote{We thank A.~Amariti for discussions on this point.} The
numerical values of the contributions, in the orthogonal and
$\boldsymbol{\omega}$ bases, are collected in
tables~\ref{tab:contributions_ort} and~\ref{tab:contributions_om}.

Comparing the two tables illustrates once again the basis-independence proved in
section~\ref{subsub:basis-indep}. Every bare contribution is four times larger
in the orthogonal basis, compensated exactly by the multiplicities, so that the
full contributions agree; the factor of four is the $(\det M)^2$ of
\eqref{eq:Hchange}. Note also that in the $\boldsymbol{\omega}$-basis every Weyl
multiplicity equals $|\mathcal{W}_{B_2}|=8$, whereas in the orthogonal basis
they take the values $2$, $4$ and $8$. In particular, the orthogonal basis
cannot be used to deduce that a solution is Weyl-fixed, and hence trivial; see
section~\ref{subsub:weyl-contrib}.
\begin{table}[h!]
\centering
\renewcommand{\arraystretch}{0.9}
    \begin{tabular}{c l c c l}
    \hline
    No. & \textbf{Bare contribution} & \textbf{Weyl mult.} & \textbf{Full mult.} & \textbf{Total contribution} \\
    \hline
    1 & $0.46155 + 0.08182i$ & 2 & 2 & $0.92311 + 0.16365i$ \\
    2 & $-0.35183 + 0.09590i$ & 2 & 2 & $-0.70367 + 0.19180i$ \\
    3 & $-0.21172 - 0.27555i$ & 2 & 2 & $-0.42345 - 0.55110i$ \\
    4 & $0.02834 - 0.01039i$ & 4 & 4 & $0.11338 - 0.04159i$ \\
    5 & $-0.56736 - 1.21546i$ & 4 & 4 & $-2.26944 - 4.86186i$ \\
    6 & $0.01847 - 0.02556i$ & 4 & 4 & $0.07389 - 0.10226i$ \\
    7 & $0.43363 + 0.05058i$ & 4 & 4 & $1.73452 + 0.20233i$ \\
    8 & $-1.42625 + 1.41295i$ & 4 & 4 & $-5.70500 + 5.65180i$ \\
    9 & $1.00310 + 1.42204i$ & 4 & 4 & $4.01241 + 5.68816i$ \\
    10 & $0.20548 + 0.12035i$ & 4 & 4 & $0.82195 + 0.48140i$ \\
    11 & $-0.01036 - 0.04996i$ & 4 & 4 & $-0.04144 - 0.19985i$ \\
    12 & $0.66571 - 1.61389i$ & 4 & 4 & $2.66287 - 6.45559i$ \\
    13 & $0.11538 + 0.02045i$ & 8 & 8 & $0.92311 + 0.16365i$ \\
    14 & $-0.08795 + 0.02397i$ & 8 & 8 & $-0.70367 + 0.19180i$ \\
    15 & $-0.05293 - 0.06888i$ & 8 & 8 & $-0.42345 - 0.55110i$ \\
    \hline
    \end{tabular}
\caption{Table of $B_2$ contributions and multiplicities in the orthogonal basis.}
\label{tab:contributions_ort}
\end{table}
\begin{table}[h!]
\centering
\renewcommand{\arraystretch}{0.9}
    \begin{tabular}{c l c c l}
    \hline
    No. & \textbf{Bare contribution} & \textbf{Weyl mult.} & \textbf{Full mult.} & \textbf{Total contribution} \\
    \hline
    1 & $0.11538 + 0.02045i$ & 8 & 8 & $0.92311 + 0.16365i$ \\
    2 & $-0.08795 + 0.02397i$ & 8 & 8 & $-0.70367 + 0.19180i$ \\
    3 & $-0.05293 - 0.06888i$ & 8 & 8 & $-0.42345 - 0.55110i$ \\
    4 & $0.00708 - 0.00259i$ & 8 & 16 & $0.11338 - 0.04159i$ \\
    5 & $-0.14184 - 0.30386i$ & 8 & 16 & $-2.26944 - 4.86186i$ \\
    6 & $0.00461 - 0.00639i$ & 8 & 16 & $0.07389 - 0.10226i$ \\
    7 & $0.10840 + 0.01264i$ & 8 & 16 & $1.73452 + 0.20233i$ \\
    8 & $-0.35656 + 0.35323i$ & 8 & 16 & $-5.70500 + 5.65180i$ \\
    9 & $0.25077 + 0.35551i$ & 8 & 16 & $4.01241 + 5.68816i$ \\
    10 & $0.05137 + 0.03008i$ & 8 & 16 & $0.82195 + 0.48140i$ \\
    11 & $-0.00259 - 0.01249i$ & 8 & 16 & $-0.04144 - 0.19985i$ \\
    12 & $0.16642 - 0.40347i$ & 8 & 16 & $2.66287 - 6.45559i$ \\
    13 & $0.02884 + 0.00511i$ & 8 & 32 & $0.92311 + 0.16365i$ \\
    14 & $-0.02198 + 0.00599i$ & 8 & 32 & $-0.70367 + 0.19180i$ \\
    15 & $-0.01323 - 0.01722i$ & 8 & 32 & $-0.42345 - 0.55110i$ \\
    \hline
    \end{tabular}
\caption{Table of $B_2$ contributions and multiplicities in the $\boldsymbol{\omega}$-basis.}
\label{tab:contributions_om}
\end{table}

\subsection{The equations and solutions for $C_2$ SYM}
\label{sub:C2}

Langlands duality relates the algebras $B_N$ and $C_N$ by exchanging long and
short roots. At the level of groups it exchanges the center with the fundamental
group, so that the simply connected form of one type corresponds to the adjoint
form of the other \cite{Goddard:1976qe}. Its field-theoretic realization is
S-duality of $\mathcal{N}=4$ SYM \cite{Montonen:1977sn}, which for
non-simply-laced algebras acts through a Hecke group rather than
$\mathrm{SL}(2,\mathbb{Z})$ \cite{Argyres:2006qr}.

For $B_2\cong C_2$ the two algebras are isomorphic, so Langlands duality acts as a
self-duality of the algebra, exchanging its two global forms
$\mathrm{USp}(4)\cong \mathrm{Spin}(5)$ and $\mathrm{SO}(5)$. Since the index
depends only on the algebra, this exchange is invisible in the final answer and
amounts, for our purposes, to a change of primitive basis.

We stress that this is a statement about bases and global forms, not about the
modular action of section~\ref{subsub:symm}. The $\mathrm{PSL}(2,\mathbb{Z})$
acting on the BAE solutions is generated by transformations of the modular
parameter $\omega$ of the $S^1\times S^3$ background, whereas S-duality acts on
the holomorphic gauge coupling $\tau_{\mathrm{YM}}$, on which the index does not
depend, and for non-simply-laced algebras is realized not by
$\mathrm{SL}(2,\mathbb{Z})$, as stated above. The two should not be conflated.

For completeness, the $\mathrm{USp(4)}$ BAEs in orthogonal basis read
\begin{subequations}\label{eq:BAE_C2}
\begin{align}
        Q_1 &=\displaystyle\prod_{\Delta}
        \left(\frac{\theta_1(-2u_1+\Delta)}{\theta_1(2u_1+\Delta)}\right)^2
        \frac{\theta_1(-u_1-u_2+\Delta)\,\theta_1(-u_1+u_2+\Delta)}
        {\theta_1(u_1+u_2+\Delta)\,\theta_1(u_1-u_2+\Delta)}=1\ ,\\
        Q_2 &=\displaystyle\prod_{\Delta}
        \left(\frac{\theta_1(-2u_2+\Delta)}{\theta_1(2u_2+\Delta)}\right)^2
        \frac{\theta_1(-u_1-u_2+\Delta)\,\theta_1(u_1-u_2+\Delta)}
        {\theta_1(u_1+u_2+\Delta)\,\theta_1(-u_1+u_2+\Delta)}=1\ ,
\end{align}
\end{subequations}
and the change of basis from the $B_2$ to the $C_2$ holonomy vector is:
\begin{equation}\label{eq:so5tousp4}
    \begin{pmatrix}y_1\\y_2\end{pmatrix}_{\mathfrak{usp}(4)}
    =\begin{pmatrix}\tfrac12&\tfrac12\\[2pt]\tfrac12&-\tfrac12\end{pmatrix}
     \begin{pmatrix}x_1\\x_2\end{pmatrix}_{\mathfrak{so}(5)}\ .
\end{equation}
Applying this matrix to the $B_2$ solutions of section \ref{sub:sol}, we get the set of 15 classes of physical solutions of the $C_2$ BAEs. In table \ref{tab:solutions_c2} we collect their numerical values. Other center-equivalent representatives can be obtained by adding the vector $(1/2, 1/2)$ or $(\omega/2, \omega/2)$. Let us also note that, in this basis, the solutions Nos. 4--12 do not have a half-rational Weyl-representative, but the solutions Nos. 13--15 do.
\begin{table}[h!]
\centering
\renewcommand{\arraystretch}{1.0}
    \begin{tabular}{c l l}
    \hline
    No. & \textbf{$u_1$} & \textbf{$u_2$} \\
    \hline
    1 & $\frac{1}{2}\omega + \frac{1}{4}$ & $\frac{3}{4}$ \\
    2 & $\frac{1}{4}\omega + \frac{1}{2}$ & $\frac{3}{4}\omega$ \\
    3 & $\frac{1}{4}\omega + \frac{1}{4}$ & $\frac{3}{4}\omega + \frac{1}{4}$ \\
    4 & $0.31893854\omega + 0.62218171$ & $0.68106147\omega + 0.87781830$ \\
    5 & $0.64727277\omega + 0.26384746$ & $0.85272723\omega + 0.73615255$ \\
    6 & $0.67685353\omega + 0.30469925$ & $0.82314647\omega + 0.19530075$ \\
    7 & $0.00418879\omega + 0.32740431$ & $0.99581122\omega + 0.17259569$ \\
    8 & $0.61862873\omega + 0.30521283$ & $0.88137127\omega + 0.69478717$ \\
    9 & $0.61014521\omega + 0.50365416$ & $0.88985480\omega + 0.99634585$ \\
    10 & $0.28438585\omega + 0.37495835$ & $0.71561415\omega + 0.12504165$ \\
    11 & $0.67534629\omega + 0.00153642$ & $0.82465372\omega + 0.99846358$ \\
    12 & $0.64679705\omega + 0.60439089$ & $0.85320296\omega + 0.89560911$ \\
    13 & $0.80168878\omega + 0.63388557$ & $\frac{3}{4}$ \\
    14 & $0.35851224\omega + 0.22275114$ & $\frac{1}{4}\omega$ \\
    15 & $0.63218483\omega + 0.44257884$ & $\frac{1}{4}\omega + \frac{3}{4}$ \\
    \hline
    \end{tabular}
\caption{Representatives of $C_2$ solutions in the orthogonal basis, one per Weyl and center orbit, at generic chemical potentials  $\omega=1.82i,\ \Delta_1=0.421+0.411i,\ \Delta_2=0.34+0.33i$.}
\label{tab:solutions_c2}
\end{table}

%%%%%%%%%%%%%%%%%%%%%%%%%%%%%%%%%%%%%%%%%%%%%%%%%%%%%%%%%%%%%%
\section{Numerical and semi-analytic results for $G_2$ SYM}
\label{sec:G2}
%%%%%%%%%%%%%%%%%%%%%%%%%%%%%%%%%%%%%%%%%%%%%%%%%%%%%%%%%%%%%%

We now adopt methods similar to those used for $B_2$ to solve the $G_2$ BAEs
numerically, and to recover one isolated solution analytically. These are the
first solutions to the BAEs with an exceptional gauge algebra, and the analytic
one is our first example of a Weyl-fixed solution that nonetheless contributes
to the index.

\paragraph{Conventions.} The minimal presentation of the $G_2$ BAEs is the
so-called unipotent basis, related to the simple-coroot ($\boldsymbol{\omega}$)
basis by the unit-determinant shear
$\big(\begin{smallmatrix}1&0\\1&1\end{smallmatrix}\big)$. As for $B_2$, this
basis minimizes the complexity heuristic \eqref{eq:min_mod}:
$\mathrm{TotalExp}_{G_2}=6$ in the unipotent basis, against $10$ in either the
$\boldsymbol{\omega}$- or the $\boldsymbol{\alpha}$-basis. In these coordinates the
six positive roots have components
\begin{equation}\label{eq:G2roots}
  \{\alpha>0\}_{G_2}=\{(1,0),\,(-1,1),\,(0,1),\,(1,1),\,(2,1),\,(1,2)\}\ ,
\end{equation}
and the Weyl group $\mathcal{W}_{G_2}\cong D_6$ (dihedral of order $12$) is
generated by the reflections in the short and long simple roots
$\alpha_1=(1,0)$ and $\alpha_2=(-1,1)$,
\begin{equation}\label{eq:G2weylgen}
  s_1=\begin{pmatrix}-1&1\\0&1\end{pmatrix}\ ,\qquad
  s_2=\begin{pmatrix}0&1\\1&0\end{pmatrix}\ ,\qquad
  (s_1s_2)^6=\id_2\ ,
\end{equation}
acting on $u=(u_1,u_2)$ by left multiplication; the remaining ten elements are
products of $s_1$ and $s_2$, listed in table~\ref{tab:G2Weyl}. Unlike $B_2$,
$G_2$ admits a single global form, its center being trivial.
\begin{table}[ht!]
\centering\small
\setlength{\tabcolsep}{6pt}
\begin{tabular}{llc@{\qquad}llc}
\toprule
word & matrix & $\det$ & word & matrix & $\det$ \\
\midrule
$e$          & $\begin{psmallmatrix}1&0\\0&1\end{psmallmatrix}$   & $+$ &
$s_2s_1s_2$      & $\begin{psmallmatrix}1&0\\1&-1\end{psmallmatrix}$  & $-$ \\
$s_1$        & $\begin{psmallmatrix}-1&1\\0&1\end{psmallmatrix}$  & $-$ &
$s_1s_2s_1s_2$   & $\begin{psmallmatrix}0&-1\\1&-1\end{psmallmatrix}$ & $+$ \\
$s_2$        & $\begin{psmallmatrix}0&1\\1&0\end{psmallmatrix}$   & $-$ &
$s_2s_1s_2s_1$   & $\begin{psmallmatrix}-1&1\\-1&0\end{psmallmatrix}$ & $+$ \\
$s_1s_2$     & $\begin{psmallmatrix}1&-1\\1&0\end{psmallmatrix}$  & $+$ &
$s_1s_2s_1s_2s_1$& $\begin{psmallmatrix}0&-1\\-1&0\end{psmallmatrix}$ & $-$ \\
$s_2s_1$     & $\begin{psmallmatrix}0&1\\-1&1\end{psmallmatrix}$  & $+$ &
$s_2s_1s_2s_1s_2$& $\begin{psmallmatrix}1&-1\\0&-1\end{psmallmatrix}$ & $-$ \\
$s_1s_2s_1$  & $\begin{psmallmatrix}-1&0\\-1&1\end{psmallmatrix}$ & $-$ &
$s_1s_2s_1s_2s_1s_2$ & $\begin{psmallmatrix}-1&0\\0&-1\end{psmallmatrix}$ & $+$ \\
\bottomrule
\end{tabular}
\caption{The twelve elements of $\mathcal{W}_{G_2}\cong D_6$ in the unipotent
basis, as reduced words in the simple reflections. The longest element is
$s_1s_2s_1s_2s_1s_2=-\id_2$.}
\label{tab:G2Weyl}
\end{table}%

\paragraph{Numerical solution set.} At $\omega=1.82i$,
$\Delta_1=0.421+0.411i$, $\Delta_2=0.34+0.33i$, we solved the $G_2$
BAEs numerically and found $29$ distinct contributing Weyl orbits, all isolated
(table~\ref{tab:G2}).

\begin{table}[ht!]
\centering\small
\setlength{\tabcolsep}{5pt}
\begin{tabular}{cll}
\toprule
No. & $u_1$ & $u_2$ \\
\midrule
1  & $0.000474982903\,\omega+0.4141520332$ & $0.004348151\,\omega+0.660787769$ \\
2  & $0.006198953\,\omega+0.1328330276$    & $0.644688392\,\omega+0.6123285316$ \\
3  & $0.00785385598\,\omega+0.6255397179$  & $0.373055277\,\omega+0.274277055$ \\
4  & $0.022467171\,\omega+0.3009477460$    & $0.644578130\,\omega+0.6399172631$ \\
5  & $0.023709212\,\omega+0.5266147703$    & $0.639736054\,\omega+0.308172071$ \\
6  & $0.0359849903\,\omega+0.7495150439$   & $0.632974689\,\omega+0.721665306$ \\
7  & $0.1292174289\,\omega+0.3488993979$   & $0.569817555\,\omega+0.074534939$ \\
8  & $0.1401701698\,\omega+0.788561197$    & $0.508993285\,\omega+0.642360415$ \\
9  & $0.1543160800\,\omega+0.4260692348$   & $0.262451648\,\omega+0.006212433$ \\
10 & $0.1581013630\,\omega+0.3032944123$   & $0.298060422\,\omega+0.570401376$ \\
11 & $0.1598380780\,\omega+0.5814901794$   & $0.2472485927\,\omega+0.144071584$ \\
12 & $0.1833232588\,\omega+0.3396004529$   & $0.453677525\,\omega+0.001374283$ \\
13 & $0.2106716176\,\omega+0.1207709052$   & $0.801752218\,\omega+0.3533771623$ \\
14 & $0.2241431781\,\omega+0.4119502823$   & $0.391009716\,\omega+0.5835043703$ \\
15 & $0.272067940\,\omega+0.246596868$     & $0.593535403\,\omega+0.407231021$ \\
16 & $0.280607363\,\omega+0.7924483100$    & $0.548505203\,\omega+0.8409195205$ \\
17 & $0.316660778\,\omega+0.7229666844$    & $0.988485478\,\omega+0.116670039$ \\
18 & $0.337716964\,\omega+0.6342198457$    & $0.652435310\,\omega+0.009553102$ \\
19 & $0.344667686\,\omega+0.9903776665$    & $0.068172276\,\omega+0.044603222$ \\
20 & $0.364181291\,\omega+0.6097762377$    & $0.679091053\,\omega+0.182581906$ \\
21 & $0.364262476\,\omega+0.9824369744$    & $0.658561238\,\omega+0.566832750$ \\
22 & $0.366825074\,\omega+0.04214209541$   & $0.969094408\,\omega+0.7292629378$ \\
23 & $0.372076970\,\omega+0.373463876$     & $0.002241007\,\omega+0.384872529$ \\
24 & $0.448459038\,\omega+0.2460371983$    & $0.646899680\,\omega+0.028781626$ \\
25 & $0.591796298\,\omega+0.2497002106$    & $0.712613002\,\omega+0.6403532800$ \\
26 & $0.602857923\,\omega+0.8917713905$    & $0.889098339\,\omega+0.438398424$ \\
27 & $0.651067934\,\omega+0.3642468021$    & $0.968559554\,\omega+0.051744344$ \\
28 & $0.670273193\,\omega+0.5281695522$    & $0.009380367\,\omega+0.358248713$ \\
29 & $\omega/2$                      & $1/2$ \\
\bottomrule
\end{tabular}
\caption{The $29$ contributing Weyl orbits of $G_2$ BAE solutions, one representative per orbit. Solution No. 29 is the fully rational orbit $(\tfrac{\omega}{2},\tfrac12)$.}
\label{tab:G2}
\end{table}

Apart from row~$29$, none of these solutions is even partially rational: no
holonomy component agrees with a rational of denominator $\leq30$ to within
$10^{-5}$. Nor is any of them pinned: across all $28$ orbits and all six
positive roots \eqref{eq:G2roots}, the pairing $\langle\alpha,u\rangle$ stays at
distance $\geq10^{-2}$ from every half-period, far above the accuracy of the
numerics. In the terminology of section~\ref{sub:special} they are therefore
$0$-partially pinned, i.e.\ generic. This is the first case in which such
solutions occur (every $B_2$ solution of section~\ref{sec:B2} is at least
$1$-pinned). Locating them analytically would require solving the 
$G_2$ system directly, which will be done elsewhere \cite{polypaper}.

\paragraph{The fully rational solution.} Exactly one of the $29$ orbits is
independent of the $\Delta_a$, namely that of
\begin{equation}\label{eq:G2rational}
  (u_1,u_2)=\left(\frac{\omega}{2},\frac12\right)\ .
\end{equation}
Its orbit has size six and consists precisely of the ordered pairs of
\emph{distinct} half-periods $\tfrac12$, $\tfrac{\omega}{2}$,
$\tfrac{1+\omega}{2}$. It is stabilized by
$\{\id_2,-\id_2\}\subset\mathcal{W}_{G_2}$, and $Q_1=Q_2=1$ on it identically in
the $\Delta_a$. 

Crucially, no root of $G_2$ vanishes on it. Evaluating
\eqref{eq:G2roots} and reducing modulo $\zz+\omega\zz$, the six positive roots
give
\begin{equation}\label{eq:G2rootvalues}
  \langle\alpha,{u}\rangle\in
  \left\{\tfrac{\omega}{2},\ \tfrac{1+\omega}{2},\ \tfrac12,\
  \tfrac{1+\omega}{2},\ \tfrac12,\ \tfrac{\omega}{2}\right\}\ ,
\end{equation}
in the order of \eqref{eq:G2roots}: each of the three nonzero half-periods is
attained exactly twice, and none of the pairings lands on the period lattice.
Hence $\mathcal{Z}(u^\star)\neq0$ and the solution contributes, despite being
Weyl-fixed. This is the evasion mechanism anticipated in
section~\ref{subsub:weyl-contrib}, and it differs from the $B_2$ one: here the
sign prefactor $(-1)^{\langle2\rho_{\mathcal{W}},e_i\rangle}$ is trivial, since
$2\rho_{\mathcal{W}}=(4,6)$ has both components even, and the evasion is due
entirely to the absence of a root landing on the period lattice.

\paragraph{Half-pinned solutions.} Finally, we searched the loci on which
exactly one coordinate sits at a nonzero half-period, the $G_2$ analogue of the
$u_2$-pinned rows of table~\ref{tab:solutions}. We found $48$ such Weyl classes,
none of them contributing: $\mathcal{Z}=0$ at every one.

\paragraph{No continuous families.} Our numerical searches yielded no continuous families of solutions for $G_2$. Although we do not provide a formal proof here or present evidence analogous to that of section~\ref{sub:num-eval}, we conjecture that this result holds analytically. We aim to establish a proof in forthcoming work \cite{polypaper}.

%%%%%%%%%%%%%%%%%%%%%%%%%%%%%%%%%%%%%%%%%
\section{Revisiting the Cardy-like limit}
\label{sec:cardy}
%%%%%%%%%%%%%%%%%%%%%%%%%%%%%%%%%%%%%%%%%

The relation between the Bethe Ansatz method and the Cardy-like limit of the
index has been discussed extensively
\cite{Choi:2018hmj,Honda:2019cio,ArabiArdehali:2019tdm,Cabo-Bizet:2019osg,Kim:2019yrz,ArabiArdehali:2019orz,Amariti:2024bsr},
though mostly for specific theories and, within those, for specific solutions
or saddles. Here we take a more systematic view, prompted by the more exotic
behavior of the $B_2$ solutions found in section~\ref{sec:B2}. 

We make two points. First, the Cardy-like limit $|\omega|\to0$ of the index\footnote{We use ``Cardy-like limit'' and ``Cardy equations'' in the
sense of this small-$|\omega|$ limit. A different high-temperature limit of the
$\mathcal{N}=1$ index, controlled by the coefficients of $\mathrm{tr}\,R$ and
$\mathrm{tr}\,R^3$, was introduced earlier in \cite{DiPietro:2014bca} and
developed in
\cite{ArabiArdehali:2015ybk,ArabiArdehali:2016vpb,DiPietro:2016ond}; for more
recent developments see, e.g.,
\cite{Cassani:2021fyv,ArabiArdehali:2021nsx,Ohmori:2021sqg} and references
therein.} of a $\mathcal{N}=1$ theory
\cite{Choi:2018hmj,Honda:2019cio,ArabiArdehali:2019tdm,Cabo-Bizet:2019osg,Kim:2019yrz} can be understood as an infinitesimal version of the Bethe Ansatz
method. Indeed the BAEs are finite-difference equations whose leading term,
as the shift $\omega$ becomes small, are precisely the saddle-point equations one writes down in the Cardy-like limit. 
Second, and less expectedly, the saddle-point equations as usually written are
not general enough to capture all saddles, because the standard asymptotic
expansion of the elliptic gamma function presupposes that no root approaches
the period lattice. This assumption is violated already by the $D_2$ solutions,
and more markedly by those of $B_2\cong C_2$.

\subsection{The Cardy-like approach}
\label{sub:cardysaddle}

We start by writing the index \eqref{eq:int_formula} as a matrix model,
\begin{subequations}\label{eq:matrixmodel}
\begin{equation}
    \mathcal{I}(\omega;\Delta_a)=
    \int_{[0,1]^{\rk\alg}}e^{-S(u;\Delta_a,\omega)}
    \prod_{i=1}^{\rk\alg}du^i\ ,
\end{equation}
with effective action
\begin{equation}
    S(u;\Delta_a,\omega)\coloneqq
    -\log\left(\kappa_\alg\,\mathcal{Z}(u;\Delta_a,\omega)\right)\ .
\end{equation}
\end{subequations}
Since $S$ scales as $\omega^{-2}$ (because of the scaling of $\mathcal{Z}$), the integral is amenable to steepest
descent as $\varepsilon\coloneqq|\omega|\to0$ at fixed phase
$\hat\omega\coloneqq\omega/\varepsilon$, giving
\begin{equation}\label{eq:cardysum}
    \mathcal{I}(\omega;\Delta_a)\ \sim\
    \sum_{\hat u\,\in\,\text{saddles}}
    \mathcal{I}_{\mathrm{Cardy}}(\hat u)\ ,
\end{equation}
i.e. an asymptotic expansion in $\varepsilon$, where the sum runs over solutions of
the \emph{saddle-point} (or \emph{Cardy}) equations
\begin{equation}\label{eq:cardy}
    \partial_iS(u;\Delta_a,\omega)\coloneqq
    \frac{\partial S(u;\Delta_a,\omega)}{\partial u^i}=0\ ,
\end{equation}
and $\mathcal{I}_{\mathrm{Cardy}}(\hat u)$ denotes the Gaussian contribution of
the saddle $\hat u$, computed in \eqref{eq:CardyIndexExpansion} below.

\subsubsection{Expansion of the saddle-point equations}
\label{subsub:cardyexp}

We assume only three properties of $S$ and of its saddles, all of which are known to hold in every case treated in the literature so far.
\begin{enumerate}
    \item[\emph{i)}] The complexified holonomy $u=(u^1,\ldots,u^{\rk\alg})$ admits an expansion in
    $\varepsilon$,
    \begin{equation}\label{eq:uexp}
        u(\varepsilon)=u^{(0)}+\varepsilon\,u^{(1)}
        +\varepsilon^2u^{(2)}+\mathcal{O}(\varepsilon^3)\ .
    \end{equation}
    This is more general than the Ansatz $u=a+b\,\omega$ with $a,b$ real,
    common in the literature (see e.g. \cite{Cabo-Bizet:2019osg,ArabiArdehali:2019orz,Amariti:2020jyx}), which
    truncates \eqref{eq:uexp} at first order with $u^{(1)}$ aligned along
    $\hat\omega$.
    \item[\emph{ii)}] $S(u;\omega)\equiv S(u;\Delta_a,\omega)$ is a Laurent series in $\varepsilon$ starting at
    order $\varepsilon^{-2}$,
    \begin{equation}\label{eq:Sexp}
        S(u;\omega)=\frac{S_{-2}(u)}{\varepsilon^2}
        +\frac{S_{-1}(u)}{\varepsilon}+S_0(u)
        +\varepsilon\,S_1(u)+\mathcal{O}(\varepsilon^2)\ .
    \end{equation}
    \item[\emph{iii)}] The dependence of $S$ on $u$ at leading order is
    polynomial (e.g.~via Bernoulli polynomials).
\end{enumerate}
We suppress the dependence on $\Delta_a$ throughout, and work to the accuracy
of \eqref{eq:Sexp}, disregarding the logarithmic and exponentially suppressed
corrections in $\varepsilon$ known to be present
\cite{Cabo-Bizet:2019osg,GonzalezLezcano:2020yeb,Amariti:2021ubd,Amariti:2020jyx}; these do not affect the
comparison below, which is order by order in $\varepsilon$.

Substituting \eqref{eq:uexp} and \eqref{eq:Sexp} into \eqref{eq:cardy} and
expanding about $u^{(0)}$, we collect terms by degree in $\varepsilon$. At
order $\varepsilon^{-2}$,
\begin{equation}\label{eq:cardyLO}
    \partial_iS_{-2}(u^{(0)})=0\ ,
\end{equation}
which fixes $u^{(0)}$ and defines the leading Hessian
\begin{equation}\label{eq:hessian}
    \mathcal{H}^{(0)}_{ij}\coloneqq\partial_i\partial_jS_{-2}(u^{(0)})\ .
\end{equation}
At order $\varepsilon^{-1}$,
\begin{equation}\label{eq:deg-1}
    \sum_{j=1}^{\rk\alg}\mathcal{H}^{(0)}_{ij}\,u^{(1)}_j
    +\partial_iS_{-1}(u^{(0)})=0\ ,
\end{equation}
and at order $\varepsilon^0$,
\begin{equation}\label{eq:deg0}
    \sum_{j=1}^{\rk\alg}\mathcal{H}^{(0)}_{ij}u^{(2)}_j
    +\tfrac12\sum_{j,k=1}^{\rk\alg}u^{(1)}_ju^{(1)}_k\,
      \partial_j\partial_k\partial_iS_{-2}(u^{(0)})\\
    +\sum_{j=1}^{\rk\alg} u^{(1)}_j\,\partial_j\partial_iS_{-1}(u^{(0)})
    +\partial_iS_0(u^{(0)})=0\ .
\end{equation}
Assuming $\mathcal{H}^{(0)}$ invertible, \eqref{eq:deg-1} determines the
first correction,
\begin{equation}\label{eq:CardyOrd1}
    u^{(1)}_i=-\sum_{j=1}^{\rk\alg}(\mathcal{H}^{(0)})^{-1}_{ij}\,
    \partial_jS_{-1}(u^{(0)})\ .
\end{equation}
More compactly, writing
$D^i_{n,m}\coloneqq\big(\varepsilon^m\,\partial_iS_n(u)\big)\big|_{\varepsilon=0}$,
the equation at order $\varepsilon^K$ reads
\begin{equation}\label{eq:cardy_compact}
    C^i_K\coloneqq\sum_{n+m=K}D^i_{n,m}=0\ .
\end{equation}

\subsubsection{Contribution of a saddle}
\label{subsub:cardycontrib}

Steepest descent gives, for each saddle,
\begin{equation}\label{eq:steepest}
    \mathcal{I}_{\mathrm{Cardy}}(\hat u)\approx
    \sqrt{\frac{(2\pi)^{\rk\alg}}{\det\big(-\mathcal{H}(\hat u)\big)}}\;
    e^{-S(\hat u;\,\omega)}\ ,
\end{equation}
with $\mathcal{H}_{ij}=\partial_i\partial_jS$ the full Hessian. We evaluate the
two factors in turn.

Substituting $\hat u\approx u^{(0)}+\varepsilon u^{(1)}$ into \eqref{eq:Sexp},
the leading condition \eqref{eq:cardyLO} removes $u^{(1)}$ from the
$\varepsilon^{-2}$ and $\varepsilon^{-1}$ terms, and $u^{(2)}$ drops out at
order $\varepsilon^0$ for the same reason; this is why first order in
\eqref{eq:uexp} suffices for an $\mathcal{O}(\varepsilon^0)$ evaluation of the
action. Using \eqref{eq:CardyOrd1}, the linear and quadratic corrections
combine into a single Gaussian term,
\begin{equation}\label{eq:Scardy}
    S(\hat u;\omega)\approx\frac{S_{-2}(u^{(0)})}{\varepsilon^2}
    +\frac{S_{-1}(u^{(0)})}{\varepsilon}+S_0(u^{(0)})
    -\tfrac12\sum_{i,j=1}^{\rk\alg}\partial_iS_{-1}(u^{(0)})\,
      (\mathcal{H}^{(0)})^{-1}_{ij}\,\partial_jS_{-1}(u^{(0)})\ .
\end{equation}
The Hessian at the saddle is $\mathcal{H}_{ij}\approx
\varepsilon^{-2}\mathcal{H}^{(0)}_{ij}$; pulling $\varepsilon^{-2}$ out of a
$\rk\alg\times\rk\alg$ determinant and taking the inverse square root gives
\begin{equation}\label{eq:prefactor}
    \frac{1}{\sqrt{\det(-\mathcal{H})}}
    \approx\varepsilon^{\rk\alg}\,
    \frac{1}{\sqrt{\det(-\mathcal{H}^{(0)})}}\ .
\end{equation}
Altogether,
{\small
\begin{equation}\label{eq:CardyIndexExpansion}
    \mathcal{I}_{\mathrm{Cardy}}\approx
    \varepsilon^{\rk\alg}
    \sqrt{\tfrac{(2\pi)^{\rk\alg}}{\det(-\mathcal{H}^{(0)})}}\,
    \exp\Bigg[\tfrac{S_{-2}(u^{(0)})}{\varepsilon^2}
    +\tfrac{S_{-1}(u^{(0)})}{\varepsilon}+S_0(u^{(0)})
    -\tfrac12\sum_{i,j=1}^{\rk\alg}\partial_iS_{-1}\,
      (\mathcal{H}^{(0)})^{-1}_{ij}\,\partial_jS_{-1}\Bigg]\ .
\end{equation}}

\subsection{The BAEs as finite-difference equations}
\label{sub:finitediff}

To compare the two methods we recall a defining property of the Bethe operators
\eqref{eq:BAops} \cite[Eq.~(2.24)]{Benini:2018mlo}: they implement a shift of
the holonomy vector $u$ by $\omega$ along the $i$-th direction,
\begin{equation}\label{eq:Qshift}
    Q_i(u;\Delta_a,\omega)\,\mathcal{Z}(u;\Delta_a,\omega)
    =\mathcal{Z}(u-\omega\,v_i;\Delta_a,\omega)\ ,
\end{equation}
where $v_i$ is the $i$-th element of the primitive basis of
section~\ref{sub:choices}. Taking logarithms and suppressing the parameters,
\begin{equation}\label{eq:logQ}
    \log Q_i(u)=S(u)-S(u-\omega v_i)\eqqcolon\nabla_iS\ ,
\end{equation}
so that the BAEs $Q_i=1$ become the finite-difference equations
\begin{equation}\label{eq:BAEdiff}
    \nabla_iS=2\pi i\,k_i\ ,\qquad k_i\in\mathbb{Z}\ ,
\end{equation}
the integers $k_i$ recording the branch of the logarithm. Comparing
\eqref{eq:BAEdiff} with \eqref{eq:cardy} already makes the relation between the
two methods plain: the Cardy equation is the derivative, while the BAE the finite
difference of step $\omega$, of the \emph{same} effective action.

\subsubsection{Expansion of the finite-difference equations}
\label{subsub:bethexp}

Writing the shift as a formal translation,
$S(u-\omega v_i) = e^{-\omega\partial_i}S(u)$, and dividing by $\varepsilon$ for
comparison with \eqref{eq:cardy},
\begin{equation}\label{eq:diffexp}
    \frac{1}{\varepsilon}\nabla_iS
    =\sum_{m\geq1}\frac{(-1)^{m+1}\varepsilon^{m-1}}{m!}\,
    \partial_i^{\,m}S(u)\ .
\end{equation}
Inserting \eqref{eq:Sexp}, the leading order reproduces \eqref{eq:cardyLO}
exactly,
\begin{equation}
    \partial_iS_{-2}(u^{(0)})=0\ ,
\end{equation}
with the same Hessian \eqref{eq:hessian}. At order $\varepsilon^{-1}$, however,
\begin{equation}\label{eq:bethedeg-1}
    \sum_{j=1}^{\rk\alg}\mathcal{H}^{(0)}_{ij}u^{(1)}_j+\partial_iS_{-1}(u^{(0)})
    -\tfrac12\,\partial_i^2S_{-2}(u^{(0)})=2\pi i\,k_i\ ,
\end{equation}
so that
\begin{equation}\label{eq:BAEOrd1}
    u^{(1)}_i=\sum_{j=1}^{\rk\alg}(\mathcal{H}^{(0)})^{-1}_{ij}
    \Big(2\pi i\,k_j-\partial_jS_{-1}(u^{(0)})
    +\tfrac12\,\partial_j^2S_{-2}(u^{(0)})\Big)\ .
\end{equation}
Comparison with the Cardy result \eqref{eq:CardyOrd1} shows that the first
subleading corrections already differ, by the branch integers $k_j$ and by the
term $\tfrac12\partial_j^2S_{-2}$, the latter being the leading finite-difference
correction to the derivative. At general order, using \eqref{eq:cardy_compact},
the finite-difference equations can be written entirely in terms of the Cardy
quantities,
\begin{equation}\label{eq:bethecompact}
    C^i_K+\sum_{p\geq2}\frac{(-1)^{p+1}}{p!}\,
    \partial_i^{\,p-1}C^i_{K-p+1}=2\pi i\,k_i\,\delta_{K,-1}\ ,
\end{equation}
with $\delta_{K,-1}$ a Kronecker delta.

\subsubsection{Contribution of a BAE solution}
\label{subsub:bethecontrib}

The contribution of an isolated solution to \eqref{eq:index-finite} is
$\mathcal{Z}H^{-1}$, with $H$ the Jacobian of the BAEs. Comparing its definition \eqref{eq:index-finite} with
\eqref{eq:logQ}, one finds that $H$ is controlled at leading order by the same
Hessian,
\begin{equation}\label{eq:Jacleading}
    H^{-1}(u)\approx\varepsilon^{\rk\alg}\,
    \big(\det\mathcal{H}^{(0)}\big)^{-1}\ ,
\end{equation}
so that it affects only the power-like prefactor, not the exponential, contributing at order
$\varepsilon^{\rk\alg}$. For the action, substituting \eqref{eq:uexp} into \eqref{eq:Sexp} and using the lower-order equations to
eliminate $u^{(1)}$ at order $\varepsilon^{-1}$ and $u^{(2)}$ at order
$\varepsilon^0$, exactly as in the Cardy case,
\begin{equation}\label{eq:Sbethe}
    S(u(\varepsilon);\varepsilon)\approx
    \frac{S_{-2}(u^{(0)})}{\varepsilon^2}
    +\frac{S_{-1}(u^{(0)})}{\varepsilon}+S_0(u^{(0)})
    +\sum_iu^{(1)}_i\partial_iS_{-1}(u^{(0)})
    +\tfrac12\sum_{i,j}u^{(1)}_iu^{(1)}_j\mathcal{H}^{(0)}_{ij}\ .
\end{equation}
Inserting \eqref{eq:BAEOrd1} and combining with \eqref{eq:Jacleading},
\begin{equation}\label{eq:BAEIndexExpansion}
    \mathcal{I}_{\mathrm{BAEs}}\approx
    \frac{\varepsilon^{\rk\alg}}{\det\mathcal{H}^{(0)}}\,
    \exp\left[\frac{S_{-2}(u^{(0)})}{\varepsilon^2}
    +\frac{S_{-1}(u^{(0)})}{\varepsilon}+S_0(u^{(0)})
    +\tfrac12\sum_{i,j}(\mathcal{H}^{(0)})^{-1}_{ij}V^-_iV^+_j\right]\ ,
\end{equation}
having defined the quantities
\begin{equation}
    V^\pm_i\coloneqq2\pi i\,k_i\pm\partial_iS_{-1}(u^{(0)})     +\tfrac12\partial_i^2S_{-2}(u^{(0)})\ .
\end{equation}

\subsection{The Cardy map and the Bethe Ansatz--Cardy conjecture}
\label{sub:cardymap}

The two expansions can now be compared order by order. At leading order they agree identically: \eqref{eq:cardyLO} is the same
equation in both cases. Hence the leading term $u^{(0)}$ of any BAE solution is
\emph{automatically} a solution of the leading Cardy equation. We call
\begin{equation}\label{eq:cardymap}
    u_{\mathrm{BAE}}\ \longmapsto\ u^{(0)}_{\mathrm{Cardy}}
    \coloneqq u^{(0)}_{\mathrm{BAE}}
\end{equation}
the \emph{Cardy map} on BAE solutions. This correspondence has been used
implicitly in the literature for some time (e.g. \cite{ArabiArdehali:2019orz,Cabo-Bizet:2019osg}) and is surely known to the
experts; we are not aware of a statement of it in this generality, though.

The map is not injective, and \eqref{eq:BAEOrd1} shows why: for a given
$u^{(0)}$, the Cardy equation \eqref{eq:CardyOrd1} determines $u^{(1)}$
uniquely, whereas the BAE determines it only up to the choice of branch
integers $k_i\in\mathbb{Z}^{\rk\alg}$. Distinct BAE solutions differing in the
$k_i$ therefore share a Cardy image, and the fibers of \eqref{eq:cardymap} are
labeled by these integers.

Comparing the contributions \eqref{eq:CardyIndexExpansion} and
\eqref{eq:BAEIndexExpansion}, a BAE solution and its image agree in
\begin{itemize}
    \item[\emph{i)}] the exponential factor
    $\exp\big[\varepsilon^{-2}S_{-2}(u^{(0)})
    +\varepsilon^{-1}S_{-1}(u^{(0)})\big]$, and
    \item[\emph{ii)}] the leading polynomial scaling $\varepsilon^{\rk\alg}$,
\end{itemize}
differing only in the $\mathcal{O}(\varepsilon^0)$ terms and in the precise
form of the Gaussian factor.

Since the asymptotic expansion in $\varepsilon$ is unique, and since evaluating
the index by the Bethe Ansatz formula and by steepest descent must give the
same answer, it is natural to expect that a more careful treatment would match
the two contribution by contribution, to all orders including the logarithmic
and exponentially suppressed terms. Were this not so, there would have to exist
Cardy saddles not in the image of \eqref{eq:cardymap}, whose contributions make
up the difference. We conjecture that this does not happen:
\begin{center}
    \emph{Every Cardy saddle lies in the image of the Cardy map.}
\end{center}
We refer to this as the \emph{Bethe Ansatz--Cardy conjecture}.
Contribution-by-contribution matching of the two expansions is then a
consequence.\footnote{The statement is to be understood for isolated solutions
only, since (\ref{eq:index-finite}) itself assumes them. Whether continuous
families, where present, contribute to the Cardy expansion in a manner
compatible with the conjecture we cannot say, simply because their contribution
to the index is not known; see section~\ref{sec:conc} for a possible route to
computing it.}

An observation is in order. The multiplicity with which a given saddle
contributes is not a consequence of the one-form center symmetry, as suggested e.g.
in \cite{GonzalezLezcano:2021nzk,Amariti:2020jyx,Amariti:2021ubd,Amariti:2023rci}: the index is
insensitive to the global form (section~\ref{subsub:index-choice}), whereas a
change of global variant would rescale the contribution by the order of the
center. The multiplicity is instead a choice-independent quantity, and
it is this that produces the finite $\log|Z(G)|$ correction to the Cardy
expansion of $\log \mathcal{I}_\text{Cardy}(\hat{u})$ at order $\varepsilon^0$. The way the number is apportioned on the Cardy side is itself
choice-dependent: part of it appears as a degeneracy of saddles, namely the
size of the orbit of solutions in the sense of section~\ref{subsub:physsol}, and part as the extra factor in the
evaluation of the Chern--Simons partition function
\cite{GonzalezLezcano:2020yeb,Amariti:2020jyx,Amariti:2021ubd,Cassani:2021fyv,ArabiArdehali:2021sfg}, which on the Bethe side is the
corresponding factor in $H$ related to the size of the chosen cocharacter lattice. Only the total is invariant.

\subsection{\texorpdfstring{Generalized saddle-point equations for $\mathcal{N}=4$ SYM}{Generalized saddle-point equations for N=4 SYM}}
\label{sub:generalized}

We now specialize to $\mathcal{N}=4$ $\alg$ SYM, for which the effective action
takes the form
\begin{equation}\label{eq:Seff}
\begin{split}
    S(u;\Delta_a,\omega)={}&\sum_{a=1}^3\bigg[\sum_{\alpha\in\Phi}
      \log\tilde\Gamma\big(\langle\alpha,u\rangle+\Delta_a;\omega,\omega\big)
      +\rk\alg\,\log\tilde\Gamma(\Delta_a;\omega,\omega)\bigg]\\
    &+\sum_{\alpha\in\Phi}\log\theta_0\big(\langle\alpha,u\rangle;\omega\big)
      +2\rk\alg\,\log(q;q)_\infty\ .
\end{split}
\end{equation}
The Cardy equations \eqref{eq:cardy} therefore become
\begin{equation}\label{eq:cardygen}
    \partial_iS=\sum_{\alpha\in\Phi}\alpha_i\left[
      \sum_{a=1}^3\tilde\Gamma'\big(\langle\alpha,u\rangle+\Delta_a;\omega,\omega\big)
      +\theta_0'\big(\langle\alpha,u\rangle;\omega\big)\right]=0\ ,
\end{equation}
where a prime denotes the logarithmic derivative with respect to the first
argument, $f'(x)\coloneqq\partial_x\log f(x;\omega,\omega)$. 

Then, to expand the saddle equations to leading order, the Cardy literature usually uses the following $|\omega|\to0$ asymptotic expansions of the elliptic gamma and theta functions,
\begin{subequations}
\begin{align}     
    \log\tilde\Gamma(x;\omega,\omega) &= \frac{i\pi}{3\omega^2}\,\mathrm{B}_3\big(\{x\}_\omega\big) - \frac{i\pi}{\omega}\,\mathrm{B}_2\big(\{x\}_\omega\big) + \mathcal{O}(1)\ , \\     \log\theta_0(x;\omega) &= -\frac{i\pi}{\omega}\,\mathrm{B}_2\big(\{x\}_\omega\big) + i\pi\,\mathrm{B}_1\big(\{x\}_\omega\big) + \mathcal{O}(1)\ , 
\end{align}
\end{subequations}
with $\mathrm{B}_n$ the Bernoulli polynomials. For the saddle equations we then use in the leading order
\begin{subequations}\label{eq:gammaexp}
\begin{align}     
    \partial_x\log\tilde\Gamma(x;\omega) &= \frac{i\pi}{\omega^2}\,\mathrm{B}_2\big(\{x\}_\omega\big) - \frac{2i\pi}{\omega}\,\mathrm{B}_1\big(\{x\}_\omega\big) + \mathcal{O}(1)\ , \\     \partial_x\log\theta_0(z;\omega) &= -\frac{2i\pi}{\omega}\,\mathrm{B}_1\big(\{x\}_\omega\big) + i\pi + \mathcal{O}(1)\ . 
\end{align}
\end{subequations}
This expansion presupposes $\{\tilde x\}_\omega\not\to\mathbb{Z}$, where
$\tilde x$ is the argument of the elliptic gamma function and $\{\cdot\}_\omega$
denotes the $\omega$-modded value of a complex number, which for
$\tilde x=a+b\omega$ with $a,b\in\mathbb{R}$ reduces to
$\{\tilde x\}_\omega=\{a\}+b\omega$ with $\{a\}\coloneqq a-\lfloor a\rfloor$; i.e.\ it presupposes that the real part of the
argument stays away from the period lattice. As we will see for $B_2\cong C_2$ in section~\ref{subsub:cardyB2}, for
non-$A$-type algebras even isolated BAE solutions can violate this assumption
under the Cardy map,\footnote{A different failure of the standard expansion, at rational values of $\tau$ and
$\sigma$ rather than at lattice values of the argument, is analyzed in
\cite{Jejjala:2021hlt}, where the domain of validity is characterized
explicitly.} and one must instead use \cite{Felder:1999vf}
\begin{subequations}\label{eq:gammaexp2}
\begin{align}     
    \log\tilde\Gamma(a\omega;\omega,\omega) &= \frac{i\pi}{6\omega}(a-1) - \log\big(1-e^{2\pi ia}\big) + \mathcal{O}(1)\ , \\     \log\theta_0(a\omega;\omega) &= -\frac{i\pi}{6\omega} + i\pi\Big(a-\frac{1}{2}\Big) + \log\big(1-e^{-2\pi ia}\big) + \mathcal{O}(1)\ ,
\end{align}
\end{subequations}
with derivatives
\begin{subequations}\label{eq:gammaexp2der}
\begin{align}     
    \partial_z\log\tilde\Gamma(a\omega;\omega,\omega) &= \frac{i\pi}{6\omega^2} - \frac{\pi}{\omega}\cot(\pi a) - \frac{i\pi}{\omega} + \mathcal{O}(1)\ , \\     \partial_z\log\theta_0(a\omega;\omega) &= \frac{\pi}{\omega}\cot(\pi a) + \mathcal{O}(1)\ .  
\end{align}
\end{subequations}
A refinement of this kind already appears in
\cite[Sec. C.2.1 CASE 2]{ArabiArdehali:2019orz} in the context of Cardy-like limit of the $A_1$ BAEs, where an asymptotic solution evading the standard assumption is exhibited. Its origin is unclear to us: for $\mathfrak{su}(2)$ there are no continuous families, and the isolated solutions are exhausted by the Hong--Liu ones, so the configuration in question does not obviously descend from a Bethe solution of either type.

Note that at order $\varepsilon^{-2}$ the two expansions agree, and only the
elliptic gamma terms contribute. This is the order at which the Cardy map of
section~\ref{sub:cardymap} is defined, and where the identification of saddles
is unambiguous. The expansions first differ at order $\varepsilon^{-1}$, through
the branch integers $k_i$ and the finite-difference correction
$\tfrac12\partial_i^2S_{-2}$ of \eqref{eq:BAEOrd1}, and the Cardy literature
does go further, to $\varepsilon^0$, where the finite $\log|Z(G)|$ term appears in $\log \mathcal{I}_\text{Cardy}(\hat{u})$.
A systematic comparison at those orders we leave to future work; in the examples below we work
at leading order throughout.

\subsection{Rank-two examples}
\label{sub:cardyrank2}

Throughout this section we use the orthogonal basis (of the holonomy vector $u=(u^1,\ldots,u^{\rk\alg})$, living on the complex torus $\mathfrak{h}/(L+\omega L)$), whose associated global form fixes the center and hence its action (on the torus) as shifts of the
$u^i$ by $1/k$, with
\begin{equation}\label{eq:centershifts}
    k=N\quad(A_{N-1})\ ,\qquad
    k=1\quad(B_N)\ ,\qquad
    k=2\quad(C_N,\,D_N)\ .
\end{equation}
We recall from section~\ref{subsub:symm} that these shifts are only ``half''
the center action, the other half being along the $\omega$-cycle of the torus;
only the former survives the Cardy-like limit $|\omega|\to 0$.

\subsubsection{\texorpdfstring{$A_2$}{A2}}
\label{subsub:cardyA2}

For $\mathfrak{su}(N)$ the saddles are the so-called $C$-center configurations, labeled
by a divisor $C\mid N$ \cite{ArabiArdehali:2019orz,Cabo-Bizet:2019osg,GonzalezLezcano:2020yeb},
\begin{equation}\label{eq:Ccenter}
    u_i=\frac{L}{N}+\frac{1}{C}
    \Big(\Big\lfloor\frac{i-1}{N/C}\Big\rfloor-\frac{C-1}{2}\Big)\ ,
    \quad i=1,\ldots,N\ ,\quad L=0,\ldots,\tfrac{N}{C}-1\ ,
\end{equation}
that is, $C$ equally spaced points along $(0,1]$ carrying $N/C$ coincident
holonomies each. Saddles differing in $L$ are related by the center
$\mathbb{Z}_N$, e.g.\ $u_i=0\xrightarrow{\ \mathbb{Z}_N\ }u_i=L/N$. These
configurations are the images under the Cardy map \eqref{eq:cardymap} of the
Hong--Liu solutions $\{m,n,r\}=\{C,\tfrac NC,0\}$ \cite{Hong:2018viz}.

Two extremes are worth isolating, both special cases of \eqref{eq:Ccenter}. For
$C=1$ all $N$ holonomies coincide, $u_i=L/N$: the ``confining configuration'',
which contributes trivially to the index since every root lands on the period
lattice. This is the image of the ``basic'' solution $\{1,N,0\}$ of
\cite{Hosseini:2016cyf}, the saddle dual to the black hole and dominant for
large enough charges \cite{Benini:2018ywd,Honda:2019cio,Choi:2018hmj}. For
$C=N$ the holonomies are maximally spread,
$u_i=\tfrac{2i-N+1}{2N}\sim\tfrac{2i+1}{2N}+\tfrac12$, the image of
$\{N,1,0\}$. The intermediate divisors interpolate, corresponding to the
``partially deconfined'' configurations of \cite{ArabiArdehali:2019orz}.

\begin{figure}[htb!]
\centering
\begin{tikzpicture}
  \node[sol, label={[sollbl,align=center]90:
    $\left(\tfrac13,0,\tfrac23\right)$\\[-3pt]
    {\scriptsize$\left(\tfrac13,0,\tfrac23\right)$}}]                      (n1) at (0,0) {};
  \node[sol, label={[sollbl,align=center]90:
    $\left(\tfrac{\omega}{3},0,\tfrac{2\omega}{3}\right)$\\[-3pt]
    {\scriptsize$\left(0,0,0\right)$}}]                                    (n2) at (3.2,0) {};
  \node[sol, label={[sollbl,align=center]0:
    $\left(\tfrac{2\omega}{3}{+}\tfrac13,0,\tfrac{\omega}{3}{+}\tfrac23\right)$\\[-3pt]
    {\scriptsize$\left(\tfrac13,0,\tfrac23\right)$}}]                      (n3) at (6.0,1.5) {};
  \node[sol, label={[sollbl,align=center,label distance=8pt]0:
    $\left(\tfrac{\omega}{3}{+}\tfrac13,0,\tfrac{2\omega}{3}{+}\tfrac23\right)$\\[-3pt]
    {\scriptsize$\left(\tfrac13,0,\tfrac23\right)$}}]                      (n4) at (6.0,-1.5) {};

  \draw[Sedge] (n1) -- node[edgelbl, above]{$S$} (n2);
  \draw[Tloop] (n1) to[out=180+24,in=180-24,looseness=20]
               node[edgelbl, left]{$T$} (n1);
  \draw[Tdir]  (n2) -- node[edgelbl, below left]{$T$} (n4);
  \draw[Tdir]  (n4) -- node[edgelbl, left]{$T$} (n3);
  \draw[Tdir]  (n3) -- node[edgelbl, above left]{$T$} (n2);
  \draw[Sedge] (n3) to[bend left=55] node[edgelbl, right]{$S$} (n4);
\end{tikzpicture}
\caption{The single $\mathrm{PSL}(2,\mathbb{Z})$ orbit of the four Hong--Liu
solutions of $A_2$, one Weyl-- and center--representative per node (upper
label), with the corresponding Cardy saddle (\ref{eq:Ccenter}) below it. Blue
edges denote $S$, red edges $T$; the latter acts as a three-cycle on the
$\omega$-dependent solutions and fixes the purely rational one. Holonomies are
written in the orthogonal basis, i.e.\ as $\mathfrak{u}(3)$ variables subject to
$\sum_iu^i=0$, and the global form is $\SU(3)$.}
\label{fig:A2Cardy}
\end{figure}
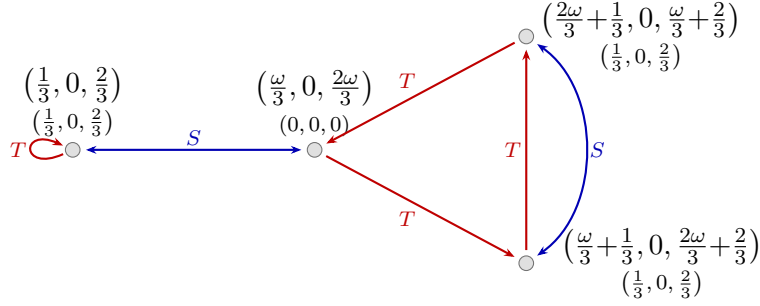

For $A_2$, the case of interest here, $N=3$ is prime, so only $C=1$ and $C=3$
occur, and both are realized: the basic solution collapses to the confining
saddle, while the remaining three Hong--Liu solutions map onto the fully spread
one. The four solutions form a single $\mathrm{PSL}(2,\mathbb{Z})$ orbit,
displayed with their images in figure~\ref{fig:A2Cardy}. %The picture is as expected, and we record it for comparison with the cases below.

\subsubsection{\texorpdfstring{$D_2$}{D2}}
\label{subsub:cardyD2}

\begin{figure}[htb!]
\centering
\begin{tikzpicture}
  \def\R{3.1}
  \node[sol, label={[sollbl,align=center]120:
    $\left(\tfrac{\omega}{2}{+}\tfrac14,\ \tfrac14\right)$\\[-3pt]
    {\scriptsize$\left(\tfrac14,\tfrac14\right)$}}]                        (n1) at (120:\R) {};
  \node[sol, label={[sollbl,align=center]60:
    $\left(\tfrac{\omega}{4}{+}\tfrac12,\ \tfrac{\omega}{4}\right)$\\[-3pt]
    {\scriptsize$\left(\tfrac12,0\right)$}}]                               (n2) at (60:\R) {};
  \node[sol, label={[sollbl,align=center]0:
    $\left(\tfrac{\omega}{4}{+}\tfrac14,\ \tfrac{\omega}{4}{+}\tfrac34\right)$\\[-3pt]
    {\scriptsize$\left(\tfrac14,\tfrac34\right)$}}]                        (n3) at (0:\R) {};
  \node[sol, label={[sollbl,align=center]-60:
    $\left(\tfrac{\omega}{4}{+}\tfrac14,\ \tfrac{3\omega}{4}{+}\tfrac14\right)$\\[-3pt]
    {\scriptsize$\left(\tfrac14,\tfrac14\right)$}}]                        (n4) at (-60:\R) {};
  \node[sol, label={[sollbl,align=center]240:
    $\left(\tfrac{\omega}{4}{+}\tfrac12,\ \tfrac{3\omega}{4}\right)$\\[-3pt]
    {\scriptsize$\left(\tfrac12,0\right)$}}]                               (n5) at (240:\R) {};
  \node[sol, label={[sollbl,align=center]180:
    $\left(\tfrac{\omega}{2}{+}\tfrac14,\ \tfrac34\right)$\\[-3pt]
    {\scriptsize$\left(\tfrac14,\tfrac34\right)$}}]                        (n6) at (180:\R) {};

  \draw[Sedge] (n1) -- node[edgelbl, above]      {$S$} (n2);
  \draw[Tedge] (n2) -- node[edgelbl, above right]{$T$} (n3);
  \draw[Sedge] (n3) -- node[edgelbl, below right]{$S$} (n4);
  \draw[Tedge] (n4) -- node[edgelbl, below]      {$T$} (n5);
  \draw[Sedge] (n5) -- node[edgelbl, below left] {$S$} (n6);
  \draw[Tedge] (n6) -- node[edgelbl, above left] {$T$} (n1);

  \begin{scope}[yshift=-5.4cm]
    \node[sol, label={[sollbl,align=center,label distance=4pt]90:
      $\left(\tfrac{\omega}{2}{+}\tfrac12,\ 0\right)$\\[-3pt]
      {\scriptsize$\left(\tfrac12,0\right)$}}]                             (m1) at (-2.8,0) {};
    \node[sol, label={[sollbl,align=center,label distance=4pt]90:
      $\left(\tfrac{\omega}{2},\ 0\right)$\\[-3pt]
      {\scriptsize$\{0,0\}$}}]                                    (m2) at (0,0) {};
    \node[sol, label={[sollbl,align=center,label distance=4pt]90:
      $\left(\tfrac12,\ 0\right)$\\[-3pt]
      {\scriptsize$\left(\tfrac12,0\right)$}}]                             (m3) at (2.8,0) {};

    \draw[Tedge] (m1) -- node[edgelbl, below]{$T$} (m2);
    \draw[Sedge] (m2) -- node[edgelbl, below]{$S$} (m3);
    \draw[Sloop] (m1) to[out=180+24,in=180-24,looseness=50]
                 node[edgelbl, left] {$S$} (m1);
    \draw[Tloop] (m3) to[out=-24,in=24,looseness=50]
                 node[edgelbl, right]{$T$} (m3);
  \end{scope}
\end{tikzpicture}
\caption{The two $\mathrm{PSL}(2,\mathbb{Z})$ orbits of the $D_2$ solutions (i.e. figure~\ref{fig:D2_orbits_ortho}),
one Weyl-- and center--representative per node (upper label), with the
corresponding Cardy saddle below it: a sextuplet (top) and a
triplet (bottom). Blue edges denote $S$, red edges $T$. Holonomies
are written in the orthogonal basis, corresponding to the global form
$\mathrm{SO}(4)$.}
\label{fig:D2Cardy}
\end{figure}
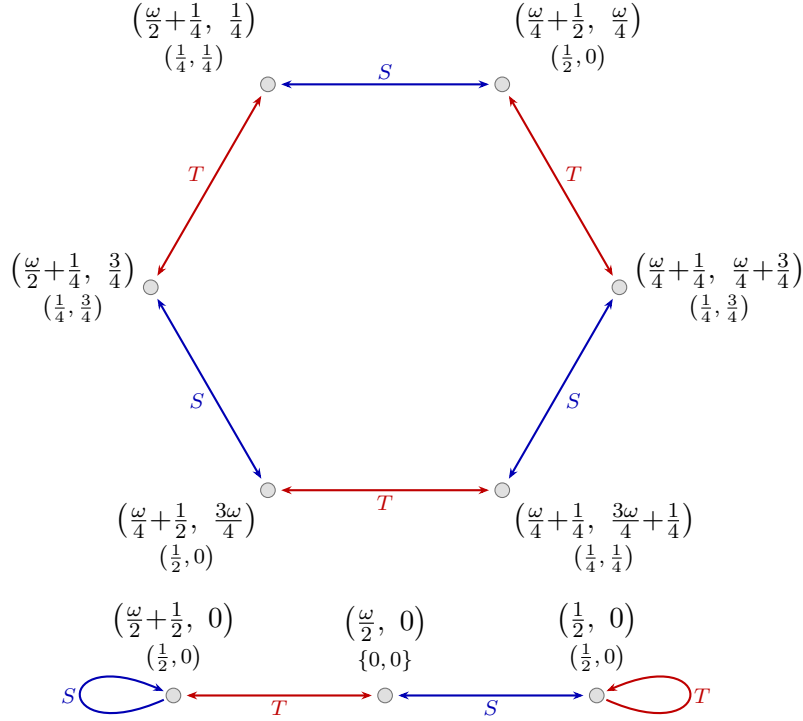

For $\mathfrak{so}(2N)$, the saddles identified in
\cite[Sec. 4.2]{Amariti:2020jyx} have all $N$ holonomies at either $u_i=0$ or
$u_i=\tfrac12$, the two being exchanged by the $\mathbb{Z}_2$ of
\eqref{eq:centershifts}; that reference does not claim the list to be
exhaustive, and further saddles were expected. Taking the $|\omega|\to0$ limit
of the fully rational $D_2$ solutions of section~\ref{sec:D2} exhibits some of
them. Alongside the  saddle $(0,0)$ and the expected
$(\tfrac12,0)$, we find the configurations $(\tfrac14,\tfrac14)$ and
$(\tfrac14,\tfrac34)$, with holonomies at quarter-periods; these lie outside
the above list and are not related to it by the center, which shifts by
$\tfrac12$ (figure~\ref{fig:D2Cardy}). Since $D_2\cong A_1\oplus A_1$ the limit
factorizes over the two summands, so even the simplest non-simple case already
produces saddles beyond those catalogued, a phenomenon we shall meet again, in
sharper form, for $B_2$ below.

\subsubsection{\texorpdfstring{$B_2\cong C_2$ and S-duality of saddles}{B2 = C2 and S-duality of Cardy saddles}}
\label{subsub:cardyB2}

\begin{figure}[htb!]
    \centering
    \includegraphics[width=\linewidth]{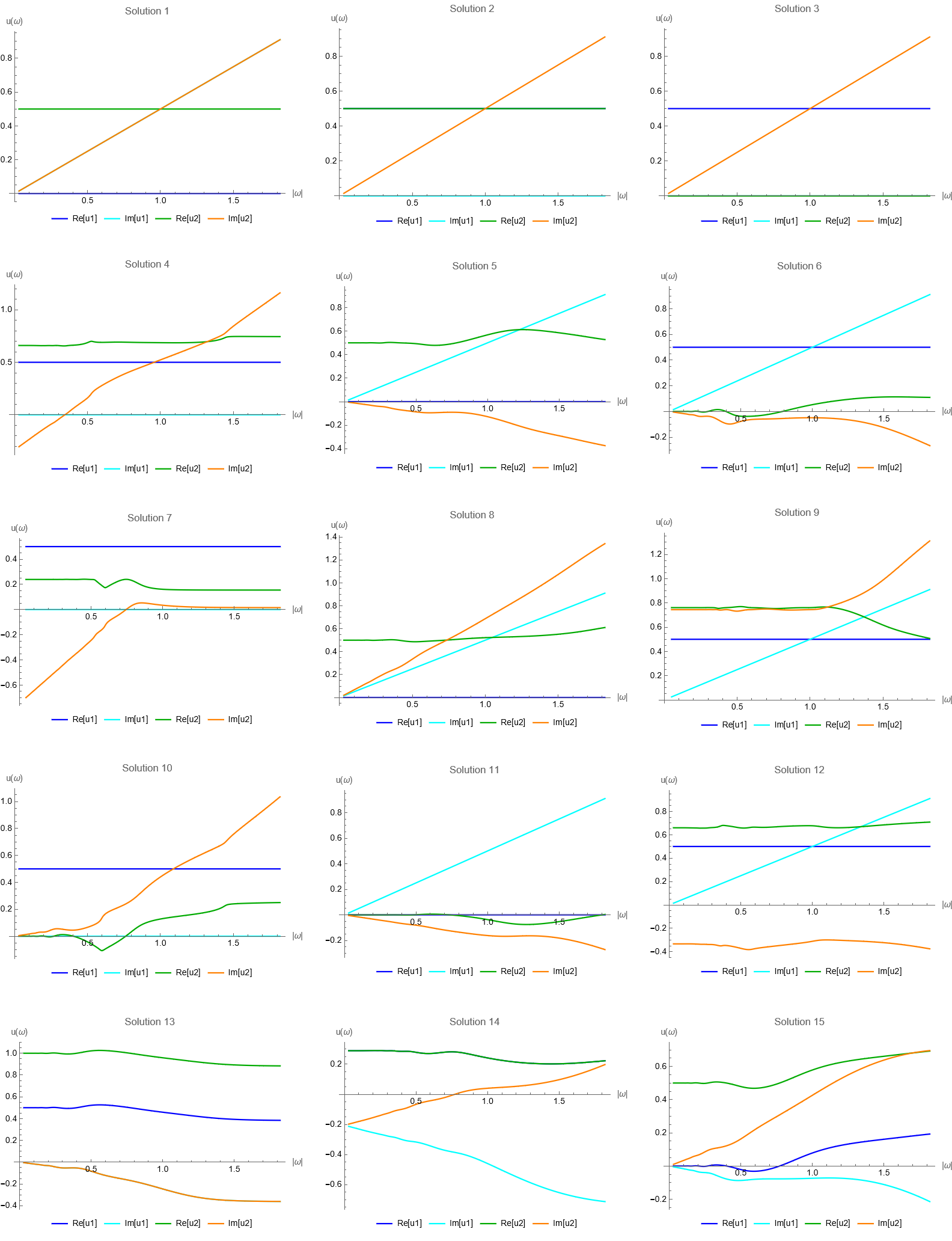}
    \caption{Numerical tracking of the Cardy limit of the $B_2$ solutions of
    section~\ref{sec:B2} along $\varepsilon\to0$ (on the horizontal axis) with $\omega=i \varepsilon$, at fixed generic values $\Delta_1=0.421+0.411i$, $\Delta_2=0.34+0.33i$.} 
    \label{fig:B2Cardy}
\end{figure}

This is the most interesting case. Since our orthogonal basis corresponds to the
$\SO(5)$ global form, the classification may be read either as
$\mathfrak{so}(2N{+}1)$ or as $\mathfrak{usp}(2N)$, at $N=2$. For
$\mathfrak{so}(2N{+}1)$ the saddles have $L$ holonomies at $u_i=0$ and $N-L$ at
$u_i=\tfrac12$, with $L=0,\ldots,N$ \cite{Amariti:2023rci}; the center being
trivial, no identification occurs. For $\mathfrak{usp}(2N)$, where saddles
related by a shift of $\tfrac12$ are identified by the $\mathbb{Z}_2$ center,
the dominant saddle depends on the $\Delta_a$, and
\cite{Amariti:2020jyx,Amariti:2021ubd,Amariti:2023rci} give:
\begin{enumerate}
    \item[\emph{i)}] $L$ holonomies at $u_i=0$ and $N-L$ at $u_i=\tfrac12$,
    called the \emph{first family} in \cite[\S2]{Amariti:2023rci}; exchanging
    $L\leftrightarrow N-L$ leaves the index unchanged, so effectively
    $L=0,\ldots,\lfloor(N-1)/2\rfloor$;
    \item[\emph{ii)}] $L$ at $u_i=0$, $L$ at $u_i=\tfrac12$, and $N-2L$ at
    $u_i=\tfrac14\xrightarrow{\ \mathbb{Z}_2\ }\tfrac34$ (the \emph{second
    family}), again with $L=0,\ldots,\lfloor(N-1)/2\rfloor$;
    \item[\emph{iii)}] two special cases of \emph{ii)}: for even $N$, the
    \emph{self-paired} saddle with $N/2$ at $u_i=0$ and $N/2$ at
    $u_i=\tfrac12$, a limiting case of both families; and $L=0$, with all $N$
    holonomies at $u_i=\tfrac14\xrightarrow{\ \mathbb{Z}_2\ }\tfrac34$.
\end{enumerate}
Tracking our analytic $B_2$ solutions of section~\ref{sec:B2} along
$|\omega|\to0$ at fixed $\Delta_a$, we find the behavior of
figure~\ref{fig:B2Cardy}.
The plots give the Cardy-like limit of each BAE solution, including the term
linear in $\varepsilon$, whose linear behavior at small $\varepsilon$ is
manifest. Using the values $\Delta_1=0.421+0.411i$, $\Delta_2=0.34+0.33i$ and
$\omega=i\varepsilon$, each limit can be written in closed form in terms of the
$\Delta_a$, as collected in table~\ref{tab:b2_cardy_solutions}; these
expressions are readily checked to be exact by substitution into the Cardy
equations \eqref{eq:cardygen}.
\begin{table}[hbt!]
    \centering
    \renewcommand{\arraystretch}{1}
    \begin{tabular}{c l l}
    \hline
    \textbf{No.} & \textbf{Numerical expansion of $(u_1, u_2)$ in $\varepsilon$} & \textbf{Analytical rewriting} \\
    \hline
    1 & $(0 + 0.5i\varepsilon, \; 0.5 + 0.5i\varepsilon)$ & $\big(\frac{\omega}{2}, \; \frac{1}{2} + \frac{\omega}{2}\big)$ \\
    2 & $(0.5, \; 0.5 + 0.5i\varepsilon)$ & $\big(\frac{1}{2}, \; \frac{1}{2} + \frac{\omega}{2}\big)$ \\
    3 & $(0.5, \; 0 + 0.5i\varepsilon)$ & $\big(\frac{1}{2}, \; \frac{\omega}{2}\big)$ \\
    4 & $(0.5, \; (0.660 - 0.333333i) + 1.0i\varepsilon)$ & $\big(\frac{1}{2}, \; 1 - \Delta_2 + \omega\big)$ \\
    5 & $(0 + 0.5i\varepsilon, \; 0.5 - 0.166667i\varepsilon)$ & $\big(\frac{\omega}{2}, \; \frac{1}{2} - \frac{\omega}{6}\big)$ \\
    6 & $(0.5 + 0.5i\varepsilon, \; 0 - 0.166667i\varepsilon)$ & $\big(\frac{1}{2} + \frac{\omega}{2}, \; -\frac{\omega}{6}\big)$ \\
    7 & $(0.5, \; (0.239 - 0.744333i) + 1.0i\varepsilon)$ & $\big(\frac{1}{2}, \; \Delta_3 + \omega\big)$ \\
    8 & $(0 + 0.5i\varepsilon, \; 0.5 + 0.666667i\varepsilon)$ & $\big(\frac{\omega}{2}, \; \frac{1}{2} + \frac{2\omega}{3}\big)$ \\
    9 & $(0.5 + 0.5i\varepsilon, \; 0.761 + 0.744333i)$ & $\big(\frac{1}{2} + \frac{\omega}{2}, \; \Delta_1 + \Delta_2\big)$ \\
    10 & $(0.5, \; 0 + 0.166667i\varepsilon)$ & $\big(\frac{1}{2}, \; \frac{\omega}{6}\big)$ \\
    11 & $(0 + 0.5i\varepsilon, \; 0 - 0.166667i\varepsilon)$ & $\big(\frac{\omega}{2}, \; -\frac{\omega}{6}\big)$ \\
    12 & $(0.5 + 0.5i\varepsilon, \; 0.660 - 0.333333i)$ & $\big(\frac{1}{2} + \frac{\omega}{2}, \; 1 - \Delta_2\big)$ \\
    13 & $(0.5 - 0.166667i\varepsilon, \; 1.0 - 0.166667i\varepsilon)$ & $\big(\frac{1}{2} - \frac{\omega}{6}, \; 1 - \frac{\omega}{6}\big)$ \\
    14 & $((0.2895 - 0.2055i) - 0.25i\varepsilon, \; u_1+0.5i\varepsilon$ & $\big(\frac{1-\Delta_1}{2} - \frac{\omega}{4}, \; \frac{1-\Delta_1}{2} + \frac{\omega}{4}\big)$ \\
    15 & $(0 - 0.166667i\varepsilon, \; 0.5 + 0.333333i\varepsilon)$ & $\big(-\frac{\omega}{6}, \; \frac{1}{2} + \frac{\omega}{3}\big)$ \\
    \hline
    \end{tabular}
    \caption{Numerical Cardy-like limit of the $B_2$ BAE solutions and their closed-form counterparts.}
    \label{tab:b2_cardy_solutions}
\end{table}
Taking the Cardy map, i.e.\ the limit $\varepsilon\to0$, of the BAE solutions
recovers the saddles listed above, and in addition yields
$\Delta_a$-dependent ones that were previously unknown. Translating to the
$\mathfrak{usp}(4)$ orthogonal basis via
\begin{equation}\label{eq:so5tousp4}
    \begin{pmatrix}y_1\\y_2\end{pmatrix}_{\mathfrak{usp}(4)}
    =\begin{pmatrix}\tfrac12&\tfrac12\\[2pt]\tfrac12&-\tfrac12\end{pmatrix}
     \begin{pmatrix}x_1\\x_2\end{pmatrix}_{\mathfrak{so}(5)}
\end{equation}
produces the corresponding $\mathfrak{usp}(4)$ saddles, both the standard and
the new $\Delta_a$-dependent ones, though only up to the action of the
$\mathbb{Z}_2$ center. The explicit S-duality of all Cardy saddles is summarized
in table~\ref{tab:usp4_so5_saddles}.
\begin{table}[hbt!]
    \centering
    \renewcommand{\arraystretch}{1}
    \begin{tabular}{c l l}
    \hline
    \textbf{No.} & \textbf{$\mathfrak{so}(5)$ Cardy saddle} & \textbf{$\mathfrak{usp}(4)$ Cardy saddle} \\
    \hline
    1 & $\big(0, \; \frac{1}{2}\big)$ & $\big(\frac{1}{4}, \; -\frac{1}{4}\big)$ \\
    2 & $\big(\frac{1}{2}, \; \frac{1}{2}\big)$ & $\big(\frac{1}{2}, \; 0\big)$ \\
    3 & $\big(\frac{1}{2}, \; 0\big)$ & $\big(\frac{1}{4}, \; \frac{1}{4}\big)$ \\
    4 & $\big(\frac{1}{2}, \; 1 - \Delta_2\big)$ & $\big(\frac{3}{4} - \frac{\Delta_2}{2}, \; -\frac{1}{4} + \frac{\Delta_2}{2}\big)$ \\
    5 & $\big(0, \; \frac{1}{2}\big)$ & $\big(\frac{1}{4}, \; -\frac{1}{4}\big)$ \\
    6 & $\big(\frac{1}{2}, \; 0\big)$ & $\big(\frac{1}{4}, \; \frac{1}{4}\big)$ \\
    7 & $\big(\frac{1}{2}, \; \Delta_3^{}\big)$ & $\big(\frac{1}{4} + \frac{\Delta_3^{}}{2}, \; \frac{1}{4} - \frac{\Delta_3^{}}{2}\big)$ \\
    8 & $\big(0, \; \frac{1}{2}\big)$ & $\big(\frac{1}{4}, \; -\frac{1}{4}\big)$ \\
    9 & $\big(\frac{1}{2}, \; \Delta_1 + \Delta_2\big)$ & $\big(\frac{1}{4} + \frac{\Delta_1 + \Delta_2}{2}, \; \frac{1}{4} - \frac{\Delta_1 + \Delta_2}{2}\big)$ \\
    10 & $\big(\frac{1}{2}, \; 0\big)$ & $\big(\frac{1}{4}, \; \frac{1}{4}\big)$ \\
    11 & $\big(0, \; 0\big)$ & $\big(0, \; 0\big)$ \\
    12 & $\big(\frac{1}{2}, \; 1 - \Delta_2\big)$ & $\big(\frac{3}{4} - \frac{\Delta_2}{2}, \; -\frac{1}{4} + \frac{\Delta_2}{2}\big)$ \\
    13 & $\big(\frac{1}{2}, \; 0\big)$ & $\big(\frac{3}{4}, \; -\frac{1}{4}\big)$ \\
    14 & $\big(\frac{1-\Delta_1}{2}, \; \frac{1-\Delta_1}{2}\big)$ & $\big(\frac{1-\Delta_1}{2}, \; 0\big)$ \\
    15 & $\big(0, \; \frac{1}{2}\big)$ & $\big(\frac{1}{4}, \; -\frac{1}{4}\big)$ \\
    \hline
    \end{tabular}
    \caption{Under the transformation from the $\mathfrak{so}(5)$ orthogonal basis to the $\mathfrak{usp}(4)$ orthogonal basis, the Cardy-like limit saddles are determined up to center action.}
    \label{tab:usp4_so5_saddles}
\end{table}

Previous analyses overlooked the $\Delta_a$-dependent saddles because they lie
outside the Ansatz usually employed,
\begin{equation}\label{eq:standardansatz}
    u_i=a_i+b_i\,\omega\ ,\qquad a_i,b_i\in\rr\ ,
\end{equation}
the $\Delta_a$ being complex. They cause no difficulty here, since the treatment
of section~\ref{sub:generalized} rests on the more general expansion
\eqref{eq:uexp}. We stress, moreover, that even the dominant $(0,0)$ saddle
\cite{Amariti:2020jyx} descends from a solution that is \emph{not} rational, namely
solution No. 11 of table~\ref{tab:solutions}; going beyond rational solutions is
therefore not optional.

Two observations close this discussion. First, from the Bethe Ansatz side all
solutions contribute on much the same footing, with the complex contributions
pairing up so that their imaginary parts cancel
(section~\ref{sub:num-eval}). We have not computed the contribution of individual
Cardy saddles, which is beyond the scope of this paper, but it is likely that
some dominate over others in a given region of the $\Delta_a$, as is typically
the case, the $(0,0)$ saddle being one example. Second, this is an explicit
instance in which the standard ansatz \eqref{eq:standardansatz} fails. Adopting
the general one in \eqref{eq:uexp}, we are able to exhibit the S-duality of the full set of
saddles, including those it misses.

%%%%%%%%%%%%%%%%%%%%%%%%%%%%%%%%%%%%%
\section{Conclusions and outlook}
\label{sec:conc}
%%%%%%%%%%%%%%%%%%%%%%%%%%%%%%%%%%%%%

We have solved the Bethe Ansatz equations of 4D $\mathcal{N}=4$ SYM for gauge
algebras outside type $A$, for the first time to our knowledge, obtaining the
complete solution set in closed form for $B_2\cong C_2$ and a numerical
treatment of $G_2$ (with one Weyl-fixed solution presented in closed form).

It is worth noting the geometric character of the special solutions we have
encountered. The partially pinned ones sit at half-periods of the torus, where
the theta-function ratios entering the Bethe operators trivialize; these are
also the zeros of $\wp'$, hence the equilibrium configurations of a system of
classical particles on the torus interacting through a Weierstrass potential proportional to $\wp$.
That the two conditions meet is not an accident, and we make the relation
precise in a companion Letter \cite{CMpaper}, where we construct a map from BAE
solutions to extrema of the \emph{untwisted} elliptic Calogero--Moser system associated
with the root system of $\alg$, valid for any semisimple gauge algebra.

Several directions suggest themselves. The most pressing is a systematic method
beyond rank two. Our treatment here is tailored to the cases at hand, and both
higher rank in $\mathcal{N}=4$ and more general $\mathcal{N}=1$ theories, in the
spirit of \cite{Benini:2018mlo}, require something better; we develop one in
\cite{polypaper}, based on recasting the BAEs as polynomial equations and
applying computational commutative algebra. We stress that such an analysis must
incorporate the caveats uncovered here, which are not artifacts of low rank: the
form of the equations depends on the choice of global structure and primitive
basis; the sign prefactor of the Bethe operators cannot in general be dropped;
Weyl-fixed solutions can contribute to the index; isolated solutions need not be
fully rational, so that an Ansatz modeled on the Hong--Liu family will miss
them; and, in the Cardy-like limit, the standard expansion of the elliptic gamma
function fails on part of the solution set, so that saddles outside the usual
classification appear, already for $D_2\cong A_1\oplus A_1$ and more markedly
for $B_2\cong C_2$. None of these is visible in type
$A$, where the usual or reduced presentation happens to be the convenient one, Weyl-fixed
solutions do not contribute to the index, and the known isolated solutions are fully
rational; all of them, however, should be expected for a generic algebra at any
rank. The same method also bears on completeness, which for $G_2$ we have not
settled, and on continuous families, which at rank two occur only for $A_2$ but
are expected to be generic higher up, as already conjecture in type $A$ in \cite[Sec. 2.1.1]{ArabiArdehali:2019orz}.

Closely related is the problem of evaluating the contribution of continuous
families, which remains open beyond the lowest ranks: a proposal for
$\mathfrak{u}(3)$ and $\mathfrak{su}(3)$ appears in \cite{Cabo-Bizet:2024kfe}, and a general treatment of the contribution of
continuous families is announced in \cite[footnote~4]{Aharony:2024ntg}, but no
prescription has so far appeared. We observe that the
finite sum \eqref{eq:index-finite} is a sum of Grothendieck residues: for an
isolated solution, $\mathcal{Z}H^{-1}$ is precisely the local residue of
$\mathcal{Z}\, \prod_{i=1}^{\rk\alg} du^i$ in \eqref{eq:int_formula} with respect to the ideal generated by the $Q_i-1$
\cite{GriffithsHarris}. When the zero locus acquires positive dimension, the
local residue is no longer defined and must be replaced by a residue
\emph{current} supported on it \cite{ColeffHerrera,Andersson:2004}, so that
schematically
\begin{equation}\label{eq:currentschematic}
    \mathcal{I}_\text{corrected}=\kappa_\alg\left(\,\sum_{\hat u\ \text{isolated}}
    \mathcal{Z}(\hat u)\,H^{-1}(\hat u)
    +\sum_{\mathcal{F}\ \text{families}}\int_{\mathcal{F}}\mathcal{Z}\,\mu_{\mathcal{F}}\right)\ ,
\end{equation}
the second sum running over the positive-dimensional families $\mathcal{F}$ and
$\mu_{\mathcal{F}}$ denoting the measure they inherit from the residue current,
which encodes the transverse behavior of the $Q_i-1$ along $\mathcal{F}$. Making
$\mu_{\mathcal{F}}$ explicit is the crux of the problem, and we hope to return
to it.

A second direction is the large-$N$ limit. The orthosymplectic solutions found
here do not visibly organize into an $N$-indexed family in the manner of the type $A$ 
Hong--Liu solutions, so the route to a $BC\!D$ analogue of the semiclassical black hole
entropy count remains open, as does the identification of the supersymmetric
black holes in $\mathrm{AdS}_5\times\mathbb{RP}^5$ whose microstates such a
count would enumerate.

Finally, the Cardy-like limit deserves further study. We have seen that the
usual saddle-point Ansatz does not capture every saddle, the missing ones being
$\Delta_a$-dependent and hence outside its reach; whether any of them dominates
in some region of parameter space we have not determined. Quantifying their
contribution to the effective action, and establishing whether the
correspondence between Bethe solutions and Cardy saddles is exact, is left to
future work.

%%%%%%%%%%%%%%%%%%%%%%%%%%%%
\section*{Acknowledgments}

We are grateful to A. Amariti, V. Bashmakov, F. Benini, P. Glorioso, C. Mascherpa, A. Nedelin, and A. Zanetti for useful discussions, and to A. Eghrari for discussions and collaboration on a related project \cite{polypaper}. Our work has been supported by Royal Society International Exchanges grant IES\textbackslash R2\textbackslash242034 and London Mathematical Society Scheme 4 grant 42366. We would like to thank the Universities of Milano and Milano-Bicocca for hospitality during various stages of this work. This work forms part of the Ph.D. project of K.K. supported by the Engineering and Physical Sciences Research Council [Studentship 2934922].

\appendix

%%%%%%%%%%%%%%%%%%%%%%%%%%%%
\section{Elliptic functions}
\label{app:ellitpic}
%%%%%%%%%%%%%%%%%%%%%%%%%%%%

We collect the conventions and identities for the elliptic functions used in
the main text. (See e.g. \cite{Felder:1999vf} for a standard reference on the subject.) All functions live on the torus
$\cc/(\zz+\omega\zz)$ with $\operatorname{Im}\omega>0$, and we set as in section \ref{sec:baes}
\begin{equation}\label{eq:fugacities}
  z=:e^{2\pi i u},\quad q=:e^{2\pi i\omega},\quad |q|<1\ ,
\end{equation}
with $q$-Pochhammer symbol $(z;q)_\infty\coloneqq\prod_{j\ge 0}(1-zq^{j})$. We write
$\theta_r(u)\equiv\theta_r(u;\omega)$ when the modulus is understood, and define
\begin{equation}\label{eq:theta0}
  \theta_0(u;\omega)\coloneqq(z;q)_\infty\,(q/z;q)_\infty ,
  \quad
  \tilde\Gamma(u;\omega)=\prod_{m,n\ge 0}\frac{1-q^{m+n+2}/z}{1-q^{m+n}z}
\end{equation}
($\tilde\Gamma$ is the elliptic gamma function at equal fugacities $p=q$). The
four Jacobi theta functions are
\begin{subequations}
    \label{eq:theta-def}
\begin{eqnarray}
  \theta_1(u;\omega)
    \!\!\!&\coloneqq&\!\!\! 2\,e^{i\pi\omega/4}\sin(\pi u)\!\prod_{n\ge1}(1-q^{n})(1-q^{n}z)(1-q^{n}/z)
     \nonumber \\ \!\!\!&=&\!\!\! i\,e^{i\pi\omega/4-i\pi u}\,(q;q)_\infty\,\theta_0(u;\omega)\ , \\
  \theta_2(u;\omega)
    \!\!\!&\coloneqq&\!\!\! 2\,e^{i\pi\omega/4}\cos(\pi u)\!\prod_{n\ge1}(1-q^{n})(1+q^{n}z)(1+q^{n}/z)
     \nonumber \\ \!\!\!&=&\!\!\! e^{i\pi\omega/4-i\pi u}\,(q;q)_\infty\,\theta_0\!\big(u+\tfrac12;\omega\big)\ , \\
  \theta_3(u;\omega)
    \!\!\!&\coloneqq&\!\!\! \prod_{n\ge1}(1-q^{n})\big(1+q^{n-\frac12}z\big)\big(1+q^{n-\frac12}/z\big)
     = (q;q)_\infty\,\theta_0\!\big(u+\tfrac12+\tfrac{\omega}{2};\omega\big)\ , \\
  \theta_4(u;\omega)
    \!\!\!&\coloneqq&\!\!\! \prod_{n\ge1}(1-q^{n})\big(1-q^{n-\frac12}z\big)\big(1-q^{n-\frac12}/z\big)
     = (q;q)_\infty\,\theta_0\!\big(u+\tfrac{\omega}{2};\omega\big)\ .
\end{eqnarray}
\end{subequations}
Here $\theta_1$ is odd in $u$ and $\theta_{2,3,4}$ are even.

% ---------------------------------------------------------------------
\subsection{Theta shifts}
\label{app:theta-shifts}

\paragraph{Quasi-periodicity of $\boldsymbol{\theta}$'s.}
Under a shift by the two periods of the torus,
\begin{equation}\label{eq:theta-quasi-1}
  \theta_1(u+1)=-\theta_1(u)\ ,\quad
  \theta_2(u+1)=-\theta_2(u)\ ,\quad
  \theta_3(u+1)=\theta_3(u)\ ,\quad
  \theta_4(u+1)=\theta_4(u)\ ,
\end{equation}
and, writing $\mu\coloneqq e^{-i\pi(\omega+2u)}=q^{-1/2}z^{-1}$,
\begin{equation}\label{eq:theta-quasi-omega}
  \theta_1(u+\omega)=-\mu\,\theta_1(u)\ ,\quad
  \theta_2(u+\omega)=\mu\,\theta_2(u)\ ,\quad
  \theta_3(u+\omega)=\mu\,\theta_3(u)\ ,\quad
  \theta_4(u+\omega)=-\mu\,\theta_4(u)\ .
\end{equation}

\paragraph{Quasi-periodicity of $\boldsymbol{\tilde{\Gamma}}$.}
The elliptic gamma function satisfies
\begin{equation}\label{eq:Gamma-quasiper}
    \tilde\Gamma(u+\omega;\omega,\omega)
    =\theta_0(u;\omega)\,\tilde\Gamma(u;\omega,\omega)\ ,
    \quad
    \tilde\Gamma(u+1;\omega,\omega)=\tilde\Gamma(u;\omega,\omega)\ ,
\end{equation}
with $\theta_0(u;\omega) = \prod_{k\geq0} (1-e^{2\pi i(u+k\omega)})(1-e^{2\pi i(-u+(k+1)\omega)})$.

\paragraph{Half-period shifts.}
The four functions are permuted by the half-periods. Under $u\to u+\tfrac12$,
\begin{equation}\label{eq:theta-half-1}
  \theta_1(u+\tfrac12)=\theta_2(u)\ ,\quad
  \theta_2(u+\tfrac12)=-\theta_1(u)\ ,\quad
  \theta_3(u+\tfrac12)=\theta_4(u)\ ,\quad
  \theta_4(u+\tfrac12)=\theta_3(u)\ ,
\end{equation}
and, with $\lambda\coloneqq e^{-i\pi(\omega/4+u)}=q^{-1/8}z^{-1/2}$, under $u\to u+\tfrac{\omega}{2}$,
\begin{equation}\label{eq:theta-half-omega}
  \theta_1(u+\tfrac{\omega}{2})=i\lambda\theta_4(u)\ ,\quad
  \theta_2(u+\tfrac{\omega}{2})=\lambda\theta_3(u)\ ,\quad
  \theta_3(u+\tfrac{\omega}{2})=\lambda\theta_2(u)\ ,\quad
  \theta_4(u+\tfrac{\omega}{2})=i\lambda\theta_1(u)\ .
\end{equation}
In particular the $\tfrac{\omega}{2}$-shift exchanges $\theta_2\leftrightarrow\theta_3$
and $\theta_1\leftrightarrow\theta_4$ up to the common factor $\lambda$, which is the
mechanism by which a ratio such as $\theta_4/\theta_2$ becomes a single Jacobi
function after the change of variables of section~\ref{app:theta-to-jacobi}. Finally,
\begin{equation}\label{eq:theta1-prime}
  \left.\frac{d\theta_1}{du}\right|_{u=0} =\pi\,\theta_2(0)\,\theta_3(0)\,\theta_4(0)\ ,
  \qquad
  \frac{d}{du}\left(\frac{\theta_1(u)}{\theta_4(u)}\right)
  =\pi\,\frac{\theta_4(0)^2\,\theta_2(u)\,\theta_3(u)}{\theta_4(u)^2}
\end{equation}
(and the analog with $2\leftrightarrow4$).

% ---------------------------------------------------------------------
\subsection{Theta functions at multiple arguments}
\label{app:theta-mult}

The theta functions obey the three-term (Riemann) identities:
\begin{subequations}\label{eq:theta-threeterm}
\begin{align}
  \theta_1(u{+}v)\,\theta_2(u{-}v)\,\theta_3(0)\,\theta_4(0)
    &= \theta_1(u)\theta_2(u)\theta_3(v)\theta_4(v)+\theta_3(u)\theta_4(u)\theta_1(v)\theta_2(v)\ ,\\
  \theta_1(u{+}v)\,\theta_3(u{-}v)\,\theta_2(0)\,\theta_4(0)
    &= \theta_1(u)\theta_3(u)\theta_2(v)\theta_4(v)+\theta_2(u)\theta_4(u)\theta_1(v)\theta_3(v)\ ,\\
  \theta_1(u{+}v)\,\theta_4(u{-}v)\,\theta_2(0)\,\theta_3(0)
    &= \theta_1(u)\theta_4(u)\theta_2(v)\theta_3(v)+\theta_2(u)\theta_3(u)\theta_1(v)\theta_4(v)\ ,\\
  \theta_2(u{+}v)\,\theta_3(u{-}v)\,\theta_2(0)\,\theta_3(0)
    &= \theta_2(u)\theta_3(u)\theta_2(v)\theta_3(v)-\theta_1(u)\theta_4(u)\theta_1(v)\theta_4(v)\ ,\\
  \theta_3(u{+}v)\,\theta_4(u{-}v)\,\theta_3(0)\,\theta_4(0)
    &= \theta_3(u)\theta_4(u)\theta_3(v)\theta_4(v)-\theta_1(u)\theta_2(u)\theta_1(v)\theta_2(v)\ .
\end{align}
\end{subequations}
Taking $v\to u$ in the
appropriate combinations gives the duplication formulae:
\begin{subequations}\label{eq:theta-duplication}
\begin{align}
  &\theta_1(2u)\,\theta_2(0)\theta_3(0)\theta_4(0) = 2\,\theta_1(u)\theta_2(u)\theta_3(u)\theta_4(u)\ ,\\
  &\theta_2(2u)\,\theta_2(0)\theta_4(0)^2= \theta_2(u)^2\theta_4(u)^2-\theta_1(u)^2\theta_3(u)^2\ ,\\
  &\theta_3(2u)\,\theta_3(0)\theta_4(0)^2 = \theta_3(u)^2\theta_4(u)^2-\theta_1(u)^2\theta_2(u)^2\ ,\\
  &\theta_4(2u)\,\theta_4(0)^3= \theta_4(u)^4-\theta_1(u)^4\ .
\end{align}
\end{subequations}

% ---------------------------------------------------------------------
\subsection{From theta to Jacobi elliptic functions}
\label{app:theta-to-jacobi}

We first introduce the parameter $m$. It is fixed by $\omega$ through
\begin{equation}\label{eq:m-def}
  m\coloneqq\frac{\theta_2(0;\omega)^4}{\theta_3(0;\omega)^4}\ ,
  \quad
  1-m=\frac{\theta_4(0;\omega)^4}{\theta_3(0;\omega)^4}\ ,
\end{equation}
the two being compatible by Jacobi's identity $\theta_3(0)^4=\theta_2(0)^4+\theta_4(0)^4$.
Let
\begin{equation}
K(m)\coloneqq\int_0^{\pi/2}\!\big(1-m\sin^2\phi\big)^{-1/2}d\phi    
\end{equation}
be the complete elliptic integral of the first kind. We have:\footnote{(\ref{eq:K-def}) corrects an evident misprint in
\cite[App.~A]{Benini:2021ano}, where the relation is printed as
$2K(m)=\theta_3(0)^2/\pi$; the factor of $\pi$ belongs in the numerator, being
fixed by $2K(m)\to\pi$ as $\omega\to i\infty$ (so that $\theta_3(0)\to 1$ in the limit).}
\begin{equation}\label{eq:K-def}
  2K(m)=\pi\,\theta_3(0;\omega)^2\ ,
  \quad
  \omega=i\,\frac{K(1-m)}{K(m)}\ .
\end{equation}%
The Jacobi elliptic functions are then defined by the dictionary
\begin{subequations}\label{eq:theta-to-jacobi}
\begin{align}
  \sn\!\big(2K(m)\,u,\,m\big) &\coloneqq m^{-1/4}\,\frac{\theta_1(u;\omega)}{\theta_4(u;\omega)} \ ,\\
  \cn\!\big(2K(m)\,u,\,m\big) &\coloneqq \left(\frac{1-m}{m}\right)^{1/4}\frac{\theta_2(u;\omega)}{\theta_4(u;\omega)} \ ,\\
  \dn\!\big(2K(m)\,u,\,m\big) &\coloneqq (1-m)^{1/4}\,\frac{\theta_3(u;\omega)}{\theta_4(u;\omega)} \ .
\end{align}
\end{subequations}
Equivalently, with the natural variable $\zeta\coloneqq 2K(m)\,u$, the functions
$\sn(\zeta,m),\cn(\zeta,m)$ have real period $4K(m)$, while $\dn(\zeta,m)$ has
period $2K(m)$; in $u$ this is period $2$ for $\sn,\cn$ and period $1$ for
$\dn$. We suppress the second argument, $\cn(\zeta)\equiv\cn(\zeta,m)$, when $m$
is fixed. Inverting one entry of \eqref{eq:theta-to-jacobi} gives the reciprocal
relation used repeatedly in the main text,
\begin{equation}\label{eq:theta-ratio-nc}
  \frac{\theta_4(u;\omega)}{\theta_2(u;\omega)}
  =\left(\frac{1-m}{m}\right)^{1/4}\frac{1}{\cn\!\big(2K(m)u\big)}\ ,
\end{equation}
i.e.\ $\theta_4/\theta_2$ is a constant times $\operatorname{nc}\coloneqq\tfrac 1 \cn$. Crucially, in a
ratio of such factors, the constant prefactor $((1-m)/m)^{1/4}$ cancels.

% ---------------------------------------------------------------------
\subsection{Jacobi (Pythagorean) identities}
\label{app:jacobi-pyth}

The Jacobi functions satisfy the three quadratic relations
\begin{equation}\label{eq:jacobi-pyth}
  \sn^2(x)+\cn^2(x)=1\ ,\quad
  \dn^2(x)+m\,\sn^2(x)=1\ ,\quad
  \dn^2(x)-m\,\cn^2(x)=1-m\ .
\end{equation}
In particular every $\cn^2$ and $\dn^2$ can be eliminated in favor of $\sn^2$,
\begin{equation}\label{eq:reduce-to-sn}
  \cn^2(x)=1-\sn^2(x)\ ,\quad
  \dn^2(x)=1-m\,\sn^2(x)\ ,
\end{equation}
which is what reduces the multiple-angle numerators
\eqref{eq:jacobi-double}--\eqref{eq:jacobi-double-sn} to polynomials in $\sn^2(x)$.

% ---------------------------------------------------------------------
\subsection{Jacobi addition, multiple-angle and shift formulae}
\label{app:jacobi-addmult}

There are a few special values of the above functions: $\sn(0)=0,\ \cn(0)=\dn(0)=1$ and $\sn(K)=1,\ \cn(K)=0,\ \dn(K)=\sqrt{1-m}$.
With the common denominator $D(x,y)\coloneqq1-m\,\sn^2(x)\sn^2(y)$, the addition formulae are:
\begin{subequations}\label{eq:jacobi-add}
\begin{align}
  \sn(x{+}y)&=\frac{\sn(x)\cn(y)\dn(y)+\sn(y)\cn(x)\dn(x)}{D(x,y)}\ ,\\
  \cn(x{+}y)&=\frac{\cn(x)\cn(y)-\sn(x)\sn(y)\dn(x)\dn(y)}{D(x,y)}\ ,\\
  \dn(x{+}y)&=\frac{\dn(x)\dn(y)-m\,\sn(x)\sn(y)\cn(x)\cn(y)}{D(x,y)}\ .
\end{align}
\end{subequations}
A convenient consequence (the product form used in the main text) is:
\begin{equation}\label{eq:cn-product}
  \cn(x{+}y)\,\cn(x{-}y)=\frac{\cn^2(y)-\sn^2(x)\dn^2(y)}{1-m\,\sn^2(x)\sn^2(y)}\ .
\end{equation}
The double-argument (multiple-angle) formulae follow from
\eqref{eq:jacobi-add} at $y=x$, with $D(x,x)=1-m\,\sn^4(x)$:
\begin{subequations}\label{eq:jacobi-double}
\begin{align}
  \sn(2x)&=\frac{2\,\sn(x)\cn(x)\dn(x)}{1-m\,\sn^4(x)}\ ,\\
  \cn(2x)&=\frac{\cn^2(x)-\sn^2(x)\dn^2(x)}{1-m\,\sn^4(x)}\ ,\\
  \dn(2x)&=\frac{\dn^2(x)-m\,\sn^2(x)\cn^2(x)}{1-m\,\sn^4(x)}\ .
\end{align}
\end{subequations}
Using the Pythagorean relations \eqref{eq:jacobi-pyth} the numerators reduce to
polynomials in $\sn^2(x)$ alone,
\begin{equation}\label{eq:jacobi-double-sn}
  \cn(2x)=\frac{1-2\sn^2(x)+m\,\sn^4(x)}{1-m\,\sn^4(x)}\ ,
  \quad
  \dn(2x)=\frac{1-2m\,\sn^2(x)+m\,\sn^4(x)}{1-m\,\sn^4(x)}\ .
\end{equation}
Finally, the quarter-period shift by $K\equiv K(m)$ acts as (with $k'\coloneqq\sqrt{1-m}$)
\begin{equation}\label{eq:jacobi-shift}
  \sn(x+K)=\frac{\cn(x)}{\dn(x)}\ ,\quad
  \cn(x+K)=-k'\,\frac{\sn(x)}{\dn(x)}\,\quad
  \dn(x+K)=\frac{k'}{\dn(x)}\ .
\end{equation}

%%%%%%%%%%%%%%%%%%%%%%%%%%%%%%%%%%%%%%%%%%%%%%%%%%%%%%%
\section{Closed-form $B_2\cong C_2$ BAE solutions}
\label{sec:closed-form}
%%%%%%%%%%%%%%%%%%%%%%%%%%%%%%%%%%%%%%%%%%%%%%%%%%%%%%%

Throughout this appendix we abbreviate $R\coloneqq\sqrt{m(1-m)}$, so that
$\big({-}(m-1)m\big)^{3/2}=R^3$.

\subsection{$x=1/2$}
\label{sub:x12}
The coefficients of the cubic \eqref{eq:B2cubic} are
{\small
\begin{subequations}\label{eq:coeffs-x12}
\begin{align}
A_3={}&(m-1)^2m^2c_1^8
 -4\sqrt{m}\,R^3\left(3c_2^2-c_3^2\sqrt{m}\right)c_1^6 \notag\\
 &-2(m-1)m^2\left(27c_2^4-22c_3^2\sqrt{m}\,c_2^2+3c_3^4m\right)c_1^4 \notag\\
 &+4m^{3/2}R\left(-27c_2^6+39c_3^2\sqrt{m}\,c_2^4-13c_3^4mc_2^2+c_3^6m^{3/2}\right)c_1^2 \notag\\
 &+m^2\left(9c_2^4-10c_3^2\sqrt{m}\,c_2^2+c_3^4m\right)^2\ ,\\[0.8em]
A_2={}&-8m\biggl[\,\frac{R^3}{m}\left(-\sqrt{m}\,c_2^2+c_3m^{1/4}c_2+c_3^2\right)c_1^6 \notag\\
 &\qquad+(m-1)\sqrt{m}\Bigl(-8\sqrt{m}\,c_2^4+11c_3m^{1/4}c_2^3
   +c_3^2(5m+11)c_2^2 \notag\\
 &\qquad\qquad\qquad\quad-c_3^3m^{3/4}c_2-2c_3^4\sqrt{m}\Bigr)c_1^4 \notag\\
 &\qquad+R\Bigl(-21\sqrt{m}\,c_2^6+39c_3m^{1/4}c_2^5+c_3^2(40m+39)c_2^4 \notag\\
 &\qquad\qquad+2c_3^3m^{3/4}c_2^3-c_3^4\sqrt{m}(7m+12)c_2^2
   -c_3^5m^{5/4}c_2+c_3^6m\Bigr)c_1^2 \notag\\
 &\qquad+18mc_2^8-45c_3m^{3/4}c_2^7-c_3^2\sqrt{m}(79m+45)c_2^6
   -23c_3^3m^{5/4}c_2^5 \notag\\
 &\qquad+2c_3^4m(16m+9)c_2^4+5c_3^5m^{7/4}c_2^3
   -c_3^6m^{3/2}(3m+5)c_2^2-c_3^7m^{9/4}c_2\biggr]\ ,\\[0.8em]
A_1={}&-16\biggl[(m-1)m\left(mc_2^4-4c_3m^{3/4}c_2^3-3c_3^2\sqrt{m}\,c_2^2
   +2c_3^3m^{1/4}c_2+c_3^4\right)c_1^4 \notag\\
 &\qquad+2c_2\sqrt{m}\,R\Bigl(2mc_2^5-13c_3m^{3/4}c_2^4-5c_3^2\sqrt{m}(m+2)c_2^3 \notag\\
 &\qquad\qquad\qquad\quad+6c_3^3m^{1/4}c_2^2+c_3^4(4m+3)c_2+c_3^5m^{3/4}\Bigr)c_1^2 \notag\\
 &\qquad+c_2^2m\Bigl(-4mc_2^6+38c_3m^{3/4}c_2^5+c_3^2\sqrt{m}(44m+29)c_2^4 \notag\\
 &\qquad\qquad+18c_3^3(m-1)m^{1/4}c_2^3-c_3^4(9m^2+28m+9)c_2^2 \notag\\
 &\qquad\qquad-2c_3^5m^{3/4}(2m+5)c_2+c_3^6m^{3/2}\Bigr)\biggr]\ ,\\[0.8em]
A_0={}&64c_2^2c_3\sqrt{m}\left(2m^{3/4}c_2^3+c_3\sqrt{m}\,c_2^2
   -2c_3^2m^{1/4}c_2-c_3^3\right) \notag\\
 &\quad\times\left(-c_1^2R+2c_2c_3m^{5/4}+2c_2^2\sqrt{m}\right)\ .
\end{align}
\end{subequations}}

\subsection{$x=\omega/2$}
\label{sub:xomega2}
{\small
\begin{subequations}\label{eq:coeffs-xom2}
\begin{align}
A_3={}&(m-1)^2m^2c_1^8+4\sqrt{m}\,R^3\left(c_2^2+c_3^2\sqrt{m}\right)c_1^6 \notag\\
 &-2(m-1)m^2\left(3c_2^4+2c_2^2c_3^2\sqrt{m}+3c_3^4m\right)c_1^4 \notag\\
 &+4m^{3/2}R\left(c_2^2-c_3^2\sqrt{m}\right)^2\left(c_2^2+c_3^2\sqrt{m}\right)c_1^2
 +m^2\left(c_2^2-c_3^2\sqrt{m}\right)^4\ ,\\[0.8em]
A_2={}&-8m\biggl[(m-1)R\left(c_2^2\sqrt{m}+3c_2c_3m^{1/4}+c_3^2\right)c_1^6 \notag\\
 &\qquad+(m-1)\sqrt{m}\Bigl(4c_2^4\sqrt{m}+11c_2^3c_3m^{1/4}+5c_2^2c_3^2(m+1) \notag\\
 &\qquad\qquad\qquad\quad+11c_2c_3^3m^{3/4}+4c_3^4\sqrt{m}\Bigr)c_1^4 \notag\\
 &\qquad-R\Bigl(5c_2^6\sqrt{m}+13c_2^5c_3m^{1/4}+7c_2^4c_3^2-2c_2^3c_3^3m^{3/4} \notag\\
 &\qquad\qquad+7c_2^2c_3^4m^{3/2}+13c_2c_3^5m^{5/4}+5c_3^6m\Bigr)c_1^2 \notag\\
 &\qquad-\sqrt{m}\left(c_2^2-c_3^2\sqrt{m}\right)^2
   \Bigl(2c_2^4\sqrt{m}+5c_2^3c_3m^{1/4} \notag\\
 &\qquad\qquad\qquad\qquad\qquad+3c_2^2c_3^2(m+1)+5c_2c_3^3m^{3/4}
   +2c_3^4\sqrt{m}\Bigr)\biggr]\ ,\\[0.8em]
A_1={}&16\biggl[-(m-1)m\left(c_2^4m+8c_2^3c_3m^{3/4}+15c_2^2c_3^2\sqrt{m}
   +8c_2c_3^3m^{1/4}+c_3^4\right)c_1^4 \notag\\
 &\qquad+2\sqrt{m}\,R\Bigl(2c_2^6m+13c_2^5c_3m^{3/4}+c_2^4c_3^2\sqrt{m}(5m+26) \notag\\
 &\qquad\qquad\quad+20c_2^3c_3^3m^{1/4}(m+1)+c_2^2c_3^4(26m+5) \notag\\
 &\qquad\qquad\quad+13c_2c_3^5m^{3/4}+2c_3^6\sqrt{m}\Bigr)c_1^2 \notag\\
 &\qquad+m\left(c_2m^{1/4}+c_3\right)^2\Bigl(4c_2^6\sqrt{m}+14c_2^5c_3m^{1/4}
   +c_2^4c_3^2(9-4m) \notag\\
 &\qquad\qquad\qquad\qquad\quad-10c_2^3c_3^3m^{3/4}+c_2^2c_3^4\sqrt{m}(9m-4) \notag\\
 &\qquad\qquad\qquad\qquad\quad+14c_2c_3^5m^{5/4}+4c_3^6m\Bigr)\biggr]\ ,\\[0.8em]
A_0={}&64c_2c_3m^{1/4}\left(c_2m^{1/4}+c_3\right)^2
   \left(2c_2^2\sqrt{m}+5c_2c_3m^{1/4}+2c_3^2\right) \notag\\
 &\quad\times\left(c_1^2R+2c_2^2\sqrt{m}+2c_2c_3m^{1/4}(m+1)+2c_3^2m\right)\ .
\end{align}
\end{subequations}}

\subsection{$x=(1+\omega)/2$}
\label{sub:x1plusomega2}
{\small
\begin{subequations}\label{eq:coeffs-x1om2}
\begin{align}
A_3={}&(m-1)^2m^2c_1^8+4\sqrt{m}\,R^3\left(c_2^2-3c_3^2\sqrt{m}\right)c_1^6 \notag\\
 &-2(m-1)m^2\left(3c_2^4-22c_3^2\sqrt{m}\,c_2^2+27c_3^4m\right)c_1^4 \notag\\
 &-4m^{3/2}R\left(-c_2^6+13c_3^2\sqrt{m}\,c_2^4-39c_3^4mc_2^2
   +27c_3^6m^{3/2}\right)c_1^2 \notag\\
 &+m^2\left(c_2^4-10c_3^2\sqrt{m}\,c_2^2+9c_3^4m\right)^2\ ,\\[0.8em]
A_2={}&8m\biggl[(m-1)R\left(\sqrt{m}\,c_2^2+c_3m^{1/4}c_2-c_3^2\right)c_1^6 \notag\\
 &\qquad+(m-1)\sqrt{m}\Bigl(2\sqrt{m}\,c_2^4+c_3m^{1/4}c_2^3-c_3^2(11m+5)c_2^2 \notag\\
 &\qquad\qquad\qquad\quad-11c_3^3m^{3/4}c_2+8c_3^4\sqrt{m}\Bigr)c_1^4 \notag\\
 &\qquad+R\Bigl(-\sqrt{m}\,c_2^6+c_3m^{1/4}c_2^5+c_3^2(12m+7)c_2^4
   -2c_3^3m^{3/4}c_2^3 \notag\\
 &\qquad\qquad-c_3^4\sqrt{m}(39m+40)c_2^2-39c_3^5m^{5/4}c_2+21c_3^6m\Bigr)c_1^2 \notag\\
 &\qquad-18m^2c_3^8+45c_2m^{9/4}c_3^7+c_2^2m^{3/2}(45m+79)c_3^6
   +23c_2^3m^{7/4}c_3^5 \notag\\
 &\qquad-2c_2^4m(9m+16)c_3^4-5c_2^5m^{5/4}c_3^3
   +c_2^6\sqrt{m}(5m+3)c_3^2+c_2^7m^{3/4}c_3\biggr]\ ,\\[0.8em]
A_1={}&-16\biggl[(m-1)m\Bigl(mc_2^4+2c_3m^{3/4}c_2^3-3c_3^2\sqrt{m}\,c_2^2 \notag\\
 &\qquad\qquad\qquad\quad-4c_3^3m^{1/4}c_2+c_3^4\Bigr)c_1^4 \notag\\
 &\qquad+2c_3\sqrt{m}\,R\Bigl(m^{3/4}c_2^5+c_3\sqrt{m}(3m+4)c_2^4
   +6c_3^2m^{5/4}c_2^3 \notag\\
 &\qquad\qquad\qquad\quad-5c_3^3(2m+1)c_2^2-13c_3^4m^{3/4}c_2
   +2c_3^5\sqrt{m}\Bigr)c_1^2 \notag\\
 &\qquad+c_3^2m\Bigl(\sqrt{m}\,c_2^6-2c_3m^{1/4}(5m+2)c_2^5
   -c_3^2(9m^2+28m+9)c_2^4 \notag\\
 &\qquad\qquad-18c_3^3(m-1)m^{3/4}c_2^3+c_3^4\sqrt{m}(29m+44)c_2^2 \notag\\
 &\qquad\qquad+38c_3^5m^{5/4}c_2-4c_3^6m\Bigr)\biggr]\ ,\\[0.8em]
A_0={}&-64c_2c_3^2m^{1/4}\left(m^{3/4}c_2^3+2c_3\sqrt{m}\,c_2^2
   -c_3^2m^{1/4}c_2-2c_3^3\right) \notag\\
 &\quad\times\left(-c_1^2R+2c_2c_3m^{1/4}+2c_3^2m\right)\ .
\end{align}
\end{subequations}}

\clearpage
\small
\bibliography{baes}
\bibliographystyle{at}

\end{document}